\documentclass[aps,11pt,prd,groupedaddress,nofootinbib,notitlepage,eqsecnum,preprintnumbers]{revtex4-2}

\usepackage[utf8]{inputenc}
\usepackage{hyperref}
\usepackage{xcolor}
\usepackage{graphicx}
\usepackage{amsmath,amssymb}
\usepackage{bm}
\usepackage{comment}
\usepackage[shortlabels]{enumitem}
\usepackage{appendix}
\usepackage{ulem}
\usepackage{tensor}

\begin{document}
\title{Gravitational waves induced by scalar-tensor mixing}

\author{\textsc{Pritha Bari$^{a,b}$}}
    \email{{pritha.bari}@{pd.infn.it}}
\author{\textsc{Nicola Bartolo$^{a,b,c}$}}
    \email{{nicola.bartolo}@{pd.infn.it}}
\author{\linebreak\textsc{Guillem Dom\`enech$^{d}$}}
\email{{guillem.domenech}@{itp.uni-hannover.de}}
\author{\textsc{Sabino Matarrese$^{a,b,c,e}$}}
    \email{{sabino.matarrese}@{pd.infn.it}}

\affiliation{$^a$ Dipartimento di Fisica e Astronomia ``G. Galilei'', Universit\`a degli Studi di Padova, via Marzolo 8, I-35131 Padova, Italy}
\affiliation{$^b$ INFN, Sezione di Padova, via Marzolo 8, I-35131 Padova, Italy}
\affiliation{$^c$ INAF, Osservatorio Astronomico di Padova, Vicolo dell’Osservatorio 5, I-35122 Padova, Italy}
\affiliation{$^d$ Institute for Theoretical Physics, Leibniz University Hannover, Appelstraße 2, 30167 Hannover, Germany.}
\affiliation{$^e$ Gran Sasso Science Institute, Viale F. Crispi 7, I-67100 L'Aquila, Italy}

\begin{abstract}
This paper explores the physics of second-order gravitational waves (GWs) induced by scalar-tensor perturbation interactions in the radiation-dominated Universe. We investigate the distinctive signatures of these GWs and their detectability compared to scalar-induced GWs. Unlike scalar-scalar induced GWs, scalar-tensor induced GWs do not present resonances or a logarithmic running in the low frequency tail in the case of peaked primordial spectra. But, interestingly, they partly inherit any primordial parity violation of tensor modes. We find that chirality in primordial GWs can lead to distinguishing effects in scalar-tensor induced GWs in the ultraviolet (UV) region. We also address a potential divergence in our GWs and explore possible solutions. This study contributes to our understanding of GWs in the early Universe and their implications for cosmology and GWs detection.
\end{abstract}

\maketitle

\section{Introduction \label{sec:intro}}
Pulsar timing array (PTA) collaborations, NANOGrav, EPTA/InPTA, PPTA, and CPTA,
have presented evidence \cite{NANOGrav:2023hvm,2023arXiv230616227A,Xu:2023wog,Reardon:2023zen} for an isotropic stochastic
gravitational wave background (SGWB). This  event represents a pivotal milestone in the field of physics, possibly heralding the advent of gravitational wave astronomy in the early Universe. This marks the second major breakthrough following the groundbreaking detection of gravitational waves (GWs) from a binary black hole merger \cite{LIGOScientific:2016aoc} in modern cosmology, which led to an increased focus within the scientific community on designing more precise GWs observations and developing robust theoretical predictions. Among the diverse range of GWs sources \cite{Buonanno:2014aza,phinney2001practical,Guzzetti:2016mkm,Kamionkowski:1993fg,Caprini:2018mtu}, it is plausible that a cosmic background of GWs permeates the universe. This GWs background represents a potential smoking gun of inflation \cite{Guth:1980zm,Starobinsky:1980te,Lyth:1998xn} and encapsulates invaluable information about the early Universe \cite{Guzzetti:2016mkm,Caprini:2018mtu,Boyle:2007zx,PhysRevD.101.123533,An:2020fff,2022arXiv220312714B,PhysRevD.104.063513},
as GWs barely interact with intervening matter. The search for the cosmic GWs background has always been a central focus in cosmology, pursued through avenues such as B-mode polarization of the cosmic microwave background (CMB) and direct detection using interferometers \cite{Maggiore:1999vm,Kamionkowski:2015yta,Guzzetti:2016mkm,Watanabe:2006qe,sakamoto2022probing, 
CMB-S4:2020lpa,Campeti:2020xwn,Flauger:2020qyi,LiteBIRD:2022cnt}.

Gravitational waves  are metric tensor perturbations that can arise from vacuum fluctuations during inflation, particularly in the context of single-field slow-roll inflation. However, within the broader framework of inflationary models, classical production mechanisms for GWs have also been explored \cite{Guzzetti:2016mkm}. Regarding non-inflationary mechanisms, a strong GWs signal can be produced by topological defects \cite{Vilenkin:1981bx,Figueroa:2012kw}, or phase-transitions \cite{Kosowsky:1991ua,Kamionkowski:1993fg}. GWs are also induced by primordial fluctuations after inflation. While at linear order in cosmological perturbation theory scalar and tensor fluctuations decouple, it is no longer the case at higher orders. For instance, the product of two spatial gradients of scalar fluctuations has a non-vanishing transverse and traceless projection that sources tensor fluctuations.  This has led to extensive research, with a significant focus on scalar perturbations as the seeds for second-order tensor perturbations, primarily because scalar perturbations dominate at the linear level.  Such a topic has indeed a long history, starting from~\cite{1967PThPh..37..831T,Matarrese:1993zf} and has been later developed in, e.g.  \cite{Matarrese_1998,Domenech:2020kqm,Espinosa_2018,Matarrese:1996pp,Bartolo:2007vp,Baumann_2007,Inomata:2019yww,Yuan_2020, Ananda_2007,Kohri:2018awv,Domenech:2019quo,2021Univ....7..398D}.

This intriguing signal from PTA results has sparked various interpretations, including inspirals of supermassive black hole binaries \cite{2023arXiv230616220A}, phase transitions \cite{2023arXiv230617205A}, and cosmic strings \cite{2023arXiv230617390K}. Among these possibilities, scalar-induced gravitational waves (SIGWs) have garnered attention and are being considered in many analyses \cite{2023arXiv230702399F,2023arXiv230617834I,2023arXiv230616227A,2023arXiv230617149F,NANOGrav:2023hvm}. These “scalar-scalar induced GWs” have also been used as probes for primordial black holes \cite{Saito_2009,Bartolo:2018evs,Bartolo:2019zvb,Garcia-Bellido:2017aan}. Analytic integral solutions for second-order GWs, induced by various quadratic combinations of cosmological perturbations, during both matter and radiation-dominated epochs, are provided in \cite{Gong:2019mui}.

Curvature (scalar) perturbations are tightly constrained by the latest Planck data on cosmic microwave background (CMB) scales, with an amplitude of $A_S=2.1\times 10^{-9}$ \cite{Planck:2018vyg}. On the other hand, the recent joint analysis using Planck2018, BICEP2/Keck2015-2018, and LIGO-Virgo-KAGRA data has placed a tight constraint on $r$ ($<0.028$) \cite{2022arXiv220800188G}, where $r=A_T/A_S$ is the so-called tensor-to-scalar perturbation ratio and $A_T$ is the amplitude of the tensor power-spectrum at CMB scales. However, these constraints would not be applicable to smaller scales without large extrapolations, leaving the amplitude of both scalar and tensor fluctuations mainly unconstrained. In fact, on small scales, there exist various well-motivated mechanisms in the literature that can generate such large perturbations, both scalar perturbations as mentioned in the induced gravitational waves scenario related to primordial black holes formation, and tensor perturbations in the context of induced matter perturbations \cite{PhysRevLett.129.091301}.

While the general expectation is that scalar-scalar induced GWs dominate the secondary GWs signal, the product of scalar-tensor and tensor-tensor also source GWs at second order. Scalar-tensor interactions may also play an important role in the wave-optics limit of the GW background \cite{2022arXiv221005718G} (see e.g. Refs.~\cite{Nakamura:1999uwi,Takahashi:2003ix,Cusin:2019rmt} for wave-optics effects in astrophysics). It is thus important to systematically study the physics of these additional GWs signals and investigate their distinct signatures and any chance at detecting them.
In this paper, we focus on gravitational waves induced by interactions between scalar and tensor perturbations in the radiation-dominated Universe, by treating such scalar-tensor interactions as a source to GWs in the early Universe. 
Understanding these interactions in cosmology might be important for a general description of cosmic GWs propagating through an inhomogenous Friedmann-Lemaître-Robertson-Walker (FLRW) universe.


Scalar-tensor induced GWs have been explored before in Refs.~\cite{Gong:2019mui, Chang:2022vlv}. In particular, Ref.~\cite{Chang:2022vlv} pointed out that, for Dirac delta primordial scalar and tensor spectra, scalar-tensor induced GWs may dominate the high-frequency regime of the total induced GWs. However, there is no general answer regarding the detectability of this signal compared to scalar-induced GWs. While GWs induced by scalar-tensor interactions are novel and intriguing in its own right, it is important to address this latter point. We improve and clarify previous studies on scalar-tensor induced GWs in several ways:
\begin{itemize}
\item We derive general formulae allowing for parity violation of primordial tensor modes. Chiral GWs violate parity symmetry and exhibit distinct handedness or helicity, where the wave's behaviour differs between left-handed and right-handed polarisations, while non-chiral ones, maintaining parity symmetry, lack this distinguishing characteristic. For models leading to primordial chiral GWs see Refs.~\cite{Obata:2016tmo,Bartolo:2017szm,Bartolo:2018elp} and the references therein. Also see Ref.~\cite{Komatsu:2022nvu} for a nice overview on the topic. We show how such parity violation is partly inherited by scalar-tensor induced GWs and discuss their distinct signatures. We also take into account finite width of primordial spectra.
\item We identify possible  infra-red divergences in the scalar-tensor induced GWs which can be traced to the fact that a constant scalar mode is naively allowed to source tensors. We propose a (rough) procedure to remove such divergences motivated by the existence of a locally inertial frame.
\item We find that for peaked primordial spectra, the low frequency tail of scalar-tensor induced GWs does not have a logarithmic running, in contrast to scalar-scalar induced GWs. There is also no resonant peak.
\end{itemize}
 
This paper is structured as follows. The next section presents the evolution equation for second-order induced GWs when the primordial perturbations are scalar and tensor. Section \ref{sec:2} discusses the general form of the power-spectrum of scalar-tensor induced GWs and explores properties of the kernel. In Section \ref{sec:3}, an example is provided with peaked scalar and tensor perturbations, considering both chiral and non-chiral waves in the primordial tensor sector. Section \ref{sec:4} explores potential strategies to circumvent the divergence in the ultraviolet (UV) region. Finally, Section \ref{sec:5} presents the conclusions and summarizes the findings of the study.

In our analysis, we assume that scalar perturbations are more pronounced on small scales than tensor perturbations, thus neglecting tensor-tensor interactions. Under the condition $A_T < A_S$, we investigate whether the scalar-tensor contribution to GWs can be distinguished from scalar-induced GWs. We identify a distinct signature of this interaction related to the chirality of GWs. Specifically, while scalar-induced GWs do not possess chirality, we demonstrate that chiral GWs, after interacting with scalar perturbations, exhibit different behaviours in the ultraviolet (UV) region for left- and right-handed waves. This characteristic can be valuable in distinguishing them from scalar-induced GWs.
We assume   $c=\hslash=M_{\rm pl}=1$ throughout this paper.

\section{Tensor modes induced by scalar-tensor interactions\label{sec:1}}

We consider a perturbed flat FLRW space-time, in the Poisson gauge, in which the metric is described by
\begin{equation}\label{eq:metric}
    ds^2= -e^{2\Phi}dt^2+a^2 e^{-2\Psi}\left(e^{\gamma}\right)_{ij}dx^idx^j,
\end{equation}
where $t$ is the coordinate time, $a(t)$ the scale factor, $\Phi$ and $\Psi$ are scalar perturbations and $\gamma_{ij}$ tensor perturbations. We neglect vector perturbations for simplicity. To get an evolution equation of the tensor perturbations sourced by a mixing of first order scalar and tensor ones, we focus on the trace-less part of the ij-th Einstein equation. We start with a general approach: we use the ADM formulation and compute the full non-linear equations, which we present in Appendix \ref{App-st1}. We then expand the non-linear equations only at linear order in $\gamma_{ij}$, keeping the full dependence on $\Psi$ and $\Phi$. In doing so, we obtain
\begin{align}\label{main}
     &\ddot{{\gamma}}_{ij}-\dot{\Phi}e^{\Phi}\dot{{\gamma}}_{ij}+3(H-\dot{\Psi})\dot{{\gamma}}_{ij}-{e^{2(\Phi+\Psi)}}{a^{-2}}\nabla^2 {\gamma}_{ij}=  {2}{a^{-2}}e^{2\Phi}\,\widehat{TT}^{ab}_{ij}\left[e^{2\Psi}\left(\mathcal{S}^{ss}_{ab}+\mathcal{S}^{sst}_{ab}\right)+\mathcal{S}^{m}_{ab}\right], 
\end{align}
where $\mathcal{S}^{ss}_{ij}$ and $\mathcal{S}^{sst}_{ij}$ are respectively given by
\begin{align}
\mathcal{S}^{ss}_{ij}=\Phi_{,ij}+\Phi_{,i}\Phi_{,j}-\Psi_{,i}\Psi_{,j}-\Psi_{,ij}+\Phi_{,i}\Psi_{,j}+\Psi_{,i}\Phi_{,j}
\end{align}
and
\begin{align}\label{S}
   \mathcal{S}^{sst}_{ij}&=-\left[\delta^{kl}{\gamma}_{ki}(\Phi-\Psi)_{,jl}+\frac{1}{2}\delta^{kl}{{\gamma}_{ki}}_{,j}(\Phi-\Psi)_{,l}+\frac{1}{2}\delta^{kl}{\gamma}_{kj,i}(\Phi-\Psi)_{,l}\right.\nonumber\\&\left.+\delta^{kl}{\gamma}_{ik}\Psi_{,j}\Phi_{,l}+\delta^{kl}{\gamma}_{ik}\Phi_{,j}\Psi_{,l}+\delta^{kl}{\gamma}_{ik}\Phi_{,j}\Phi_{,l}-\delta^{kl}{\gamma}_{ik}\Psi_{,j}\Psi_{,l}\right]\,,
\end{align}
and the contribution from the energy-momentum tensor of a perfect fluid matter, namely
\begin{align}
\mathcal{S}^{m}_{ij}=\partial_i V\partial_j V\,,
\end{align}
where $V$ is the scalar component of the linear perturbation in the spatial velocity of the perfect fluid. In Eq.~\eqref{main}, $\widehat{TT}^{ab}_{ij}$ is the transverse-traceless projector which can be found, e.g., in \cite{Domenech:2021ztg}, and latin indices are raised and lowered with the spatial background metric, which at leading order is $\delta_{ij}$. We present all the details in Appendix \ref{App-st1}.

Let us focus on the leading order terms in scalar-tensor interactions. In this case, Eq.~\eqref{main} becomes, in conformal time $d\eta=dt/a$,
\begin{equation}\label{st-eq}
    {{\gamma}}''_{ij}+2{\cal H}{{\gamma}}'_{ij}-\nabla^2 {\gamma}_{ij}={4\Phi}\nabla^2 {\gamma}_{ij}+4{\Phi}'{{\gamma}}'_{ij}\,,
\end{equation}
where $'=d/d\eta$ and we used that in the absence of anisotropic stress we have  $\Phi=\Psi$.
In what follows we will treat the right hand side of \eqref{st-eq} as a source term. To do so, we will follow a perturbative expansion and split
\begin{align}
\gamma_{ij}=\gamma^{(0)}_{ij}+\gamma^{(1)}_{ij}+...\,,
\end{align}
where $\gamma^{(0)}_{ij}$ is the solution to the homogeneous equation and $\gamma^{(1)}_{ij}$ are the scalar-tensor induced GWs. Note that inside such source term in \eqref{st-eq} there is a bare $\Phi$, namely without gradients or time derivatives. We will later show that this term is problematic for sufficiently flat primordial scalar spectrum and leads to potential divergences.

Now, we decompose scalar and tensor perturbations into their Fourier modes, we respectively have
\begin{align}\label{eq:expansionfouriertensor}
     {\gamma}_{ij}(\bm{x},\eta)&= \frac{1}{(2\pi)^3} \int d^3{k} \, e^{i\bm{k}.\bm{x}} {\gamma}_{\bm{k},\,\sigma} (\eta)\,\epsilon^{\sigma}_{ij}(\bm{\hat{k}})\,,\\
     \Phi(\bm{x},\eta)&=\frac{1}{(2\pi)^3} \int d^3{k} \,e^{i\bm{k}.\bm{x}} \Phi_{\bm{k}}(\eta)\,,\label{eq:expansionfourierscalar}
\end{align}
where $\epsilon^{\sigma}_{ij}(\bm{\hat{k}})$ are the (transverse-traceless) polarization tensors. To be compatible with the reality condition of the Fourier expansion \eqref{eq:expansionfouriertensor}, we work with left and right handed polarization tensors where $\sigma=\rm R, \rm L$ represents the polarization index. We also choose the normalisation given by
\begin{align}
\epsilon^{\sigma*}_{ij}(\bm{\hat{k}})\epsilon^{\sigma'ij}(\bm{\hat{k}})=2\delta_{\sigma \sigma'}\,.
\end{align}
In Fourier space Eq. \eqref{st-eq} becomes
\begin{align}\label{st-eq-eta-fins}
&{\gamma}_{\bm{k},\lambda}''+2\mathcal{H}{\gamma}_{\bm{k},\lambda}'+k^2{\gamma}_{\bm{k},\lambda}=\mathcal{S}_{st,\lambda}(\bm{k},\eta) ,
\end{align}
where we defined
\begin{align}\label{Sstlambda}
\mathcal{S}_{st,\,\lambda}(\bm{k},\eta) =-2\sum_\sigma\int \frac{d^3k_1}{(2\pi)^3} &\Phi^{\rm p}_{\bm{k}-\bm{k}_1}{\gamma}^{{\rm p},\,\sigma}_{\bm{k}_1}\epsilon^{\sigma}_{ij}(\bm{\hat{k}}_1) \epsilon^{ij*}_{\lambda}(\bm{\hat{k}}) \nonumber\\&\times\left[k_1^2 T_{{\gamma}}(k_1\eta) T_\Phi(c_s|\bm{k}-\bm{k}_1|\eta)-T'_{{\gamma}}(k_1\eta) T'_\Phi(c_s|\bm{k}-\bm{k}_1|\eta)\right]\,,
\end{align}
$c_s$ is the sound speed of scalar fluctuations, which for radiation domination is $c_s=1/\sqrt{3}$, and we abused notation and dropped the superscript “1” in $\gamma^{(1)}_{\bm k}$ in the left hand side of \eqref{st-eq-eta-fins} and used the homogeneous solutions for $\gamma^{(0)}_{\bm k}$ and $\Phi_{\bm k}$ inside the integrand. We split such homogeneous solutions into a primordial (initial) value and a transfer function, namely
\begin{align}\label{eq:gammaT}
     {\gamma}^{(0)}_{\bm{k},\,\sigma}(\eta)={\gamma}^{\rm p}_{\bm{k},\,\sigma} \,T_{{\gamma}}(k\eta)
     &={\gamma}^{\rm p}_{\bm{k},\,\sigma} \sqrt{\frac{\pi}{2k\eta}} J_{1/2}(k\eta)\,,
\end{align}
and
\begin{align}\label{eq:gammaT}
     \Phi_{\bm{k}}(\eta)&=\Phi^{\rm p}_{\bm{k}} \,T_\Phi(c_sk\eta)=\Phi^{\rm p}_{\bm{k}}\, 2^{\frac{3}{2}}\,\Gamma[5/2]\left({c_sk\eta}\right)^{-\frac{3}{2}}J_{3/2}\left({c_sk\eta}\right)\,,
\end{align}
where the superscript “p” refers to primordial and  $T_{{\gamma}}(k\eta)$ and $T_\Phi(c_sk\eta)$ are respectively the transfer functions for the homogeneous solution to tensor and scalar modes in radiation domination. $J_\alpha(x)$ is the Bessel function of the first kind of order $\alpha$. We note that $T_{{\gamma}}(k\eta)$ is the same for both polarizations, unless there are  parity violating terms in the gravity sector after inflation. 

Applying Green's method, the solution to Eq. \eqref{st-eq-eta-fins} reads
\begin{align}\label{sol}
    {\gamma}_{\bm{k},\,\lambda}(\eta)&={\gamma}^{(0)}_{\bm{k},\,\lambda}(\eta)+ {\gamma}^{(1)}_{\bm{k},\,\lambda}(\eta)+ ...\,,
\end{align}
with
\begin{align}\label{eq:gamma1sol}
{\gamma}^{(1)}_{\bm{k},\,\lambda}(\eta)&=\int_0^\eta d\tilde{\eta}\,  S_{st,\,\lambda}(\bm{k},\eta)\, G(\eta, \tilde{\eta})\,,
\end{align}
where $S_{st,\lambda}(\bm{k},\eta)$ is given by \eqref{Sstlambda}, ... refers to higher order solutions and $G(\eta, \tilde{\eta})$ is the Green's function for the tensor modes, namely
\begin{align}\label{eq:greens}
   G(\eta, \tilde{\eta})&=\frac{y_1(\tilde{\eta})y_2(\eta)-y_2(\tilde{\eta})y_1(\eta)}{y_1(\tilde{\eta})y'_2(\tilde{\eta})-y_2(\tilde{\eta})y'_1(\tilde{\eta})}\,.
\end{align}
In Eq.~\eqref{eq:greens} $y_1$ and $y_2$ being the two homogeneous solutions for $\gamma^{(0)}_k$. Concretely, if we take $y_1$ to be given by the “growing mode” \eqref{eq:gammaT}, $y_2$ is given by “decaying mode” which reads as in \eqref{eq:gammaT} but with $Y_{1/2}(x)$, the Bessel function of the second kind, instead of $J_{1/2}(x)$. Note that the first term of Eq. \eqref{sol} is the usual first-order (primordial) GWs, whereas the second one is for the modulated (scalar-tensor induced) GWs. In the next section we present analytical formulae for the kernel and the power-spectrum of scalar-tensor induced GWs.

\section{Scalar-Tensor Induced GW Spectrum and Kernel Function}\label{sec:2}

Let us derive a general formula to calculate the spectrum of scalar-tensor induced GWs. We aim to compute the two point function of tensor modes, namely from Eq.~\eqref{sol}
\begin{equation}\label{pow}
       \langle  {\gamma}_{\bm{k},\,\lambda}{\gamma}_{\bm{k'},\,\lambda'}\rangle= \langle  {\gamma}^{(0)}_{\bm{k},\,\lambda}{\gamma}^{(0)}_{\bm{k}',\,\lambda'}\rangle+\langle  {\gamma}^{(1)}_{\bm{k},\,\lambda}{\gamma}^{(1)}_{\bm{k}',\,\lambda'}\rangle+...\,,
   \end{equation}
where $...$ denotes two point functions involving higher order solutions of ${\gamma}_{\bm{k},\,\lambda}$, e.g. ${\gamma}^{(2)}_{\bm{k},\,\lambda}$. In this respect, it is important to note that we are neglecting the contribution $\langle  {\gamma}^{(0)}_{\bm{k},\,\lambda}{\gamma}^{(2)}_{\bm{k}',\,\lambda'}\rangle$ which naively would be of the same order as $\langle  {\gamma}^{(1)}_{\bm{k},\,\lambda}{\gamma}^{(1)}_{\bm{k}',\,\lambda'}\rangle$. However, computing ${\gamma}^{(2)}_{\bm{k},\,\lambda}$ involves solving second order equations for $\Phi$, using the solution for ${\gamma}^{(1)}_{\bm{k},\,\lambda}$ and computing the third order components in the source term of ${\gamma}_{\bm{k},\,\lambda}$. This is out of the scope of this paper. A complete discussion of the corrections to GWs up to third order in the perturbations can be found in \cite{Chen:2022dah}. Here we focus on fully understanding the solution ${\gamma}^{(1)}_{\bm{k},\,\lambda}$ and the corresponding spectral density. 

We write the two point correlation of scalar-tensor induced GWs in terms of a dimensionless power-spectrum which we call $\Delta^2_{{\gamma_1}}(k)$. Namely, we have that
   \begin{equation}\label{pow}
       \langle  {\gamma}^{(1)}_{\bm{k},\,\lambda}(\eta){\gamma}^{(1)}_{\bm{k'},\,\lambda'}(\eta)\rangle= (2\pi)^3 \delta_{\lambda \lambda'}\delta^3(\bm{k}+\bm{k}') \frac{2\pi^2}{k^3}\Delta^2_{{\gamma_1},\,\lambda}(k)\,,
   \end{equation}
where it should be noted that both $\delta_{\lambda \lambda'}$ and $\delta^3(\bm{k}+\bm{k}')$ follow from the contraction of the polarisation tensors in Eq. \eqref{eq:gamma1sol} and keeping in mind that the two-point function of the primordial $\Phi$ and ${\gamma}^{\rm p}_{\bm{k},\,\sigma}$ are written as
 \begin{align}
    \langle\Phi^{\rm p}_{\bm{k}}\,\Phi^{\rm p}_{\bm{k'}}\rangle&= (2\pi)^3 \delta^3(\bm{k}+\bm{k}') \frac{2\pi^2}{k^3}\Delta^2_{\Phi}(k)\,, \\
   \langle {\gamma}^{\rm p}_{\bm{k},\,\sigma}{\gamma}^{\rm p}_{\bm{k}',\,\sigma'}\rangle &= (2\pi)^3 \delta^3(\bm{k}+\bm{k}') \delta_{\sigma \sigma'}\frac{2\pi^2}{k^3}\Delta^{2}_{{\gamma}_0,\,\sigma}(k)\,.
   \end{align} Doing so, we arrive at \begin{align}\label{finchiral}
       \Delta^2_{{\gamma}_1,\,\lambda}(k)&= \frac{k^3}{\pi}\sum_\sigma \int 
       d^3{k}_1\frac{\Delta^2_{\Phi}(|\bm{k}-\bm{k}_1|) \Delta^{2}_{{\gamma}_0,\,\sigma}(k_1)}{k_1^3 
       |\bm{k}-\bm{k}_1|^3} \epsilon^{ij,\sigma}(\bm{\hat{k}}_1) \epsilon_{ij}^{\lambda*}(\bm{\hat{k}})\epsilon^{mn,\,\sigma}(-\bm{\hat{k}}_1) \epsilon^{\lambda *}_{mn}(-\bm{\hat{k}})\nonumber\\
       &\times \left( \int_0^\eta d\tilde{\eta} \,  G(\eta, \tilde{\eta})\,\left[k_1^2 T_{{\gamma}}(k_1\tilde{\eta} ) T_\Phi(c_s|\bm{k}-\bm{k}_1|\tilde{\eta} )-T'_{{\gamma}}(k_1\tilde{\eta} ) T'_\Phi(c_s|\bm{k}-\bm{k}_1|\tilde{\eta} )\right]\right)^2\,.
   \end{align}
To perform the integrals, it is convenient to work with the variables given by
\begin{align}
v=k_1/k\quad{,}\quad u=|\bm{k}-\bm{k}_1|/k\quad{\rm and}\quad x=k\eta\,.
\end{align}
Further using the properties of the polarization tensors, namely $\epsilon^{(\lambda)*}_{ij}(\bm{\hat{k}})=\epsilon^{(\lambda)}_{ij}(-\bm{\hat{k}})$, and $\epsilon^{\lambda}_{ij}(-\bm{\hat{k}})=\epsilon^{-\lambda}_{ij}(\bm{\hat{k}})$, we derive a compact expression for the right and left polarizations of the induced GWs respectively given by
\begin{align}\label{pow-R}
     \Delta^2_{{\gamma}_1, \,\rm R/L}(k)
       &=\frac{1}{32}\int_0^\infty dv \int_{|v-1|}^{v+1} \,
      \frac{du}{v^6u^2} \,\Delta^2_{\Phi}(uk)\, \overline{\mathcal{I}^2(x,u,v)}\, \nonumber\\
       &\times \left[ \left((v+1)^2-u^2\right)^4\,\Delta^{2}_{{\gamma}_0, \,\rm R/L}(vk)+\left((v-1)^2-u^2\right)^4\,\Delta^{2}_{{\gamma}_0,\, \rm L/R}(v k)\right]\,,
\end{align}
where we used a slash “$/$” in the subscript of $\Delta^2_{{\gamma}_1,\,\lambda}$ to differentiate between the case of right and left polarization, and we defined
\begin{align}\label{kernel-st}
       \mathcal{I}(x,u,v)= \frac{\pi}{2\sqrt{x}}\int_0^x  \,d\tilde{x}\,  \,\tilde{x}^{3/2}\,&\left(J_{1/2}(\tilde{x})Y_{1/2}(x)-J_{1/2}(x)Y_{1/2}(\tilde{x})\right)\,\nonumber\\&\times\left[v^2 T_{{\gamma}}(v\tilde{x}) T_\Phi(c_su\tilde{x})-\frac{d}{d\tilde x}{T}_{{\gamma}}(v\tilde{x}) \frac{d}{d\tilde x}{T}_\Phi(c_su\tilde{x})\right]\,.
\end{align}
An overline in $\mathcal{I}$ in Eq.~\eqref{pow-R} denotes oscillation average. By averaging over multiple wavelengths, we can  improve the accuracy of parameter estimation, by mitigating the effects of the rapid oscillations. We also note that, in what follows, we formally take the upper limit of the time integral \eqref{kernel-st} to infinity as the GW frequencies of interest enter the horizon well inside the radiation dominated universe, that is $k\tau_{\rm eq}\gg 1$ where $\tau_{\rm eq}$ is the (conformal) time of radiation-matter equality. The contribution from large values of the conformal time  is therefore negligible. In \S~\ref{sec:4} we present the full time dependence of the kernel. In this case, integrating and taking the oscillation average, we can define the following quantity
 \begin{align}\label{oscav}
           \overline{\mathcal{I}_\infty^2(u,v)}&\equiv x^2\times\overline{\mathcal{I}^2(x\to\infty,u,v)}
       \nonumber\\&=\frac{9}{2^7}  \left(\frac{v}{c_s u}\right)^2\left[\pi^2(1-s^2)^2 \Theta\left(1-s^2\right)+\Big(2s+(1-s^2)\log \left|\frac{1+s}{1-s}\right|\Big)^2\right]\,,
       \end{align}
where the subscript $\infty$ refers to the limit $x\to\infty$, we multiplied $\overline{\mathcal{I}^2(x\to\infty,u,v)}$ by $x^2$ to subtract the typical decay of sub-horizon tensor modes, i.e., $\gamma\propto 1/a$, so that $\overline{\mathcal{I}_\infty^2(u,v)}$ is time independent and for convenience we defined
\begin{align}
s=\frac{v^2+c_s^2 u^2-1}{2c_suv}\,.
\end{align}
The oscillation average is taken because GW detectors measure the time average of the GW background. Note that, in contrast to scalar-scalar induced GWs, in Eq.~\eqref{oscav} only the variable $u$ is multiplied by $c_s$ as it corresponds to the scalar mode, but not $v$ which is related to the momentum of the tensor mode. 

The spectral density of scalar-tensor induced GWs is then given by
\begin{align}\label{omega1}
      \Omega^{\rm {st-ind}}_{\rm {GW,\,R/L\, c}}(k)&= \frac{1}{12}\left(\frac{k}{\mathcal{H}}\right)^2  \Delta^{2}_{{\gamma}_1,\,R/L}(k)\nonumber\\&=\frac{1}{384} \int_0^\infty dv \int_{|v-1|}^{v+1}  \,
      \frac{du}{v^6u^2} \,\Delta^2_{\Phi}(uk)\, \overline{\mathcal{I}_\infty^2(u,v)}\, \nonumber\\
       &\hspace*{10mm}\times \left[ \left((v+1)^2-u^2\right)^4\,\Delta^{2}_{{\gamma}_0,\, \rm R/L}(vk)+\left((v-1)^2-u^2\right)^4\,\Delta^{2}_{{\gamma}_0, \,\rm L/R}(vk)\right]\,,
\end{align}
where we used that in the radiation dominated universe ${\cal H}=1/\eta$. The subscript “c” in \eqref{omega} denotes evaluation at a time where GWs are deep inside the horizon so that they behave as radiation. It is interesting to note from Eq.~\eqref{omega1} that if there is primordial parity violation of GWs, such parity violation is inherited by the scalar-tensor induced GWs and it is smeared between the polarizations. For completeness, we also give the formula for the total spectral density which is the sum of both polarizations, namely
\begin{align}
\label{omega-tot}
        \Omega^{\rm {st-ind}}_{\rm {GW},\, c}(k)&=  \sum_{\lambda=\rm R,\rm L}   \Omega^{\rm {st-ind}}_{\rm {GW},\, \lambda}(k,\eta)\nonumber\\&=\frac{1}{12} \int_0^\infty dv \int_{|v-1|}^{v+1}  \,
      \frac{du}{v^2u^2} \,\Delta^2_{\Phi}(uk)\, \overline{\mathcal{I}_\infty^2(u,v)}\left(\Delta^{2}_{{\gamma}_0, \,\rm R}(vk)+\Delta^{2}_{{\gamma}_0, \,\rm L}(vk)\right)\, \nonumber\\
       &\hspace*{30mm}\times \left[ \frac{(1+v^2-u^2)^2}{v^2}+ \Big(1+\Big(\frac{1+v^2-u^2}{2v}\Big)^2\Big)^2\,\right]\,,
\end{align}
which is of course parity symmetric. In the case of no primordial parity violation we simply take $\Delta^{2}_{{\gamma}_0, \,\rm R}=\Delta^{2}_{{\gamma}_0, \,\rm L}=\Delta^{2}_{{\gamma}_0}$ in Eq.~\eqref{omega-tot}. To evaluate the amplitude of the GWs spectral density today, we use \cite{Domenech:2021ztg}
\begin{align}
\Omega^{\rm {st-ind}}_{\rm GW,\, R/L,\, 0}h^2&=1.62\times 10^{-5}\left(\frac{\Omega_{\rm rad,\, 0}h^2}{4.18\times 10^{-5}}\right)\left(\frac{g_{\rho}(T_c)}{106.75}\right)\left(\frac{g_{s}(T_c)}{106.75}\right)^{-4/3}\Omega^{\rm {st-ind}}_{\rm {GW,\, R/L,\, c}}\,.
\end{align} 
Note that if one wants to use the curvature perturbation ${\cal R}$, one has that
\begin{align}
\Delta_\Phi^2=\left(\frac{3(1+w)}{5+3w}\right)^2\Delta_{\cal R}^2=\frac{4}{9}\Delta_{\cal R}^2\,,
\end{align}
where in the last step we use that for radiation domination $w=1/3$.
This completes the general derivation of the kernel and the spectral density of scalar-tensor induced GWs for general primordial parity of tensor modes. This is one of the new results of our work. 

\subsection{General behaviour of the kernel and differences with scalar-scalar induced GWs \label{sec:behaviourkernel}}
We proceed to show the general properties of the kernel \eqref{oscav} and its differences with the case of scalar-scalar induced GWs. First, let us examine the infrared ($k \to 0$) behaviour of our GWs.  In this limit, $u \sim v \sim 1/k \gg 1$. Hence, $s$ takes the value 
\begin{align}
    s&=\frac{1+c_s^2}{2c_s}\,.
\end{align}
In contrast, the scalar-induced gravitational waves (SIGWs) exhibit a different behaviour, as $s$ approaches $1$ (see Appendix \ref{App-sigw}). In that scenario, a logarithmic running arises, which is not observed in scalar-tensor-induced GWs due to the fact that $s$ never reaches the value of $1$ in IR. This logarithmic running has been regarded as a distinctive characteristic of SIGWs, setting them apart from primordial GWs in the infrared (IR) region. Notably, our findings demonstrate that such a feature is absent in our case.

In the opposite limit ($k \to \infty$), i.e. the UV one,  $v\to 1$ and $u\to 0$, which corresponds to the large wavelength limit for the scalars, we observe that the variable $s$ approaches $u$. As a result, the kernel exhibits the behaviour
\begin{align}\label{eq:divergence1}
\overline{\mathcal{I}_\infty^2(u \to 0,v \to 1)} \sim \frac{1}{u^2}\,.
\end{align}
Consequently, the integral in Eq. \eqref{omega-tot} becomes proportional to $1/u^4$. The other $k\to\infty$ limit, corresponding to a long wavelength tensor modes, leads to $u\to 1$ and $v\to 0$. In this case we have $s \to 1/v$, which does not lead to any divergence. 

In Eq. \eqref{oscav}, another notable feature occurs when $s=\pm 1$. In the case of  SIGWs, this scenario can lead to a logarithmic resonance. However, in our case, the presence of the term $(1-s^2)$ prevents such resonance from occurring. In fact, for $s=\pm 1$, or $v=\pm (1-c_s u)$, we have 
\begin{align}
     \overline{\mathcal{I}_\infty^2}&=\frac{9}{2^5}  \left(1-\frac{1}{c_s u}\right)^2\,.
\end{align}
This diverges in the $u \to 0$ limit.

We would like to emphasize that the kernel remains finite throughout the entire integration region being considered (except for the strict limit $u \to 0$). However, the fact that the integrand grows unboundedly for small $u$ can lead to artificial enhancements in the GWs spectrum. In particular, if both primordial scalar and tensor spectra are flat, the integral does not converge. If the tensor spectra is flat and primordial scalar peaked, say at $u=k_p/k$, then the GW spectrum grows unboundedly in the UV where $k\gg k_p$. If the scalar spectrum is flat and the tensor spectrum peaked at $v=k_p/k$, then the GW spectrum diverges at $k=k_p$ where $v=1$ and $u=0$. For these reasons, we will not consider the aforementioned cases and focus only on peaked spectra for which no such divergences occur. We later focus on identifying the source of the divergence and propose a solution in \S~\ref{sec:4}.

\begin{figure}
           \centering
           \includegraphics[width=0.49\linewidth]{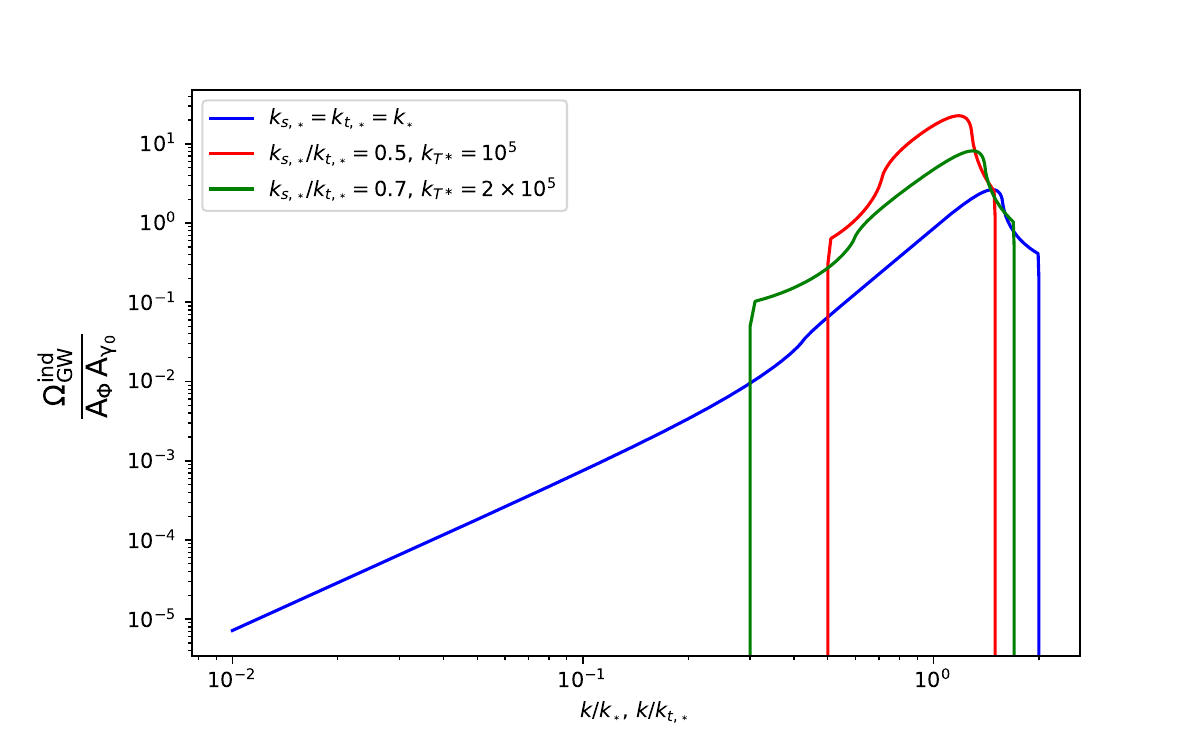}
           \includegraphics[width=0.49\linewidth]{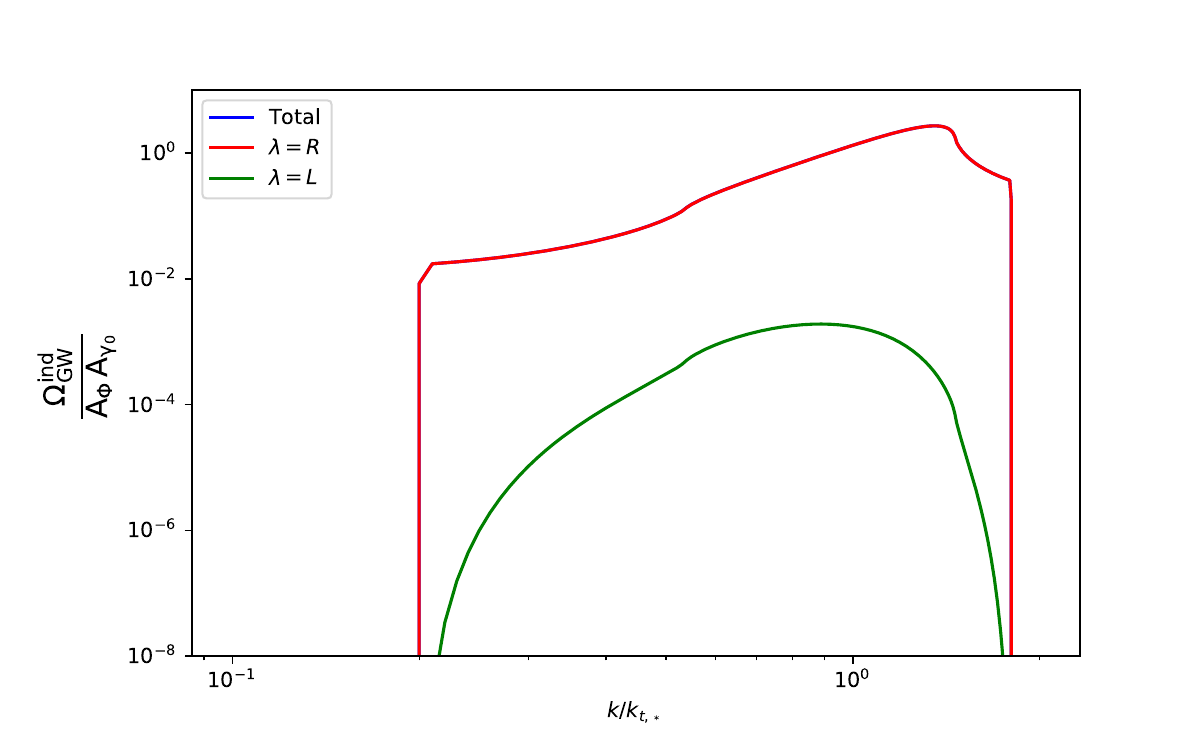}
           \caption{Left: Properly normalised density parameter for monochromatic scalar and non-chiral tensors, having same ($k_*$) and different peak locations ($k_{s,*},k_{t,*}$).
           The two dips in the second case comes from the  first Heaviside theta; $0.3<k_*/k_{t,*}<1.7$ for the green curve for example. As can be seen, different  peaks contribution is for a small range due to two thetas. Right: Same for the case with chiral primordial GWs. Here $k_{s,*}/k_{t,*}=0.8,\,k_{t,*}=10^5 \rm Mpc^{-1}$.
}\label{Omegadiffpeak}
\end{figure}

\section{Scalar-tensor induced GWs from peaked sources}\label{sec:3}
Now, we proceed to demonstrate the effect using a specific choice of input scalar and tensor perturbations: peaked sources. We do so for simplicity and because enhancements of primordial scalar and tensor fluctuations during inflation often lead to peaked primordial spectra \cite{Garcia-Bellido:2016dkw,Namba:2015gja,Thorne_2018,Shiraishi:2016yun}. We first consider Dirac delta primordial spectra and later discuss the effects of a finite width.

Let us take a Dirac delta source located at $k_{s/t,*}$, for both scalar and tensor primordial power-spectra,
       \begin{align}\label{monosamesource}
           \Delta^2_{\Phi}(k)=A_{\Phi} \,\delta\left(\ln{\frac{k}{k_{s,*}}}\right)\quad{\rm and}\quad
           \Delta^2_{{\gamma}_0,\,R/L}(k)=A_{{\gamma}_0,\,R/L}\, \delta\left(\ln{\frac{k}{k_{t,*}}}\right)\,.
       \end{align}
Note that in general we may have $k_{s,*}\neq k_{t,*}$. This kind of peaked scalar sources can be relevant for primordial black hole formation \cite{Saito_2009,Wang:2019kaf,Byrnes:2018txb}. Below, we illustrate the impact of this particular choice on the final spectra in two scenarios: when the primordial GWs exhibit chirality and when they do not. 

\begin{figure}
       \centering
       \includegraphics[width=0.49\linewidth]{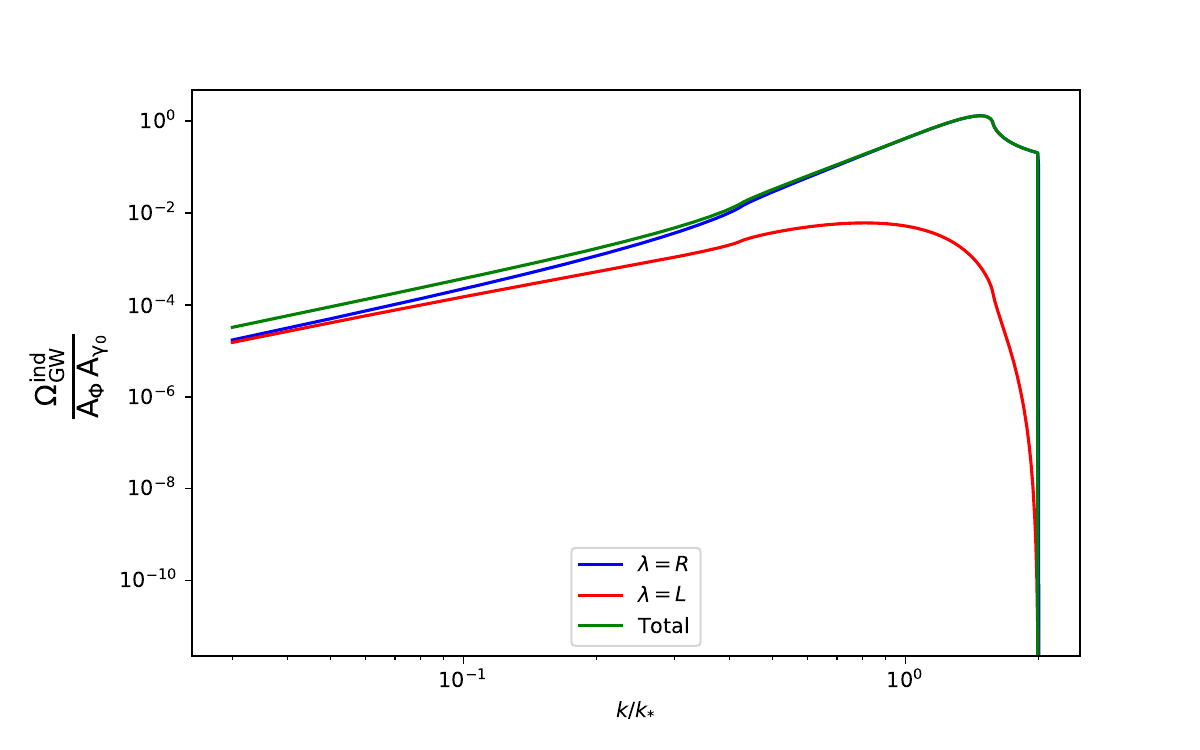}
    \includegraphics[width=0.49\linewidth]{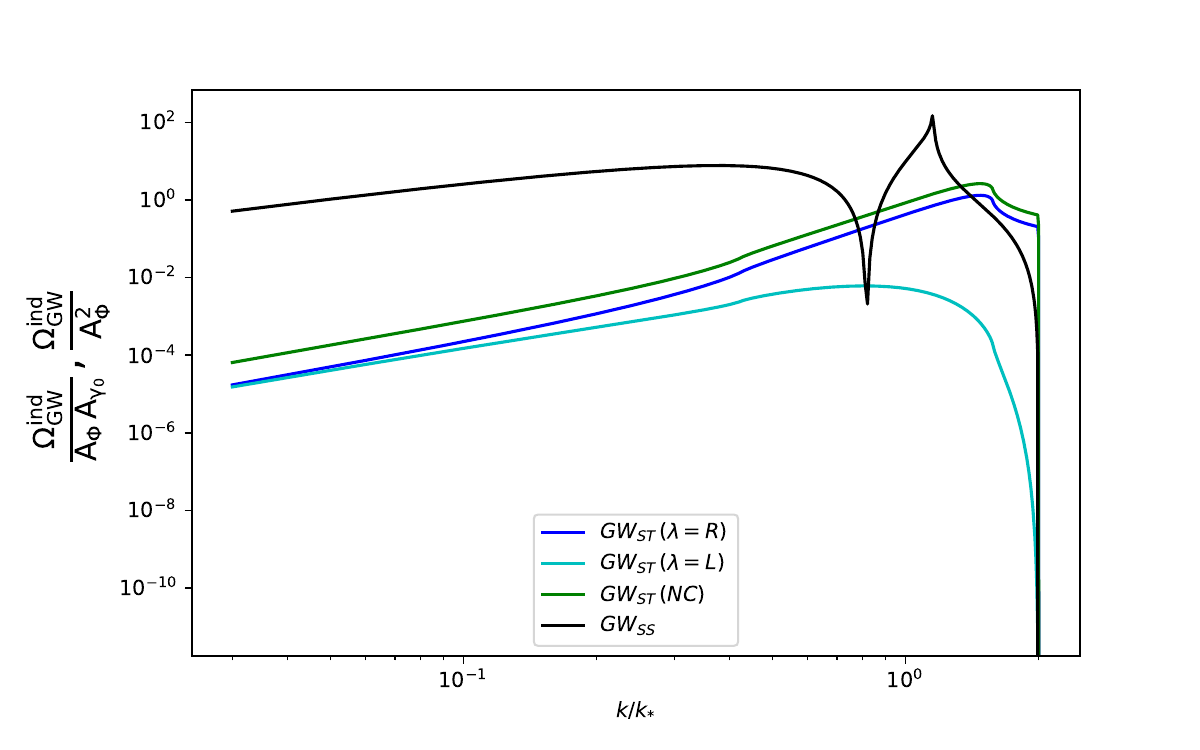}
       \caption{Left: Properly normalised total spectral density for induced GWs from monochromatic scalar and chiral GWs mixing (green), spectral density for scalar-tensor induced  right- (blue) and left-handed (red) chiral GWs,  when primordial GWs are right-handed. Right: Comparison of the gravitational wave density parameter induced by: monochromatic scalars only (black), monochromatic scalar and non-chiral tensors (green),  and  scalar-tensor induced right- (blue) and left-handed (cyan) chiral GWs,  when primordial GWs are right-handed. NC signifies non-chiral.}
       \label{st-ss}
   \end{figure}
\subsection{Non-chiral primordial GWs}

We first consider the case when $A_{{\gamma}_0,\,R}=A_{{\gamma}_0,\,L}=A_{{\gamma}_0}$. Then, we consider two different cases:
\begin{enumerate}
 \item \textbf{Peaks at the same location:} The simplest possibility, also considered in \cite{Chang:2022vlv}, is that both peaks are at the same location, namely $k_{s,*}= k_{t,*}=k_*$. In that case, we have
    \begin{align}\label{monomono}
      \Delta^2_{{\gamma}_1}(k)= A_{\Phi}\,A_{{\gamma}_0}\,
     \left(\frac{k}{k_*}\right)^2\left[1+\frac{k^4}{16k_*^4}+ \frac{3k^2}{2k_*^2}\right]\overline{\mathcal{I}^2}_{u=v=k_*/k}\,\Theta(2k_*-k)\,.
   \end{align}
   Fig. \ref{Omegadiffpeak}, left panel, shows the GWs energy density for this case. We report a small difference with Fig.~2 of \cite{Chang:2022vlv}: the high frequency part of the spectrum in \cite{Chang:2022vlv} presents some wiggles, while we find no such feature in Eq. \eqref{monomono}. Unfortunately, comparison is not so straightforward because we are not considering tensor-tensor induced GWs and we are using a different prescription for the metric perturbations (Eq. \eqref{eq:metric}). It is interesting to note though that the precise form of the scalar-tensor and tensor-tensor mixings \eqref{main} depends on how one expands the metric (i.e. exponential or linear in $\gamma_{ij}$ and $\Phi$). Ultimately, the different forms should be equivalent up to terms proportional to the linear equations of motion. 
\item \textbf{Peaks at different locations:} In general it is possible that $k_{s,*}$ and $k_{t,*}$ are two independent parameters and so $k_{s,*}\neq k_{t,*}$. If so, we find
   \begin{align}\label{diffpeak}
          \Delta^2_{{\gamma}_1}(k)&= A_{\Phi}\,A_{{\gamma}_0}\,
       \frac{k^2}{k_{s,*}k_{t,*}} \left[\frac{(k^2+k_{t,*}^2-k_{s,*}^2)^2}{k^2k_{t,*}^2}+ \frac{(4k^2k_{t,*}^2+(k^2+k_{t,*}^2-k_{s,*}^2)^2)^2}{16k^4k_{t,*}^4}\right]\nonumber\\&
       \times \overline{\mathcal{I}^2}_{v=k_{t,*}/k, u=k_{s,*}/k}\,\Theta(k_{s,*}-|k_{t,*}-k|)\,\Theta(k_{t,*}+k-k_{s,*})\,.
   \end{align}
   Naturally, the range of wavenumber of the induced waves increases with a decreasing separation of the two different peaks, which can be seen from the two Heaviside thetas, and the left panel of Fig.~\ref{Omegadiffpeak}. In other words, for Dirac delta separate peaks, the scalar-tensor induced GWs have an IR and UV cut-off.
\end{enumerate}
  
 By looking at both panels of Fig. \ref{Omegadiffpeak} , we see that they exhibit an enhancement in the induced GWs between the two dips that are determined by the first Heaviside theta function in Eq. \eqref{diffpeak}, even though the scalar and tensor source peaks are located at different positions. 
 This enhancement is related to the fact that we have a very large value of the integrand at small $u$. We have been able to obtain a solution in case of monochromatic primordial perturbations only because in this case the momenta acquire a single value. For other shapes of source primordial spectra, we have to find a way to avoid this problem.

It is interesting to note that, contrary to the scalar-scalar induced GWs, the scalar-tensor GWs spectrum for Dirac delta spectra have a finite amplitude at the cut-off $k=2k_*$. This may look suspicious at first because one expects the GW spectrum to be continuous. However, the sharp cut-off is due to the Dirac delta. Once we consider a log-normal peak the GW spectrum exponentially vanishes near the cut-off, as we shall show later. 
\begin{figure}
    \centering
    \includegraphics[width=0.49\linewidth]{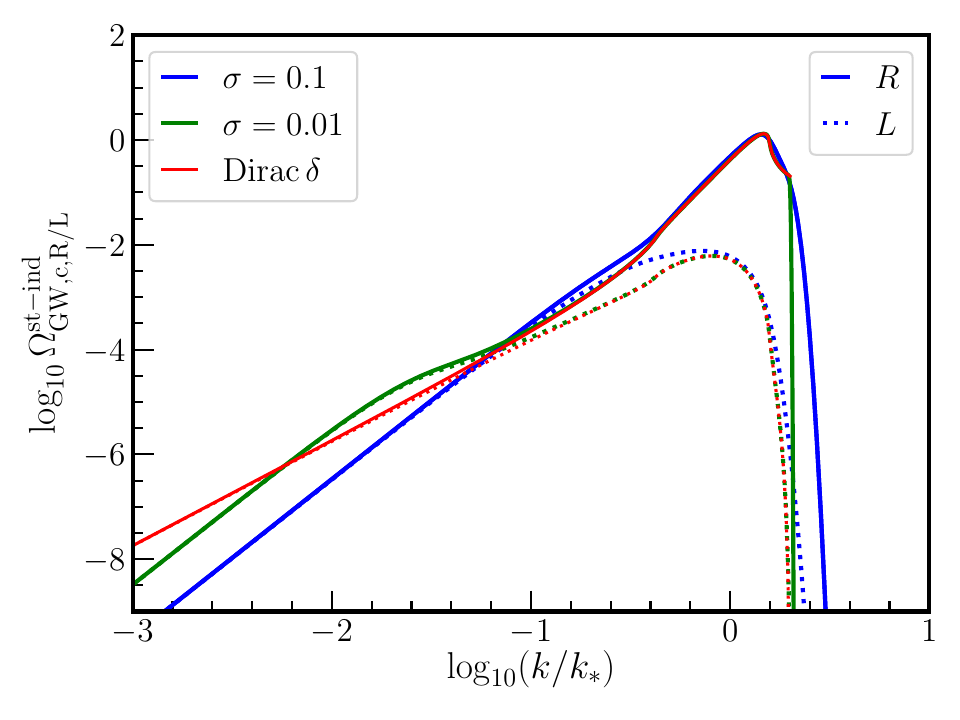}
    \includegraphics[width=0.49\linewidth]{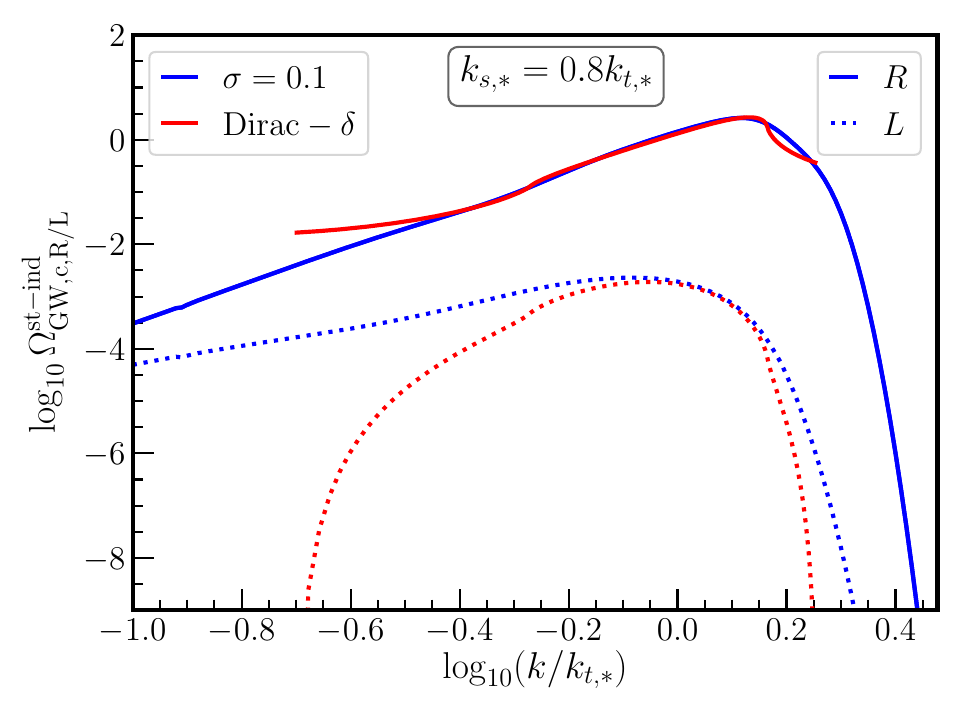}
    \caption{Scalar-tensor induced GWs from log-normal primordial scalar and tensor fluctuations. Here we only consider right-handed primordial tensors to illustrated the difference between $\Omega_{\rm GW,R}^{\rm st-ind}$ (solid lines) and $\Omega_{\rm GW,L}^{\rm st-ind}$ (dashed lines). For non-chiral primordial tensor one has to sum both contributions and multiply by 2. Left: Scalar-tensor induced GWs for same peak position for a Dirac delta (red) and log-normals with width $\sigma=0.01$ (green) and $\sigma=0.1$ (blue). Right: Different peaks positions with $k_{s,*}=0.8 k_{t,*}$. We show in blue a log-normal with $\sigma=0.1$ and in red the Dirac delta case.}\label{Lognormalall}
\end{figure}
\subsection{Chiral primordial GWs}

         For   primordial GWs, which have a delta-function peak in only one of the polarizations at the same wave-number as the primordial scalars (Eq. \eqref{monosamesource}), the present-day total spectral density of the induced GWs can be obtained from Eq. \eqref{omega-tot}
\begin{align}\label{omegachiralmono}
     \Omega^{\rm {st-ind}}_{\rm {GW},\,R/L,\,c}(k,\eta)&= \frac{1}{768} A_{\Phi}\,A_{{\gamma}_0,\,\rm R/L}\, 
     \left(\frac{k}{k_*}\right)^6\left[1+32\left(\frac{k_*}{k}\right)^4+48\left(\frac{k_*}{k}\right)^2\right]\nonumber\\&\times \overline{\mathcal{I}^2}_{u=v=k_*/k}\,\Theta(2k_*-k)\,.
\end{align}

The left panel of Fig. \ref{st-ss} displays the spectral density of induced gravitational waves resulting from the interaction between peaked scalar and peaked chiral gravitational waves. In this and the right panel,  it can be observed that when the primordial chiral gravitational waves induce gravitational waves of the same polarization, the peak of the spectrum, which is situated in the UV region, is more pronounced than that of the opposite polarization. In the IR region, however, we have an unpolarized induced wave. This could be attributed to the choice of peaked sources. Since the IR region is located far away from the peak of the GWs signal, there is effectively no detectable difference in the behaviour of the polarizations.  Although only the case with the right-handed primordial gravitational waves are displayed, the same applies to the left-handed ones. The trend is also manifested in the right panel of  Fig. \ref{Omegadiffpeak}, which exhibits the same scenarios but with different peak locations of scalar and tensor perturbations.

\subsection{Scalar-tensor induced GWs from log-normal peaks}

We now consider a more realistic situation where the peaks in the primordial spectra have a finite width. We do so by considering a log-normal spectrum, namely
\begin{align}\label{eq:lognormalpeak}
\Delta^2_{\Phi}(k)=\frac{{A_{s}}}{\sqrt{2\pi}\sigma}\exp\left[-\frac{\ln^2(k/k_{s,*})}{2\sigma^2}\right]\quad{\rm and}\quad \Delta^2_{{\gamma}_1,\,R/L}(k)=\frac{{A_{\gamma_0,\,R/L}}}{\sqrt{2\pi}\sigma}\exp\left[-\frac{\ln^2(k/k_{t,*})}{2\sigma^2}\right],
\end{align}
where we will consider for simplicity that they have the same logarithmic width $\sigma$ but in principle they can differ. We consider again the two possibilities: same peak location and different peak location. Interestingly, we find that in both cases the amplitude of the scalar-tensor induced GW spectrum is not very sensitive to the width of the primordial spectra. The spectral shape of course changes: it broadens for broader peaks.
\begin{figure}
    \centering
    \includegraphics[width=0.49\linewidth]{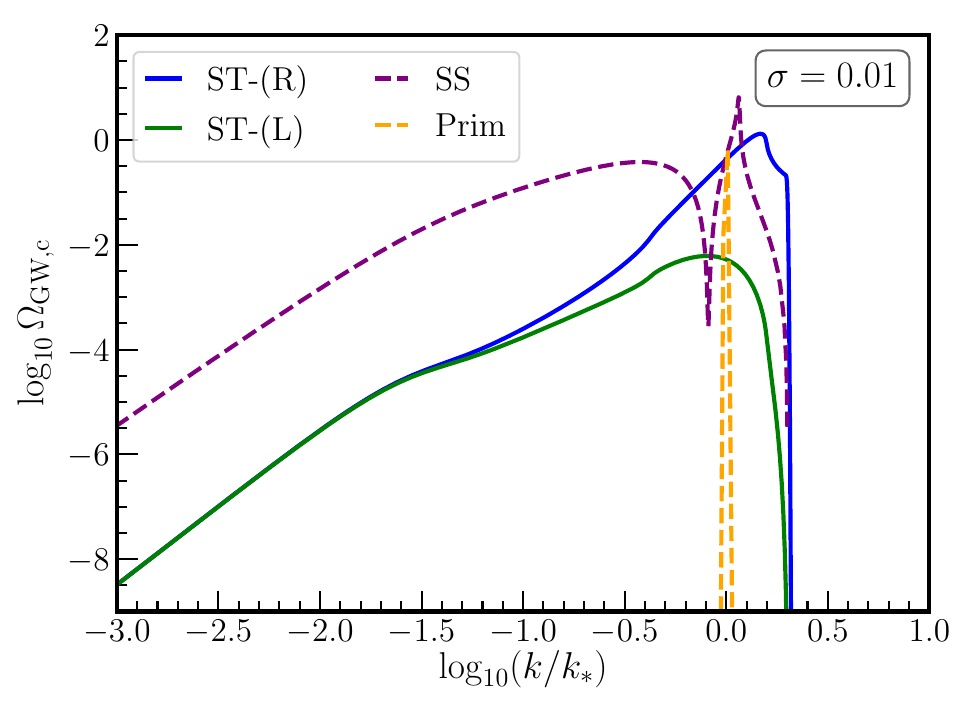}
    \includegraphics[width=0.49\linewidth]{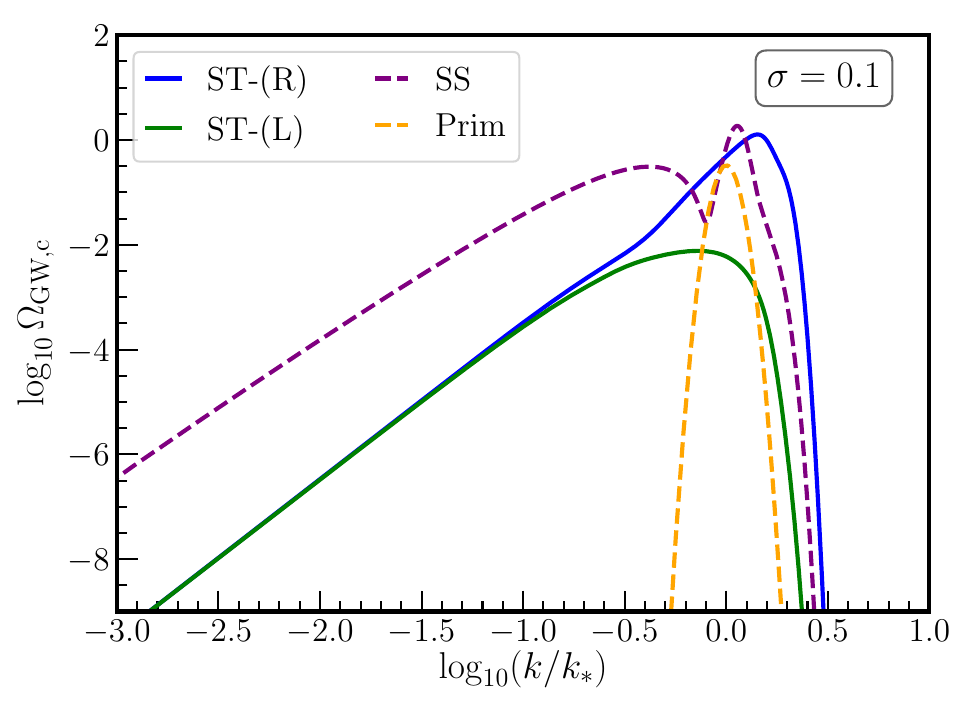}
    \caption{Collection of GWs spectrum from log-normal peaked Eq. \eqref{eq:lognormalpeak} primordial scalar and tensor spectra. For illustration purposes we assume only right-handed primordial tensor modes and we normalised the amplitude of the GW spectra setting $A_{\gamma_0}=A_\Phi=1$. In orange we show the primordial tensor spectrum, in purple the scalar-scalar induced GWs and in green and blue respectively the left and right handed scalar-tensor induced GWs. Left: GWs spectra for $\sigma=0.01$. Right: Same for $\sigma=0.1$. Scalar-tensor induced GWs may dominate in the high frequency part of the GW spectrum and extend the parity violation beyond frequencies covered by the primordial tensor spectrum.}\label{Lognormalall2}
\end{figure}
For the log-normal we compute the scalar-tensor induced GWs numerically. We show our results in Figs.~\ref{Lognormalall} and \ref{Lognormalall2}. For simplicity, we only considered only right polarization for the primordial tensor modes. However, one can consider the case of non-chiral primordial tensor modes by summing both lines in Figs.~\ref{Lognormalall} and \ref{Lognormalall2} corresponding to scalar-tensor induced GWs and multiplying by 2. Since we are plotting in logarithmic scale, the change in amplitude is not significant.

In Fig.~\ref{Lognormalall} we show the case of same peak position on the left and different peak position on the right. See how as one gets closer to the peak all lines are similar to the Dirac delta case. Also, note that for a finite width primordial spectra there is no sharp cut-offs in the scalar-tensor induced GWs, as expected. Since we are considering finite width of the primordial scalar and tensor spectra, we can also show and compare all contributions (primordial tensor, scalar-scalar and scalar-tensor induced GWs). We do so in Fig.~\ref{Lognormalall2} where we show our results for $\sigma=0.01$ (left) and $\sigma=0.1$ (right). See how in both cases the scalar-tensor induced GWs dominate the spectrum near the cut-off, even for $\sigma=0.1$. In this way, we extend the results of \cite{Chang:2022vlv} and show that even in the case of not too sharp and not too broad primordial spectrum, the scalar-tensor induced GW eventually have the potential to show-up in the high frequency part of the spectrum.

\subsection{Future prospects for scalar-tensor induced GWs}


\begin{figure}
    \centering
    \includegraphics[width=0.49\linewidth]{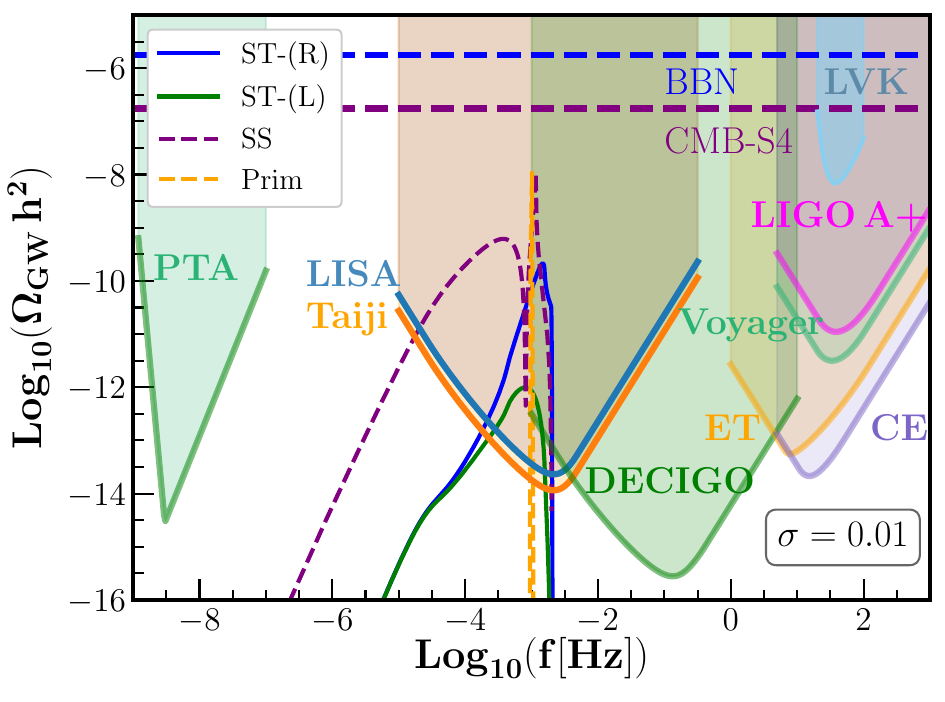}
    \includegraphics[width=0.49\linewidth]{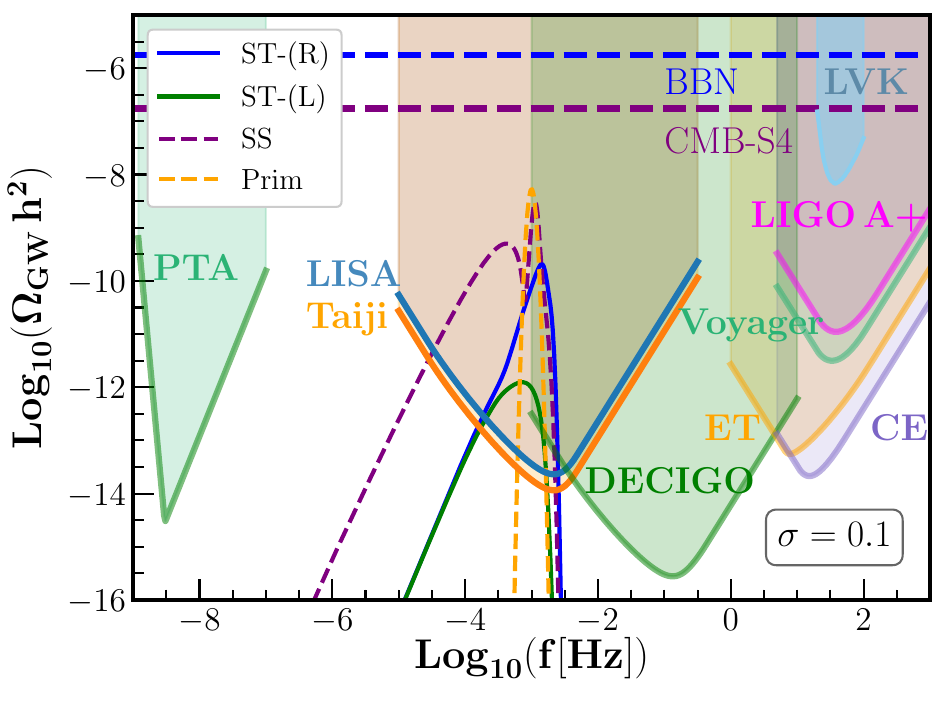}
    \caption{GW spectrum from primordial tensor modes (orange dashed), scalar-scalar induced GWs (purple dashed) and scalar-tensor left (solid green) and right (solid blue) handed induced GWs for log-normal spectra (Eq. \eqref{eq:lognormalpeak}) with $A_{ \gamma_{0,R}}=10^{-3}$ and $A_\Phi=10^{-2}$, $f_*=10^{-3}\,\rm Hz$. The left figure is for $\sigma=0.01$ and the right for $\sigma=0.1$. See how the scalar-tensor induced GWs extend primordial tensor parity  violation to the high frequency part of the spectrum. Also note how the low frequency part of the spectrum quickly becomes non-chiral. We include the power-law integrated sensitivity curves~\cite{Thrane:2013oya} for PTA, LISA, Taiji \cite{Barke:2014lsa,Ruan:2018tsw}, DECIGO, Einstein Telescope (ET), Cosmic Explorer (CE), Voyager and LIGO A+ experiments. The sensitivity curves can be found in \cite{ce,A+,voyager,Schmitz:2020syl}. We also plot the upper bounds on the GW background from the LIGO/Virgo/KAGRA collaboration \cite{KAGRA:2021kbb}. The horizontal thick long dashed lines qualitatively present the current constraint from BBN \cite{Cyburt:2004yc,Arbey:2021ysg,2023arXiv230112299G} (in blue) and future constraints from CMB-S4 experiments (in purple) \cite{2016arXiv161002743A,Arbey:2021ysg}. \label{Lognormalall2sensitivty}}
\end{figure}


After demonstrating two examples of scalar-tensor-induced spectra without encountering divergence issues, our attention now shifts to the detectability of these spectra. As stated earlier, we make the assumption that $A_{\gamma_0}<A_\Phi$, which enables us to neglect the tensor-tensor contribution and results in an effect that is subdominant compared to scalar-scalar induced GWs. 

In the right panel of Fig. \ref{st-ss}, it is evident that there exists only a limited range of scales 
 (approximately \(k/k_* \in [1.34, 2]\)), where the scalar-induced GWs do not surpass the scalar-modulated ones. Detecting the modulated waves amidst the dominance of the former requires identifying a characteristic that can distinguish our effect. As observed in previous section, non-chiral primordial waves lack such a property. However, when scalar modulation affects chiral primordial GWs, a disparity in the energy density between left and right circularly polarized waves becomes apparent. This distinction offers a potential avenue for detecting and studying the modulated waves.

 Fig. \ref{Lognormalall2sensitivty} provides a comprehensive comparison between the primordial  GWs, the realistic log-normal case, and the sensitivities of various probes. It is evident from the plot that while scalar-tensor induced GWs can dominate over SIGWs in the high-frequency range, this dominance is limited to a small range of scales. We observe the same behaviour for $\sigma=0.1$ and $\sigma=0.01$ in Fig. \ref{Lognormalall2}. On the other hand, the behaviour of the different-parity induced waves presents a distinguishing characteristic that sets them apart from SIGWs, particularly in the UV scales. This parity-violating behaviour of the scalar-tensor induced waves can be observed in Fig. \ref{Lognormalall2sensitivty}, where it extends beyond the peak of the primordial tensor spectrum. While this effect is not very significant for $\sigma \sim 0.1$, it becomes important for $\sigma \lesssim 0.01$. We would like to clarify that the figure shown in Fig. \ref{Lognormalall2sensitivty} provides a quantitative description of the power-spectrum shape of the induced GWs, rather than the detectability of their chiral properties. It is important to note that planar detectors typically do not have the capability to directly detect the chirality of GWs, unless specific methods are employed, such as leveraging the motion of the solar system with respect to the cosmic reference frame (as discussed in \cite{Domcke:2019zls}, see also the refs. therein). However, studies have shown that by cross-correlating the output of multiple detectors, such as LISA and Taiji, it is possible to detect and study the parity violation in the stochastic gravitational wave background \cite{Orlando:2020oko,Seto:2020zxw}. While we mention the potential of using this chirality to distinguish our induced GWs, the actual detectability of the same requires further detailed analysis beyond the scope of this work. In the low-frequency range, however, the induced waves consistently remain unpolarized, as depicted in the corresponding figures. 
\begin{figure}
    \centering
    \includegraphics[width=0.55\linewidth]{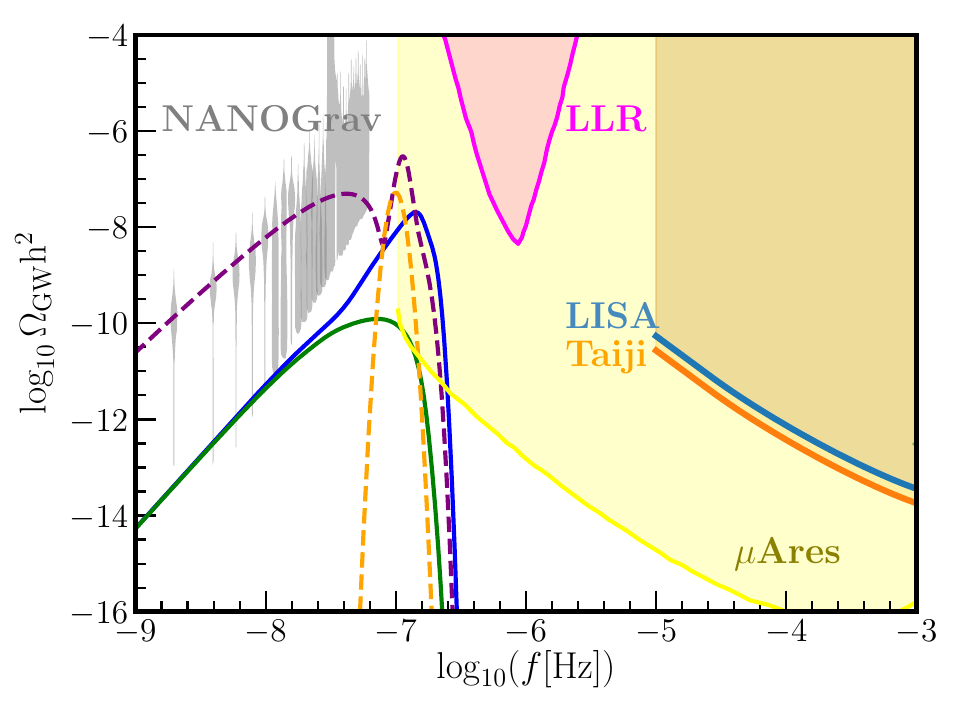}
    \caption{GWs spectrum from primordial tensor modes (orange dashed), scalar-scalar induced GWs (purple dashed) and scalar-tensor left (solid green) and right (solid blue) handed induced GWs for log-normal spectra (Eq. \eqref{eq:lognormalpeak}) with $A_{ \gamma_{0,R}}=10^{-2}$ and $A_\Phi=10^{-1}$, $f_*=10^{-7}\,\rm Hz$. Gray violins indicate recent NANOGrav results \cite{NANOGrav:2023hvm}.  We also include future sensitivity of $\mu$-Ares \cite{Sesana:2019vho} and Lunar Laser Ranging in magenta from Ref.~\cite{2022PhRvL.128j1103B}. See also \cite{Fedderke:2021kuy} for other ideas to detect $\mu\rm Hz$ GWs. Also included are the sensitivity curves of the LISA and Taiji detectors, to show that these detectors are unable to detect the peak of SIGWs, which could potentially explain the observations made by the NANOGrav collaboration.\label{fig:Nanograv}}
\end{figure}
 Upon examining Fig. 10 in \cite{Salehian:2020dsf} and Fig. 3 in \cite{Machado:2018nqk}, we observe a similar behaviour in the induced GWs as depicted in our Fig. \ref{st-ss}. It is noteworthy that these studies investigate GWs production mechanisms involving chiral dark photons, which are entirely distinct from our approach. Based on these observations, a hypothesis emerges, suggesting that there is a distinct polarization behaviour present in the UV region of induced GWs. It says that the polarized primordial component makes  the  peak of the induced GWs of the same polarization more enhanced compared to that of  the oppositely polarized GWs, while the IR region remains unpolarized. However, the thorough investigation and verification of this hypothesis are left for future endeavours.

In Fig. \ref{fig:Nanograv}, we present the recent results from the NANOGrav \cite{NANOGrav:2023hvm}, which may have detected the stochastic gravitational wave background using pulsar timing arrays (PTAs). The SGWB observed by PTAs can be considered as the IR tail of the SIGWs \cite{2023arXiv230617834I}. We leave a detailed analysis of our scalar-tensor induced GWs signal with the new PTA data \cite{NANOGrav:2023hvm,2023arXiv230616227A} for future work. Here, we demonstrate an example where the peak of the SIGWs lies in the $0.1\mu \rm Hz$ range, which is currently beyond the sensitivity range of existing detectors.  However, there has been proposals for a future detectors in this frequency range \cite{Sesana:2019vho,2022PhRvL.128j1103B}. If such a detector, preferably with better sensitivity and ability to detect chirality, is realized, it would be capable of detecting the peak, where it can be distinguishable from our signal based on the latter's chirality properties.

\section{Origin of potential divergences and possible solutions\label{sec:4}}

In \S~\ref{sec:behaviourkernel} we have anticipated that the momentum integral in  scalar-tensor induced GWs contains red potential divergence for vanishing scalar mode momentum.\footnote{We note that in practice there are always IR and UV cut-offs on the primordial scalar spectrum, respectively related to the start and end of inflation. Therefore, for physical primordial spectra the divergence appears as an artificial enhancement of the scalar-tensor induced GW spectrum.} The divergence in the integrand appears as $1/u^4\sim (k/q)^4$ for $q\ll k$, where here $k$ and $q$ respectively are the internal tensor and scalar mode momentum (note that the because $q\ll k$ we have that the internal tensor mode momentum is $|\bm{k}-\bm{q}|\sim k$ so there is not much difference between the external and internal mode in this limit). It should also be noted that we did not encounter such issues in \S~\ref{sec:3}, as we considered scalar-tensor induced GWs from peaked primordial spectra. But, as soon as we consider a relatively flat scalar primordial spectra, the divergence shows up. For flat scalar and tensor primordial spectra the integral divergences for all $k$. It is important to stress that the potential divergence simply appears in the solution of the second order tensor modes. But, as a purely classical calculation, such term must somehow cancel when dealing with observables. It is thus plausible that such problematic terms disappear when including higher order contributions. Nevertheless, for practical purposes, in this section we identify the source of such potential divergence and propose a well-motivated way to circumvent it. We leave a deeper study for future work.

We proceed as follows. We first derive the general time-dependent kernel to show that divergence is not completely associated with the limit $x=k\tau\to\infty$ and that artificial enhancements appear for scalar modes with wavelengths larger than that of tensor modes. We will see that this is independent on whether scalar modes are superhorizon or subhorizon. In fact, we show that removing the contribution from superhorizon scalar modes does not solve the problem. We then provide an argument based on a local inertial frame to bypass the problem.

\subsection{General time dependent kernel}
We start by rewriting the kernel \eqref{kernel-st} in a more practical form for the general integrations. We rewrite the Bessel functions, $J_\alpha(x)$, in terms of spherical Bessel functions, $j_\alpha(x)$, as
\begin{align}\label{eq:kernelsphericalB}
{\cal I}(x,u,v)=\frac{v^2}{x}\left({ I}_y(x,v,u)\sin x +{ I}_j(x,v,u)\cos x \right)\,,
\end{align}
where we defined
\begin{align}\label{eq:IjIydef}
{I}_{j,y}(x,v,u)\equiv\int_{x_i}^x d\tilde x \,{\tilde x^{2}}
\left\{
\begin{aligned}
j_{0}(\tilde x)\\
y_{0}(\tilde x)
\end{aligned}
\right\}
\left(j_{0}(v\tilde x)j_{0}(c_su\tilde x)-j_{2}(v\tilde x)j_{2}(c_su\tilde x)\right)\,.
\end{align}
In this way, we have that the first term in Eq.~\eqref{eq:kernelsphericalB} corresponds to the “growing mode” of a free tensor mode in a FLRW background and the second one to the “decaying mode”. We will now compute the functional form of ${I}_j(x)$ and ${I}_y(x)$ for general $x$. Without loss of generality we will take $x_i\to 0$, which does not influence the present discussion. After several integrations, similar to those in Ref.~\cite{Kohri:2018awv}, we obtain that
\begin{align}
{I}_j(x,v,u)=&\frac{3(1-s^2)}{8c_suv}G_0[{\rm Si} [x]]+\frac{3(1-s^2)}{8c_suvx}G_1[\cos x]\nonumber\\&+\frac{3}{8(c_suvx)^2}G_2[\sin x]+\frac{3}{16(c_suvx)^3}G_3[\cos x]+\frac{9}{16(c_suv)^3x^4}G_4[\sin x]\,,
\end{align}
and
\begin{align}
{I}_y(x,v,u)=&-\frac{3}{8c_suv}\left(2s+(1-s^2)\left(\log\left|\frac{1+s}{1-s}\right|+F_0[{\rm Ci} |x|]\right)\right)+\frac{3(1-s^2)}{8c_suvx}F_1[\sin x]\nonumber\\&-\frac{3}{8(c_suvx)^2}F_2[\cos x]-\frac{3}{16(c_suvx)^3}F_3[\sin x]-\frac{9}{16(c_suv)^3x^4}F_4[\cos x]\,.
\end{align}
In the expressions above we have defined
\begin{align}
F_0&[{\rm Ci} |x|]\nonumber\\&=  {\rm Ci}\left|(1 + c_s u - v) x\right| +
 {\rm Ci}\left|(1 - c_s u + v) x\right| - {\rm Ci}\left|(1 + c_s u + v) x\right|-{\rm Ci}\left|(1 - c_s u - v) x\right|\,,
\end{align}
and
\begin{align}
&F_1[\sin x]\nonumber\\&=  \frac{\sin\left[(1 + c_s u - v) x\right]}{1+c_su-v}+\frac{\sin\left[(1 - c_s u + v) x\right]}{1-c_su+v}  -\frac{\sin\left[(1 + c_s u + v) x\right]}{1+c_su+v}-\frac{\sin\left[(1 - c_s u - v) x\right]}{1-c_su-v}\,.
\end{align}
Then, $G_0[{\rm Si}[x]]$ is obtained by replacing ${\rm Ci}|x|$ by ${\rm Si}[x]$ in $F_0[{\rm Ci} |x|]$. Similarly, $G_1[\cos x]$ is obtained by replacing $\sin x$ by $\cos x$ in $F_1[\sin x]$. The other functions $F_2,F_3,F_4$ and $G_2,G_3,G_4$  are suppressed when $c_suvx\gg1$. For the interested reader, we write them explicitly in Appendix~\ref{app:expressionsgeneralkernel}. It is straightforward to check that we recover Eq. \eqref{oscav} in the limit of $x\to\infty$ and we take the oscillation average.\footnote{The function $F_0[{\rm Ci} |x|]$ with ${\rm Ci}[x]$ vanish when $x\to\infty$. The other function gives
\begin{align}
G_0[{\rm Si}[x\to\infty]]=\pi\Theta(1-|c_su-v|)-\pi\Theta(1-(c_su+v))\,.
\end{align}
It is only non-vanishing when $|c_su-v|<1<c_su+v$ which corresponds to $s^2<1$.
}

\begin{figure}
\includegraphics[width=0.49\columnwidth]{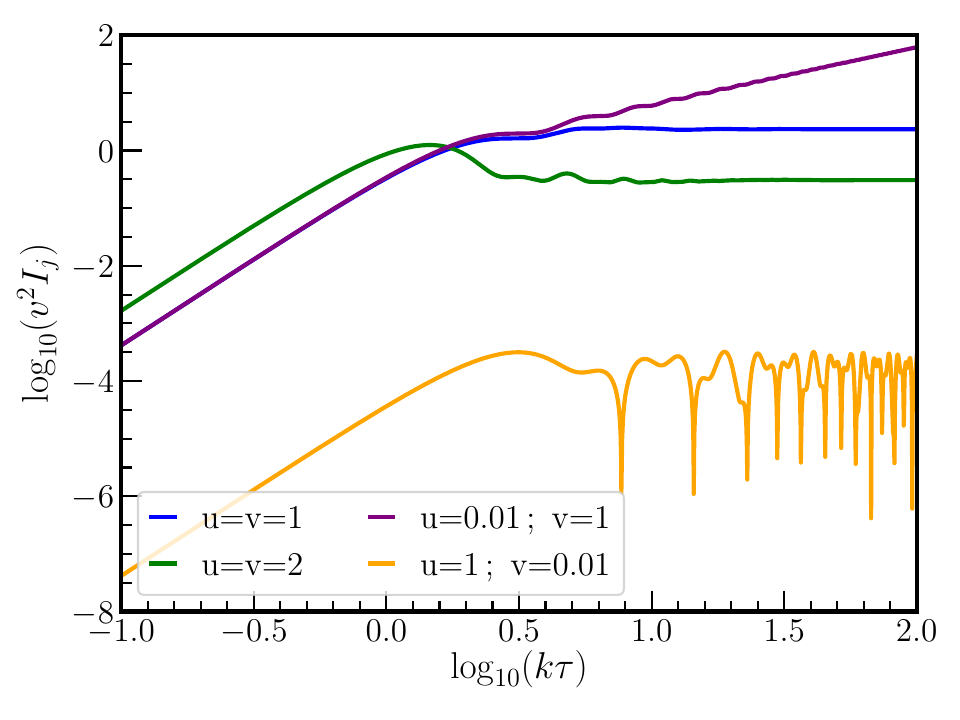}
\includegraphics[width=0.49\columnwidth]{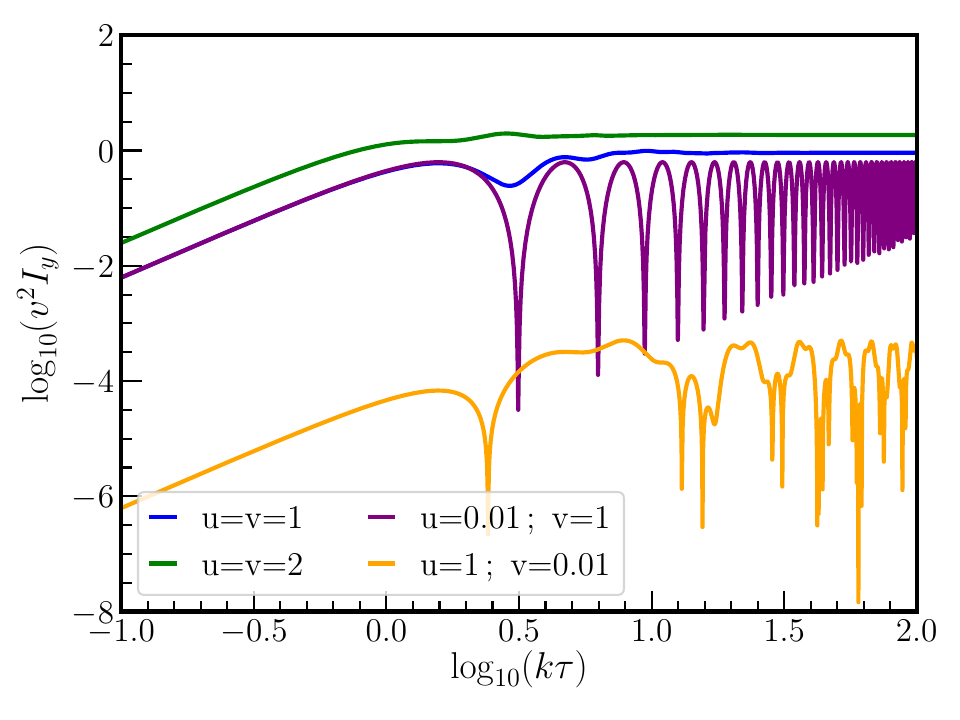}
\caption{Left: Behaviour of kernel functions $I_j$, Right: The same for $I_y$ \eqref{eq:IjIydef}, for various values of the scalar momenta: in blue $(u=1,v=1)$, in green $(u=2,v=2)$, in purple $(u=0.01,v=1)$ and in orange $(u=1,v=0.01)$. See how for $I_j$ on the left figure as we lower the value of $u$ the function $I_j$ grows with $x$. Instead, $I_y$ on the right figure approaches a constant (or oscillations around a constant) as $x\tau>1$. \label{fig:IJIY}}
\end{figure}

\subsubsection{Long wavelength scalar modes}

With the full time dependence of the kernel, we can study two limits involving a subhorizon induced tensor mode ($x\gg1$) and a long wavelength scalar mode ($u\ll1$): (i) the scalar mode is superhorizon at the time $x$, i.e. $ux\ll1$, and (ii) the scalar mode is subhorizon at time $x$, namely $ux\gg 1$. Note that the limit $u\ll1$ corresponds to the regime where $u\sim0$, $v\sim 1$ and $s\sim 0$ in the kernel. For case (i), where $ux\ll 1$, we find that
\begin{align}\label{eq:IjIy}
{ I}_j(x,ux\ll1)\approx\frac{x-\cos x\sin x}{2}+{\cal O}(u^2,u^2x^2)\quad,\quad
{ I}_y(x,ux\ll1)\approx -\frac{\sin^2x}{2}+{\cal O}(u^2,u^2x^2)\,.
\end{align}
Note that the integral $I_j$ in Eq. \eqref{eq:IjIy} is diverging for $x\to \infty$. This is one of the sources of the divergence in the naive oscillation average presented in Eq.~\eqref{oscav}. Nevertheless, even if we consider the total time dependence of the kernel before doing the oscillation average, the situation does not improve. For instance, the kernel Eq. \eqref{eq:kernelsphericalB} in this limit leads us to
\begin{align}\label{eq:gammacase1}
{\cal I}(x,ux\ll1,x\gg1)\approx \frac{1}{2}\left(\cos x-\frac{\sin x}{x}\right)\,.
\end{align}
Looking at Eq. \eqref{eq:gammacase1}, we see that in addition to the regular ``growing'' mode ($\sin x/x$) there is a non-decaying mode ($\cos x$). As we shall shortly show, after the mode $u$ enters the Hubble horizon, the integrals $I_{j,y}(x)$ approach a constant. However, the constant will be larger for smaller $u$ modes, since these modes had more time to grow. This can be seen in case $(ii)$, where $ux\gg1$, in which the integrals $I_j$ and $I_y$ respectively yield
\begin{align}\label{eq:IjIy2}
{ I}_j(u\ll1,ux\gg1)\approx\frac{3\pi}{8c_su}-\frac{3c_s\pi u}{32}\quad,\quad
{ I}_y(u\ll1,ux\gg1)\approx -\frac{3}{4}+\frac{c_s^2u^2}{16}\,.
\end{align}
We see that despite the fact that after the mode $u$ enters the Hubble horizon, the kernel \eqref{eq:kernelsphericalB} behaves as a regular kernel for tensor modes in the sense that there is a growing and decaying mode, coefficient of the decaying mode, that is $I_j$ given in Eq. \eqref{eq:IjIy2} diverges for small $u$. This is the same divergence that we found in Eq. \eqref{eq:divergence1} for $\overline{{\cal I}_\infty^2(u\ll1, v\sim1)}\sim 1/u^2$. However, the total integrand diverges as $1/u^4$. We will come back to the remaining $1/u^2$ at the end of the section.

In passing, we mention that we find no such behaviour when the wavenumber of the primordial tensor mode vanishes, i.e. when $v\to 0$, $u\to 1$ and $s\to -\infty$. Explicitly, in the limit where $x\gg1, ux\gg 1, vx\gg1$, we find that 
\begin{align}
{ I}_j(v\ll1,x\gg1)\approx 0 \quad,\quad
{ I}_y(v\ll1,x\gg1)\approx\frac{1}{1 - c_s^2} + \frac{(5 - c_s^2) v^2}{5 (1- c_s^2)^3}\,.
\end{align}
The integrand of the momentum integral for scalar-tensor induced GWs, Eq.~\eqref{omega-tot}, vanishes for $v\to0$.\footnote{The $c_s=1$ case has to be treated separately but one can check that it is regular as well in the limit $v\to 0$.} Thus, the only issue are scalar modes with wavelengths longer than the tensor mode.

\subsubsection{Removing superhorizon scalar mode contribution}

To illustrate the fact that the issue is more subtle, let us show that even if we remove the strange non-decaying mode in Eq. \eqref{eq:gammacase1} by subtracting the contribution from superhorizon scalar modes, the divergence is not cured. The problematic term in the equations of motion for induced tensor modes Eq. \eqref{st-eq} is the term that has $\Phi$ in front of $\nabla^2\gamma_{ij}$. For analytical purposes, we remove the contribution of superhorizon $\Phi$ coming from the mentioned term by introducing the following correction terms in Eq.~\eqref{eq:IjIydef},   which is to be subtracted to the coefficients of the growing and decaying modes $I_{j,y}$
\begin{align}\label{eq:correction}
{ C}_{j,y}(x,v)\equiv \int_{x_i}^x d\tilde x \,{\tilde x^{2}}
\left\{
\begin{aligned}
j_{0}(\tilde x)\\
y_{0}(\tilde x)
\end{aligned}
\right\}
j_{0}(v\tilde x)\,,
\end{align}
where we used that $T_\Phi(x\ll1)\sim 1$. We will later cut the integral near sound horizon crossing, that is at $\alpha c_sux=1$ where $\alpha$ is a free parameter. After integration, we find that
\begin{align}\label{eq:Cj}
{C}_j(x)=\frac{\cos (x) \sin (v x)-v \sin (x) \cos (v x)}{v \left(v^2-1\right)}&\,,
\end{align}
and
\begin{align}\label{eq:Cy}
{C}_y(x)=\frac{\sin (x) \sin (v x)+v \cos (x) \cos (v x)}{v \left(v^2-1\right)}-\frac{1}{v^2-1}&\,.
\end{align}

We can see that the correction terms $C_j$ and $C_y$ cancel the leading order terms of the non-decaying modes we found in Eq.~\eqref{eq:IjIy}. To see this, we take the limit $v\sim 1$ in Eqs. \eqref{eq:Cj} and \eqref{eq:Cy}, which respectively yields
\begin{align}
{C}_j(ux\ll1,v\sim 1)\approx\frac{x-\cos x\sin x}{2}\approx  { I}_j(x,ux\ll1)&\,,
\end{align}
and
\begin{align}
{C}_y(ux\ll1,v\sim 1)\approx -\frac{\sin^2x}{2}\approx { I}_y(x,ux\ll1)&\,.
\end{align}
We can now follow the corrected kernel until the scalar mode enters the sound Hubble horizon, i.e. we evaluate $C_{j/y}$ at $\alpha c_sux=1$ where $\alpha\sim O(1)$ free parameter to be fixed shortly. The corrected kernel is then given by
\begin{align}\label{eq:correctedkernel}
{\cal I}(ux\gg1)=\frac{v^2}{x}\left(\left[{ I}_y(x)-C_y(x=1/(\alpha c_su))\right]\sin x +\left[{ I}_j(x)-C_j(x=1/(\alpha c_su)\right]\cos x \right)\,.
\end{align}
In the limit where the divergence appeared, that is for $ux\gg1$, $u\ll1$ and $v\sim 1$, we find that
\begin{align}
{C}_j(ux\gg1,v\sim 1)\approx\frac{1}{2 \alpha  c_s u}-\frac{1}{4} \sin \left(\frac{2}{\alpha  c_s
   u}\right)&\,,
\end{align}
and
\begin{align}
{C}_y(ux\gg1,v\sim 1)\approx-\frac{1}{2} \sin ^2\left(\frac{1}{\alpha  c_s u}\right)&\,.
\end{align}
By requiring that the divergence in $I_j$ is cancelled by that in $C_j$ we fix $\alpha$ to
\begin{align}
\alpha=\frac{4}{3\pi}\approx0.42\,.
\end{align}
This value of $\alpha$ means that we subtract a constant $\Phi$ a little bit after sound Hubble horizon crossing, i.e. at $uc_sx\approx 2.3$.

The results of this section show that the divergence in the kernel can be made finite in the limit $x\to\infty$ (as in Eq.~\eqref{oscav}) by subtracting the superhozion contribution of a constant $\Phi$. Nevertheless, the coefficient of the kernel become constant which implies that the integrand for scalar-tensor induced GWs Eq. \eqref{omega-tot} still diverges as $1/u^2$ for $u\ll1$. This means that the divergences for long wavelength scalar modes are not entirely due to their superhorizon contribution but in fact some subhorizon as well. In the next subsection we identify the full source of the divergence.

\subsection{Source of the potential divergence and possible solutions}

To understand the source of the divergence, let us split the gravitational potential $\Phi$ into a short and a long wavelength contribution, namely
\begin{align}
\Phi=\Phi_{\rm short}+\Phi_{\rm long}\,,
\end{align}
where $\partial_k\Phi_{\rm long}\ll \partial_k \gamma_{ij}$. In other words, if $L_\Phi$ is the characteristic length of $\Phi_{\rm long}$ and $L_\gamma$ the characteristic length of $\gamma_{ij}$, we have that $\partial_k\Phi_{\rm long}\times L_\gamma\sim {\cal O}(\epsilon)\,\Phi_{\rm long}$ where we defined $\epsilon=L_\gamma/L_\Phi$. In that case, the equations of motion for scalar-tensor induced tensor modes \eqref{st-eq} read
\begin{equation}\label{stnow}
    {{\gamma}}''_{ij}+2{\cal H}{{\gamma}}'_{ij}-\nabla^2 {\gamma}_{ij}={4(\Phi_{\rm short}+\Phi_{\rm long})}\nabla^2 {\gamma}_{ij}+4(\Phi_{\rm short}'+\Phi_{\rm long}'){{\gamma}}'_{ij}\,.
\end{equation}
For the time derivatives we have that $\Phi_{\rm long}'\ll \Phi_{\rm short}'$ so we can neglect $\Phi_{\rm long}'$. However, in front of $\nabla^2 {\gamma}_{ij}$ we have directly $\Phi_{\rm short}+\Phi_{\rm long}$. In radiation domination, the amplitude of $\Phi_{\rm short}$ has decayed more than that of $\Phi_{\rm long}$ and therefore, it appears that the amplitude $\Phi_{\rm long}$ could be the dominant contribution and have a big impact on $\gamma_{ij}^{(1)}$. For instance, if we consider that $\Phi_{\rm long}={\rm constant}$, we see that the right hand side of Eq. \eqref{stnow} has now a term proportional only to $\nabla^2\gamma_{ij}$. If we think of it as a “source”, then $\gamma_{ij}$ is resonating with itself and leading to divergences. We will see this when computing the general kernel.


However, our intuition from the equivalence principle tells us that long wavelength modes cannot affect significantly the physics of short wavelength modes. We can use the fact that $\partial_k\Phi_{\rm long}\ll \partial_k \gamma_{ij}$ to consider that $\Phi_{\rm long}\approx {\rm constant}$ from the point of view of $\gamma_{ij}$. We then rescale the spatial coordinates as
\begin{align}
dx_i \to (1-2\Phi_{\rm long}) dx_i\quad{\rm so\,\,that}\quad \nabla^2\to (1+4\Phi_{\rm long})\nabla^2\,.
\end{align}
This coordinate transformation cancels the constant factor of $\Phi_{\rm long}\nabla^2 {\gamma}_{ij}$ in the right hand side of Eq. \eqref{stnow} and we are left with
\begin{equation}\label{stnow2}
    {{\gamma}}''_{ij}+2{\cal H}{{\gamma}}'_{ij}-\nabla^2 {\gamma}_{ij}\approx {4\Phi_{\rm short}}\nabla^2 {\gamma}_{ij}+4\Phi_{\rm short}'{{\gamma}}'_{ij}\,.
\end{equation}
Thus, we conclude that long wavelength scalar modes cannot affect the local generation of tensor modes. Although in the strict sense $\Phi_{\rm long}$ is not a constant, it shows that if we view the scalar-tensor mixing in the right hand side of Eq.~\eqref{st-eq} as a local source, then we must remove the contribution from long wavelength scalar modes. However, if we are dealing with propagation effects, the amplitude of $\Phi$ is very real as it can be seen in the gravitational lensing of gravitational waves  and time-delay effects (for the latter case see, e.g. the analysis in~\cite{Bartolo:2018rku}).

If we go to the next to leading order in the gradient expansion, we roughly expect that 
\begin{align}\label{eq:philongexp}
\Phi_{\rm long}\approx {\rm constant}+{\cal O}\left(\epsilon^2\right)\Phi_{\rm long}\,,
\end{align}
where $\epsilon^2=L_\gamma^2/L_\Phi^2\approx q^2/k^2$, where $q$ is the wavenumber of $\Phi$ and $k$ the wavenumber of $\gamma_{ij}$. The fact that it starts at second order in $\epsilon$ is because of symmetry: the ``amplitude" of $\Phi$ cannot depend on the direction of $\bm{q}$. This is consistent with the superhorizon expansion of $\Phi$, i.e. $T_\Phi(x\ll1)=1+{\cal O}(x^2)$. This argument is also supported by the existence of a local inertial frame. In the so-called Conformal Fermi Coordinates, see e.g. Ref.~\cite{Dai:2015rda} for the local expansion of the metric (although in a very different context), the scalar piece of the spatial component of the metric expanded around the local Fermi frame reads
\begin{align}
g_{ij}(x_F)=a^2(\eta_F)\left[\delta_{ij}-\frac{1}{3}R^F_{ikjl}x_F^kx_F^l\right]\,,
\end{align}
where the subscript “F” refers to evaluation at the Fermi frame. We then have that at leading order in $\Phi_F$
\begin{align}
R^F_{ikjl}x_F^kx_F^l=\frac{1}{2}\left(x_{i,F}x^k_F \partial_k\partial_j \Phi_F+x_{j,F}x^k_F \partial_k\partial_i \Phi_F-\partial_i\partial_j \Phi_F x_F^k x_{k,F}-\delta_{ij}x^k_{F}x^l_F \partial_k\partial_l \Phi_F\right)\,.
\end{align}
Since $\partial_i\partial_j \Phi_F\sim q^2 \Phi_F$ and $x_F\lesssim L_\gamma\sim 1/k$ (i.e. the expansion is valid on the surroundings of the point $x_F$ which are smaller than the tensor wavelength), this is consistent with Eq. \eqref{eq:philongexp} since
\begin{align}
R^F_{ikjl}x_F^kx_F^l\approx {\cal O}(\epsilon^2)\Phi_F\,.
\end{align}

However, note that although we understand the dependence on $\epsilon$, the coefficient that would enter in Eq.~\eqref{stnow} is not determined. To do that, we need a careful treatment of these subtle coordinate transformations and their relation between gauge transformations. This is out of the scope of this paper and we leave it for future work. Instead, we use the above arguments to propose a \textit{phenomenological} solution. We interpolate between $P_\Phi$ and $P_{\nabla^2\Phi}$ for short and long wavelength scalar modes by including in the integral Eq. \eqref{omega1} the following function of $u$:
\begin{align}
f(u)\equiv\frac{u^4}{d^4+u^4}\,.
\end{align}
This function goes from $f(u\gg1)\sim 1$ to $f(u\ll1)\sim (u/d)^4$ and cures any divergence. This also removes the need to subtract superhorizon contributions as they are naturally suppressed. The factor $d$ reflects the uncertainty in the right moment for the transition between short and long wavelength scalar modes which should be $d\sim {\cal O}(0.1)- {\cal O}(1)$. Thus, we propose the regular form of the scalar-tensor induced GWs as
\begin{align}\label{omega}
      \Omega^{\rm {st-ind}}_{\rm {GW,\,R/L}}(k)&=\frac{1}{384} \int_0^\infty dv \int_{|v-1|}^{v+1}  \,
      \frac{du}{v^6u^2} \,\Delta^2_{\Phi}(uk)\, f(u)\,\overline{\mathcal{I}_\infty^2(u,v)}\, \nonumber\\
       &\hspace*{10mm}\times \left[ \left((v+1)^2-u^2\right)^4\,\Delta^{2}_{{\gamma}_0,\, \rm R/L}(vk)+\left((v-1)^2-u^2\right)^4\,\Delta^{2}_{{\gamma}_0, \,\rm L/R}(vk)\right]\,.
\end{align}
For example, if we take $d=1$, $\Delta^2_\Phi=A_\Phi$ and $\Delta^2_{\gamma,\,R/L}=A_{\gamma,\,R/L}$ we have
\begin{align}\label{eq:stindflat}
\Omega^{\rm {st-ind}}_{\rm {GW,\,R/L,\,c}}(k)\approx A_\Phi\left(0.48A_{\gamma,\,R/L}+0.043A_{\gamma,\,L/R}\right)\,,
\end{align}
which is a sensible result if one compares it to the scalar-scalar induced GWs for flat primordial spectrum, which gives $\Omega^{\rm {ss-ind}}_{\rm {GW}}\approx 0.82\,A_\Phi^2$ \cite{Kohri:2018awv,Balaji:2022dbi}. Eq.~\eqref{eq:stindflat} also tells us that if we have primordial tensor parity violation with a scale invariant spectrum, the scalar-tensor induced GWs are also scale invariant and mostly parity violating. 

For a Dirac delta spectrum, we have that $u=v\geq 1/2$, so that the correction from $f(u)$ is bounded below by $f(u)\geq 2^{-4}/(d^4+2^{-4})$. If we choose $d=1$, the suppression factor at $u=v=1/2$ is $\sim 0.06$, quickly becoming $1$ for $u=v>1$. However, we do not expect large corrections for the Dirac delta case with the same peak location, which means that the factor $d$ should be $O(0.1)$, giving us $f(u) \gtrsim 0.99$, and barely modifying the prediction in the Dirac delta case. This also implies that the suppression effect should only be important when the scalar mode has at least a wavelength ten times larger than the tensor mode.

Note that one may argue that there should be a gauge in which such divergences do not occur and the amplitude of curvature perturbation does not appear directly into the equations of motion for tensor modes. If that is the case, that particular gauge might be more appropriate for calculations. Unfortunately, we have found no standard gauge choice in which that occurs. For example, the third order action in Maldacena's paper in the flat and comoving gauges contain these kind of couplings. So if such gauge exist is by no means a trivial transformation. And, importantly, we would then enter into the discussion of the well-known gauge dependence of the energy density of GWs at second order in perturbation theory. For these reasons, we propose a method within the Poisson gauge which, besides some uncertainty, should provide a good estimate for the correct scalar-tensor induced GWs spectrum.

\section{Summary and discussion}\label{sec:5}

While the focus of GWs hunt has traditionally been on first-order GWs generated during primordial inflation, there is also a growing interest in the second-order effects, specifically scalar-induced GWs (SIGWs). These SIGWs arise from the nonlinear interactions between scalar perturbations and GWs, and their detection could provide valuable insights into the early Universe and inflationary models. In this context, we are investigating a nonlinear interaction between first-order scalar perturbations and first-order tensor perturbations. Instead of directly generating gravitational waves from zero, these interactions modulate the existing waves. While similar modulated waves have been studied before \cite{Gong:2019mui,Chang:2022vlv} our analysis generalizes, for the first time, to the case of general chirality of primordial GWs \cite{Obata:2016tmo,Bartolo:2017szm,Bartolo:2018elp,Komatsu:2022nvu}, hoping to address the detectability of this effect compared to SIGWs. Our work aims to explore the distinctive features and implications of chiral modulated gravitational waves.

We have presented an analysis of both chiral and non-chiral peaked gravitational waves, modulated by a peaked scalar perturbation. Our investigation aimed to determine the extent to which these modulated waves are buried under the dominant SIGWs. Interestingly, while non-chiral waves are expected to surpass SIGWs only in a limited range of wave numbers, we have discovered a distinguishing feature in the case of chiral waves. This feature enables us to differentiate between the two induced polarization modes, providing a potential avenue for detecting and characterizing the modulated waves amidst the dominant SIGWs. We also anticipate the possibility of a universal unpolarised IR behaviour of induced GWs. However, further investigation is needed to confirm and fully understand this effect. 

Another significant finding in this study is the identification of a potential divergence in the momentum integral of the kernel. This divergence emerges due to the existence of exceptionally long wavelengths in the scalar modes, compared to that of the   tensor modes.  It is important to note that such divergences do not occur in the SIGWs case, as it involves a source term that includes the gradient of the gravitational potential $\Phi$. However, in the current scenario, the divergence arises from the multiplication of the gravitational potential $\Phi$ with the gradient of the tensor mode \footnote{To avoid any confusion, we want to emphasize that our divergence is unrelated to the renormalization process in Quantum Field Theories (QFTs). Instead, it is a purely classical calculation.}. We observe this potential divergence for the first time in the context of scalar-modulated tensor modes, originating from the long-wavelength $\Phi$ modes. To address this divergence issue, a possible solution involves separating the long and short modes of the scalar field and absorbing the long modes into a new background. This approach can help mitigate the divergence problem. However, to fully resolve this issue, further investigation is required, and it will be explored in future work. A factor with a free parameter is introduced in the kernel to address the divergence, but its effectiveness and implications need to be thoroughly investigated in subsequent studies.

\section*{Acknowledgements}
We would like to thank J.L.~Bernal, A.~Caravano, E.~Komatsu, A.~Ricciardone, F.~Schmidt and M.~Sasaki for valuable discussions.  G.D. is supported by the DFG under the Emmy-Noether program grant no. DO 2574/1-1, project number 496592360. Calculations of the scalar-scalar induced GW spectrum have been done with  \href{https://github.com/Lukas-T-W/SIGWfast/releases}{\textsc{SIGWfast}} \cite{2022arXiv220905296W}. P.B.  acknowledges funding from Italian Ministry of Education, University and Research (MIUR) through the ``Dipartimenti di eccellenza'' project Science of the Universe. P.B., N.B., and S.M. acknowledge support from the COSMOS network (www.cosmosnet.it) through the ASI (Italian Space Agency) Grants 2016-24-H.0, 2016-24-H.1-2018 and 2020-9-HH.0. For N.B. and S.M., it is a pleasure to recall that some preliminary notes about the GWs propagation including scalar and tensor perturbations at second-order, on which the starting of this project was based, were already shared with Lev Kofman during a visit at the Physics Department of Padova. 

\appendix 
   \section{Evolution equation of scalar-tensor induced gravitational waves}\label{App-st1}
We expand the three-space metric in Eq. \eqref{eq:metric}, keeping the tensor terms up to the linear order, as 
\begin{align}
    h_{ij}&=a^2 e^{-2\Psi} (\delta_{ij}+\gamma_{ij})\,,\\
   h^{ij}&=a^{-2} e^{2\Psi} (\delta^{ij}- \gamma^{ij})\,.
\end{align}
With these metric components, the extrinsic curvature reads \cite{Salopek:1990jq}
\begin{equation}\label{K}
    K_{ij}=-\frac{1}{2N}\,\dot{h}_{ij}\,.
\end{equation}
To get an evolution equation of the tensor perturbations sourced by a mixing of first order scalar and tensor ones, we focus on the trace-less part of the ij-th Einstein equation. It can be written as  below \cite{Salopek:1990jq},  defining the  trace-less part of Eq. \eqref{K} as $\overline{K}_{ij}= K_{ij}-\frac{1}{3}K h_{ij}$,  
\begin{equation}\label{tr}
    \frac{\partial \overline{K}^i_k}{\partial t} =-N^{|i}_{|k}+\frac{1}{3} N^{|l}_{|l}\,\delta^i_k+N\,(K\overline{K}^i_k +^{(3)}\overline{R}^i_k-8\pi G\, \overline{S}^i_k)\,,
\end{equation}
where $^{(3)}\overline{R}^i_k$ is the Ricci tensor associated with the three-metric $h_{ik}$, and $\overline{S}^i_k$ is the space-space contribution of the matter energy-momentum tensor.
Vertical
bars denote three-space covariant derivatives with connection
coefficients determined from $h_{ij}$. Trace and trace-less parts of $ K_{ij}$ are, respectively,
\begin{align}
    K&=-\frac{1}{2N}\,h^{ij}\dot{h}_{ij}=-\frac{3}{N}\,(H-\dot{\Psi})\,,\\
    \overline{K}^i_k&=h^{ij}\overline{K}_{jk}\,,\nonumber\\
   & = -\frac{1}{2N}\,\dot{\gamma}^i_k\,.
\end{align}
We obtain the rest of the terms of the the trace-less ij-th equation Eq. \eqref{tr} from $h_{ij}$, 
\begin{align}
    N^{|i}_{|k}&=\gamma^{ij} N_{|jk}= \frac{Ne^{2\Psi}}{a^2}\left[\Phi^{,i}_{,k}+\Phi^{,i}\Phi_{,k}-\Phi_{,m}\Psi^{,m}\delta^i_k+\Phi^{,i}\Psi_{,k}+\Psi^{,i}\Phi_{,k}+\frac{1}{2}\Phi_{,m}\gamma^{i,m}_k-\frac{1}{2}\Phi_{,m}{\gamma^{im}}_{,k}\right.\nonumber\\&\left.-\frac{1}{2}\Phi_{,m}\gamma^{m,i}_k+\gamma^{mt}\Phi_{,m}\Psi_{,t}\delta^i_{k}-\gamma^{ij}\Phi_{,jk}-\gamma^{ij}\Phi_{,j}\Phi_{,k}-\Psi_{,k}\Phi_{,j}\gamma^{ij}-\Phi_{,k}\Psi_{,j}\gamma^{ij}\right]\,,\\
    N^{|l}_{|l}&= \frac{Ne^{2\Psi}}{a^2}\left[\nabla^2\Phi-\Psi^{,l}\Phi_{,l}+\Phi^{,l}\Phi_{,l}-\Phi_{,m}\gamma^{lm}_{,l}+\gamma^{mt}\Phi_{,m}\Psi_{,t}-\Phi_{,ml}\gamma^{lm}-\Phi_{,m}\Phi_{,l}\gamma^{lm}\right]\,,\\
    \tensor[^{(3)}]{R}{_{jk}}&=\nabla^2\Psi (\delta_{jk}+\gamma_{jk} )-\frac{1}{2}\nabla^2 \gamma_{jk}-\frac{1}{2} \gamma_{jk,m}\Psi^{,m}+\Psi_{,j}\Psi_{,k}+\Psi_{,jk}+\frac{1}{2}\gamma^m_{j,km}+\frac{1}{2}\gamma^m_{k,jm} \nonumber\\&-\gamma^{mt}_{,m}\Psi_{,t}\delta_{jk}-\gamma^{mt}\Psi_{,tm}\delta_{jk}-\Psi^{,\lambda}\Psi_{,\lambda}\delta_{jk}-\Psi^{,\lambda}\Psi_{,\lambda}\gamma_{jk}-\frac{1}{2}\Psi_{,\lambda}\gamma^{\lambda}_{j,k}-\frac{1}{2}\Psi_{,\lambda}\gamma^{\lambda}_{k,j}+\Psi_{,\lambda}\Psi_{,t}\gamma^{\lambda t}\delta_{jk}\,,\\
      \tensor[^{(3)}]{R}{^i_k}&=\gamma^{ij}\,  \tensor[^{(3)}]{R}{_{jk}}= \frac{e^{2\Psi}}{a^2}\left[\nabla^2\Psi \delta^i_{k}-\frac{1}{2}\nabla^2 \gamma^i_{k}-\frac{1}{2}\gamma^i_{k,m}\Psi^{,m}+\Psi^{,i}\Psi_{,k}+\Psi^{,i}_{,k}+\frac{1}{2}\gamma^{mi}_{,km}+\frac{1}{2}\gamma^{m,i}_{k,m}\right.\nonumber\\&\left. -\gamma^{mt}_{,m}\Psi_{,t}\delta^i_{k}-\gamma^{mt}\Psi_{,tm}\delta^i_{k}-\Psi^{,\lambda}\Psi_{,\lambda}\delta^i_{k}-\frac{1}{2}\Psi_{,\lambda}\gamma^{i\lambda}_{,k}-\frac{1}{2}\Psi_{,\lambda}\gamma^{\lambda,i}_{k}\right.\nonumber\\&\left.+\gamma^{\lambda t}\Psi_{,t}\Psi_{,\lambda}\delta^i_{k}-\Psi_{,jk}\gamma^{ij}-\Psi_{,k}\Psi_{,j}\gamma^{ij} \right]\,,\\
       \tensor[^{(3)}]{R}{}&=  \tensor[^{(3)}]{R}{^i_i}=\frac{e^{2\Psi}}{a^2}\left[4\nabla^2\Psi-2\Psi^{,i}\Psi_{,i}+\gamma^{mi}_{,im} -3 \gamma^{m t}_{,m}\Psi_{,t}-3\gamma^{mt}\Psi_{,tm}-\Psi_{,m}\gamma^{im}_{,i}\right.\nonumber\\&\left.+3\gamma^{m t}\Psi_{,t}\Psi_{,m}-\Psi_{,ji}\gamma^{ij}-\Psi_{,i}\Psi_{,j}\gamma^{ij}\right]\,,\\
        \tensor[^{(3)}]{\overline{R}}{^i_k}&=  \tensor[^{(3)}]{R}{^i_k}-\frac{1}{3} \tensor[^{(3)}]{R}{}\, \delta^i_k= \frac{e^{2\Psi}}{a^2}\left[-\frac{1}{3}\nabla^2\Psi\delta^i_k-\frac{1}{2}\nabla^2\gamma^i_k-\frac{1}{2}\Psi^{,m}\gamma^{i}_{k,m}+\Psi^{,i}\Psi_{,k}+\Psi^{,i}_{,k}\right.\nonumber\\&\left.+\frac{1}{2}\gamma^{mi}_{,km}+\frac{1}{2}\gamma^{m,i}_{k,m}+\frac{1}{3}\gamma^{m t}\Psi_{,t}\Psi_{,m}\delta^i_{k}+\frac{1}{3}\gamma^{m t}\Psi_{,mt}\delta^i_{k} -\frac{1}{3}\Psi^{,m}\Psi_{,m}\delta^i_{k}
   - \frac{1}{2}\gamma^{im}_{,k}\Psi_{,m} \right.\nonumber\\&\left.- \frac{1}{2}\gamma^{m,i}_{k}\Psi_{,m}-\gamma^{ij}\Psi_{,jk}-\gamma^{ij}\Psi_{,k}\Psi_{,j}-\frac{1}{3}\gamma^{m t}_{,mt}\delta^i_{k}+\frac{1}{3}\gamma^{mt}_{,t}\Psi_{,m}\delta^i_{k}-\frac{1}{3}\nabla^2(\Phi-\Psi)\delta^i_{k}\right]\,.
\end{align}
Hence Eq. \eqref{tr} becomes
\begin{align}\label{main-app}
     &\ddot{\gamma}^i_k-\dot{\Phi}e^{\Phi}\dot{\gamma}^i_k+3(H-\dot{\Psi})\dot{\gamma}^i_k-\frac{e^{2(\Phi+\Psi)}}{a^2}\nabla^2 \gamma^i_k= 2 \frac{e^{2(\Phi+\Psi)}}{a^2}\left[\Phi^{,i}_{,k}+\Phi^{,i}\Phi_{,k}-\Psi^{,i}\Psi_{,k}-\Psi^{,i}_{,k}\right.\nonumber\\&\left.+\Phi^{,i}\Psi_{,k}+\Psi^{,i}\Phi_{,k}-\frac{2}{3}\Phi^{,l}\Psi_{,l}\delta^i_k-\frac{1}{3}\Phi^{,l}\Phi_{,l}\delta^i_k+\frac{1}{3}\Psi^{,l}\Psi_{,l}\delta^i_k\right]+\mathcal{S}\,, 
\end{align}
where $\mathcal{S}$ contains scalar-tensor mixed terms like $\sim \Phi(\gamma_{ij})$, and contribution from the matter component of the Universe, $\overline{S}^i_k$,
\begin{align}\label{S}
   \mathcal{S}&=2\frac{e^{2(\Phi+\Psi)}}{a^2}\left[-\gamma^{ij}(\Phi-\Psi)_{,jk}+\frac{1}{3}\gamma^{lm}(\Phi-\Psi)_{,lm}\,\delta^i_k-\frac{1}{2}{\gamma^{im}}_{,k}(\Phi-\Psi)_{,m}+\frac{1}{2}\gamma^{i,m}_{k}(\Phi-\Psi)_{,m}\right.\nonumber\\&\left.-\frac{1}{2}\gamma^{m,i}_{k}(\Phi-\Psi)_{,m}+\frac{1}{3}{\gamma^{lm}}_{,l}(\Phi-\Psi)_{,m}\delta^i_k-\gamma^{ij}\Psi_{,k}\Phi_{,j}-\gamma^{ij}\Phi_{,k}\Psi_{,j}-\gamma^{ij}\Phi_{,k}\Phi_{,j}\right.\nonumber\\&\left.\frac{1}{3}\gamma^{lm}\Phi_{,l}(\Phi+2\Psi)_{,m}\delta^i_k+\gamma^{ij}\Psi_{,k}\Psi_{,j}-\frac{1}{3}\gamma^{lm}\Psi_{,l}\Psi_{,m}\delta^i_k-\frac{1}{2}\gamma^{m,i}_{k,m}-\frac{1}{2}\gamma^{im}_{,km}+\frac{1}{3}\gamma^{lm}_{,lm}\delta^i_k\right]\nonumber\\&+16\pi G e^{2\Phi}\overline{S}^i_k\,.
\end{align}
These equations lead to Eq. \eqref{main}, after  collecting scalar-scalar and scalar-tensor terms.

\section{calculation of the kernel function}
The dimension-less power-spectrum for each polarization $\lambda$ of the correction to GWs, $\gamma_1(\bm{k},\eta)$, i.e. the second term on the right hand side of Eq. \eqref{sol}, is given by Eq. \eqref{finchiral}, where we have  used the definition of the two-point function of $\Phi$ and $\gamma^{(\sigma)}_{\bm{k}}$ 
    \begin{align}
       \langle\Phi_{\bm{k}-\bm{k}_1}(0)\Phi_{\bm{k'}-\bm{k}'_1}(0)\rangle&= (2\pi)^3 \delta^3(\bm{k}-\bm{k}_1+\bm{k}'-\bm{k}'_1) \frac{2\pi^2}{|\bm{k}-\bm{k}_1|^3}\Delta^2_{\Phi}(|\bm{k}-\bm{k}_1|)\,, \\
   \langle \gamma^{(\sigma)}_{\bm{k}_1}(0)\gamma^{(\sigma')}_{\bm{k}'_1}(0))\rangle &= (2\pi)^3 \delta^3(\bm{k}_1+\bm{k}'_1) \delta_{\sigma \sigma'}\frac{2\pi^2}{k_1^3}\Delta^{(\sigma)2}_{\gamma_0}(k)\,.
   \end{align}
Contraction of the polarisation tensors give
\begin{align}\label{pow-R-app}
     \Delta^2_{\gamma_1, \rm R/L}(k)&= \frac{k^3}{\pi}\int 
       d^3{k}_1\frac{\Delta^2_{\Phi}(|\bm{k}-\bm{k}_1|) }{k_1^3 
       |\bm{k}-\bm{k}_1|^3} \left[4 \cos^8{\theta/2}\,\Delta^{(\sigma)2}_{\gamma_0, \rm R/L}(k_1)+4 \sin^8{\theta/2}\,\Delta^{(\sigma)2}_{\gamma_0, \rm L/R}(k_1)\right]\nonumber\\
       &\times \left( \int_0^\eta d\tilde{\eta}   G(\eta, \tilde{\eta})\left[k_1^2 T_{\gamma}(k_1\tilde{\eta} ) T_\Phi(|\bm{k}-\bm{k}_1|\tilde{\eta} )-T'_{\gamma}(k_1\tilde{\eta} ) T'_\Phi(|\bm{k}-\bm{k}_1|\tilde{\eta} )\right]\right)^2,
\end{align}
which leads to Eq. \eqref{pow-R}. The Green's function is given by \begin{align}\label{G}
        G(x, \tilde{x})&=\frac{\pi}{2k}\,\tilde{x} \,\sqrt{\frac{\tilde{x}}{x}}\,\left(J_{1/2}(\tilde{x})Y_{1/2}(x)-J_{1/2}(x)Y_{1/2}(\tilde{x})\right)\,.
   \end{align}
Here the kernel has been written as,  taking the upper limit of the time integral to be $\infty$,
\begin{align}\label{kernel-st-app}
       &\mathcal{I}= \int_0^\infty k \,d\tilde{x}\,  G(x, \tilde{x})\,\left[v^2 T_{\gamma}(v\tilde{x}) T_\Phi(c_s u\tilde{x})-\dot{T}_{\gamma}(v\tilde{x}) \dot{T}_\Phi(c_s u\tilde{x})\right]\,,\nonumber\\
       &=  \frac{\pi}{4} \frac{v}{c_su}\,\frac{1}{x}\, \Bigg\{-\cos{x}\left(1-P^0_2(\cos{m})\right)\Theta\left(v+c_s u-1\right)\Theta\left(1-\left|v-c_s u\right|\right)\nonumber\\
       &-\frac{2}{\pi}\sin{x}\left[(Q^0_0(\cosh{n})-Q^0_2(\cosh{n}))\Theta\left(1-v-c_s u\right)\right.\nonumber\\&\left.-(Q^0_0(\cos{m})-Q^0_2(\cos{m}))\Theta\left(v+c_s u-1\right)\Theta\left(1-\left|v-c_s u\right|\right)\right]\Bigg\}\,,
       \end{align}
       where $2uvc_s\cos{m}=v^2+c_s^2u^2-1$ and $2uvc_s\cosh{n}=1-v^2-c_s^2u^2$, and $P^l_m, Q^l_m$ are the associated Legendre polynomials of the first and second kind.
       It gives the oscillation average of the kernel squared ($\cos{m}=s=-\cosh{n}$)
       \begin{align}\label{oscav-app}
           \langle\mathcal{I}^2\rangle
       &=\frac{9}{2^7x^2}  \left(\frac{v}{c_su}\right)^2\left[\pi^2(1-s^2)^2 \Theta\left(1-|s|\right)+\Big(2s+(1-s^2)\log \left|\frac{1+s}{1-s}\right|\Big)^2\right]\,.
       \end{align}
\section{Useful formulae for the calculation of the kernel function in the induced gravitational waves  integral}\label{App-st2}
      We here write down some formulae used to calculate Eq. \eqref{kernel-st}
      with $ G(x, \tilde{x})$ defined in Eq. \eqref{G}, and using
      \vspace{-3mm}
      \begin{align}
       T_\Phi(x)&=2^{3/2}\Gamma (5/2)\left(\frac{x}{\sqrt{3}}\right)^{-3/2}J_{3/2}\left(\frac{x}{\sqrt{3}}\right)\,,\nonumber\\
       \dot{T}_\Phi(ux)&= -\frac{3}{x}j_2 (ux/\sqrt{3})=-\frac{3^{5/4}}{x\sqrt{x}}\sqrt{\frac{\pi}{2u}}J_{5/2} (ux/\sqrt{3})\,, \nonumber\\       T_{\gamma}(x)&=\sqrt{\frac{\pi}{2x}} J_{1/2}(x)\,,\nonumber\\
       \dot{T}_{\gamma}(vx)&= -v j_1(vx)=-\sqrt{\frac{\pi v}{2x}} J_{3/2} (vx)\,,
   \end{align}
   the kernel turns out to be, for radiation domination ($c_s=1/\sqrt{3}$)
     \begin{align}
       \mathcal{I}&=  \int_0^\infty  d\tilde{x}  \left(\frac{\pi}{2}\right)^2 3^{5/4}\sqrt{\frac{v}{u}}\frac{1}{\sqrt{x \tilde{x}}}\left(J_{1/2}(\tilde{x})Y_{1/2}(x)-J_{1/2}(x)Y_{1/2}(\tilde{x})\right)\nonumber\\&\times
       \left[\sqrt{3}\frac{v}{u}J_{1/2}(v\tilde{x})J_{3/2} (u\tilde{x}/\sqrt{3})-J_{3/2}(v\tilde{x})J_{5/2} (u\tilde{x}/\sqrt{3})\right]\,.
       \end{align}
        Using the recurrence relation $2n/z J_n(z)=J_{n-1}(z)+J_{n+1}(z)$, we have, putting $n=3/2$,
        \begin{align}
       \mathcal{I}&=  \int_0^\infty d\tilde{x}  \left(\frac{\pi}{2}\right)^2 v\sqrt{\frac{v}{u/\sqrt{3}}}\sqrt{\frac{\tilde{x}}{x}}\left(J_{1/2}(\tilde{x})Y_{1/2}(x)-J_{1/2}(x)Y_{1/2}(\tilde{x})\right)\nonumber\\&\times
       \left[J_{1/2}(v\tilde{x})J_{1/2} (u\tilde{x}/\sqrt{3})-J_{5/2}(v\tilde{x})J_{5/2} (u\tilde{x}/\sqrt{3})\right],\nonumber\\\label{finkernel}
       &=  \left(\frac{\pi}{2}\right)^2 v\sqrt{\frac{v}{u/\sqrt{3}}}\frac{1}{\sqrt{x}}\Bigg\{Y_{1/2}(x)\int_0^\infty d\tilde{x} \sqrt{\tilde{x}}J_{1/2}(\tilde{x}) \left[J_{1/2}(v\tilde{x})J_{1/2} (u\tilde{x}/\sqrt{3})-J_{5/2}(v\tilde{x})J_{5/2} (u\tilde{x}/\sqrt{3})\right]\nonumber\\
       &-J_{1/2}(x) \int_0^\infty d\tilde{x} \sqrt{\tilde{x}}Y_{1/2}(\tilde{x})
       \left[J_{1/2}(v\tilde{x})J_{1/2} (u\tilde{x}/\sqrt{3})-J_{5/2}(v\tilde{x})J_{5/2} (u\tilde{x}/\sqrt{3})\right]\Bigg\}\,.
       \end{align}
       Using the formulae given in \cite{1985JMP....26..633G}, we have 
       \begin{align}
          &\int_0^\infty d\tilde{\tau} \sqrt{\tilde{x}}\,J_{1/2}(\tilde{x}) J_{1/2}(v\tilde{x})J_{1/2} (u\tilde{x}/\sqrt{3}) = \begin{cases}
             \sqrt{\frac{\sqrt{3}}{2\pi vu}} & \quad  |v-\frac{u}{\sqrt{3}}|<1<v+\frac{u}{\sqrt{3}}\\ 
             0 & \quad  |v-\frac{u}{\sqrt{3}}|>1 \\
            & \quad  \text{or}\, v+\frac{u}{\sqrt{3}}<1
          \end{cases}\nonumber\\
          &\int_0^\infty d\tilde{x} \sqrt{\tilde{x}}\,J_{1/2}(\tilde{x}) J_{5/2}(v\tilde{x})J_{5/2} (u\tilde{x}/\sqrt{3})= \begin{cases}
             \sqrt{\frac{\sqrt{3}}{2\pi vu}}\,P^0_2(\cos{m})\\ \text{for}\, |v-\frac{u}{\sqrt{3}}|<1<v+\frac{u}{\sqrt{3}}\\ 0 \\ \text{for}\, 1< |v-\frac{u}{\sqrt{3}}|\, \text{or}\, 1>v+\frac{u}{\sqrt{3}}
          \end{cases}\nonumber\\
           &\int_0^\infty d\tilde{x} \sqrt{\tilde{x}}\,Y_{1/2}(\tilde{x}) J_{1/2}(v\tilde{x})J_{1/2} (u\tilde{x}/\sqrt{3}) = \begin{cases}
             -\frac{1}{\pi}\sqrt{\frac{2\sqrt{3}}{\pi vu}}\,Q^0_0(\cos{m}) \\ \text{for}\,  |v-\frac{u}{\sqrt{3}}|<1<v+\frac{u}{\sqrt{3}}\\ 
             \frac{1}{\pi}\sqrt{\frac{2\sqrt{3}}{\pi vu}}\,Q^0_0(\cosh{n}) \\ \text{for}\, 1>v+\frac{u}{\sqrt{3}}
          \end{cases}\nonumber\\
           &\int_0^\infty d\tilde{x} \sqrt{\tilde{x}}\,Y_{1/2}(\tilde{x}) J_{5/2}(v\tilde{x})J_{5/2} (u\tilde{x}/\sqrt{3}) = \begin{cases}
             -\frac{1}{\pi}\sqrt{\frac{2\sqrt{3}}{\pi vu}}\,Q^0_2(\cos{m}) \\ \text{for}\,  |v-\frac{u}{\sqrt{3}}|<1<v+\frac{u}{\sqrt{3}}\\ 
             \frac{1}{\pi}\sqrt{\frac{2\sqrt{3}}{\pi vu}}\,Q^0_2(\cosh{n}) \\ \text{for}\, 1>v+\frac{u}{\sqrt{3}}
          \end{cases}
       \end{align}
        where $2uv/\sqrt{3}\cos{m}=v^2+u^2/3-1$ and $2uv/\sqrt{3}\cosh{n}=1-v^2-u^2/3$, and
         \begin{align}
           P^0_2(\cos{m})&=\frac{3\cos^2{m}-1}{2}\,,\\
           Q^0_0(\cos{m})&=\frac{1}{2}\ln{\frac{1+\cos{m}}{1-\cos{m}}}\,,\\
           Q^0_2(\cos{m})&= \frac{3\cos^2{m}-1}{4}\ln{\frac{1+\cos{m}}{1-\cos{m}}}-\frac{3\cos{m}}{2}\,.
       \end{align}
       Applying all these, we have Eq. \eqref{kernel-st-app}.
       Remembering that $\langle \cos^2{x}\rangle=\langle \sin^2{x}\rangle=1/2$, we have, for the oscillation average, Eq. \eqref{oscav}. The same procedure can be applied to the scalar induced GWs.

     \section{Scalar-induced tensor perturbations}\label{App-sigw}
Since we have assumed $A_{\gamma_0} < A_\Phi$, the SIGWs are expected to have larger amplitudes compared to our modulated GWs. To assess the magnitude of this difference and evaluate the prospects of detecting our effect, we compare the amplitudes of SIGWs to those of our modulated GWs.
To get the SIGWs, we need the matter contribution in Eq. \eqref{S}. Previously it was ignored as we considered linear scalar-linear tensor and it contains no linear tensor if not some anisotropic stress. Now,
\begin{align}
     \overline{S}^i_k&=h^{ij}\overline{S}_{jk}\nonumber\\
     &= S^i_k-\frac{1}{3}(\delta^i_k -\gamma^{ij}\gamma_{jk})(\delta^n_p -\gamma^{mn}\gamma_{np})S^p_{n}\,,
\end{align}
where
\begin{align}
    S^i_{k}&= (\rho+P)u^i u_j+P\delta^i_j\nonumber\\&=(\overline{\rho}+\overline{P})V^2+3(\overline{P}+\delta P_1+\delta P_2)-\overline{P}\gamma^{mn}\gamma_{mn}\,.
\end{align}
Hence,
\begin{align}
     \overline{S}^i_k&=(\overline{\rho}+\overline{P})(v^iv_k-\frac{1}{3}v^2\delta^i_k)
     +\overline{P}(\gamma^{ij}\gamma_{jk}+\frac{1}{3}\gamma^{mn}\gamma_{mn}\delta^i_k)\,.
\end{align}
Here $\overline{\rho},\overline{P}$ are the background energy density and pressure respectively. The perturbation in pressure in first and second orders are given by $\delta P_1, \delta P_2$ respectively.
For scalar-scalar interaction, we need only the first term. As we are taking transverse trace-less component, there is no first order contribution. The evolution equation becomes
\begin{align}
     \ddot{\gamma}^i_k+3H\dot{\gamma}^i_k+\frac{k^2}{a^2} \gamma^i_k&=\frac{4}{a^2}\Phi^{,i}\Phi_{,k}+16\pi G(\overline{\rho}+\overline{P})V^iV_k\,,
\end{align}
where 
\begin{equation}
    V^i=-\frac{2}{8\pi G a^2(\overline{\rho}+\overline{P})}\partial^i(\Phi'+\mathcal{H}\Phi)\,.
\end{equation}
In terms of conformal time and Fourier space, it is
\begin{align}
\gamma^{''}_{\bm{k}}+2\mathcal{H}\gamma'_{\bm{k}}+k^2\gamma_{\bm{k}}&=4\int \frac{d^3\bm{k_1}}{(2\pi)^3} \Phi_{\bm{k}}(0)\Phi_{\bm{k}-\bm{k_1}}(0) \epsilon^{k}_i(\bm{\hat{k}}) k_1^i k_{1k}\, \left[T_\Phi(k_1\eta)T_\Phi(|\bm{k}-\bm{k_1}|\eta)\right.\nonumber\\&\left. +\frac{1}{2}\left(\mathcal{H}T_\Phi(k_1\eta)+T'_\Phi(k_1\eta)\right)\left(\mathcal{H}T_\Phi(|\bm{k}-\bm{k_1}|\eta)+T'_\Phi(|\bm{k}-\bm{k_1}|\eta)\right)\right]\,.
\end{align}
    The sum of polarization states is
   \begin{align}
      \sum_{\lambda=+,\times}(\epsilon^{ik}_{(\lambda)}(\bm{\hat{k}})k_{1\, i}k_{1\, k})^2&=k_1^4 \Big(1-\Big(\frac{1+x^2-y^2}{2x}\Big)^2\Big)^2=k_1^4\sin^4{\theta}\,.
   \end{align}
  Defining the kernel as
  \begin{align}
      I(k_1, |\bm{k}-\bm{k}_1|)&= \int_0^\eta d\tilde{\eta}\,   G(\eta, \tilde{\eta}) \left[T_\Phi(k_1\eta)T_\Phi(|\bm{k}-\bm{k_1}|\eta) \right.\nonumber\\&\left.+\frac{1}{2} \left(\mathcal{H}T_\Phi(k_1\eta)+T'_\Phi(k_1\eta)\right)\left(\mathcal{H}T_\Phi(|\bm{k}-\bm{k_1}|\eta)+T'_\Phi(|\bm{k}-\bm{k_1}|\eta)\right)\right]\,,
  \end{align}
   the total GWs power-spectrum for both polarization is ($\tau=k\eta$)
   \begin{align}\label{sigwpow}
      &\Delta^2_{\gamma_1}(k)= \frac{8k^3}{\pi} \int d^3\bm{k_1}  \frac{\Delta^2_{\Phi}(k_1)\Delta^2_{\Phi}(|\bm{k}-\bm{k_1}|)}{k^3|\bm{k}-\bm{k_1}|^3} \sum_{\lambda=+,\times}(\epsilon^{ik}_{(\lambda)}(\bm{\hat{k}})k_{1\, i}k_{1\, k})^2 \langle I^2 \rangle\,,\nonumber\\
      &=\frac{8k^3}{\pi} \int d^3\bm{k_1}  \frac{\Delta^2_{\Phi}(k_1)\Delta^2_{\Phi}(|\bm{k}-\bm{k_1}|)}{k^3|\bm{k}-\bm{k_1}|^3}k_1^4\sin^4{\theta}\langle I^2 \rangle\,,\nonumber\\
      &=16 k^2 \int_0^\infty dv \int_{|v-1|}^{v+1} du \, \frac{v^2}{u^2}\, \Big(1-\Big(\frac{1+v^2-u^2}{2v}\Big)^2\Big)^2\Delta^2_{\Phi}(uk)\Delta^2_{\Phi}(vk) \left( \int_0^\tau  d\tilde{\tau} \, G(\tau, \tilde{\tau})\right.\nonumber\\&\left. \times \left[ T_\Phi(v\tilde{\tau})T_\Phi(u\tilde{\tau}) +\frac{1}{2}\left(\mathcal{H}T_\Phi(v\tilde{\tau})+k\dot{T}_\Phi(v\tilde{\tau})\right)\left(\mathcal{H}T_\Phi(u\tilde{\tau})+k\dot{T}_\Phi(u\tilde{\tau})\right)\right]\right)^2\,.
   \end{align}
   The time integral is
    \begin{align}
       I&=  \int_0^\tau  d\tilde{\tau} \, \frac{3\pi^2}{8k\sqrt{uv}}\sqrt{\frac{3\tilde{\tau}}{\tau}}\, \left(J_{1/2}(\tilde{\tau})Y_{1/2}(\tau)-J_{1/2}(\tau)Y_{1/2}(\tilde{\tau})\right)\nonumber\\&\times
       \left[J_{1/2}(v\tilde{\tau}/\sqrt{3})J_{1/2} (u\tilde{\tau}/\sqrt{3})+2J_{5/2}(v\tilde{\tau}/\sqrt{3})J_{5/2} (u\tilde{\tau}/\sqrt{3})\right]\,,
   \end{align}
    and its oscillation average reads (taking the upper limit of the integration as infinity)
     \begin{align}
           \langle I^2\rangle&= \frac{3^4}{2^5k^2} \left(\frac{1}{uv\tau}\right)^2\Bigg\{\Theta\left(\frac{u+v}{\sqrt{3}}-1\right)\frac{9\pi^2}{4}n^4
       +\left(\frac{3n^2}{2}\ln{\frac{1+n}{1-n}}-3n\right)^2\Bigg\}\,.
       \end{align}
       Eq. \eqref{sigwpow} becomes 
        \begin{align}
      &\Delta^2_{\gamma_1}(k)=\frac{3^4}{2} \left(\frac{1}{\tau}\right)^2  \int_0^\infty dv \int_{|v-1|}^{v+1} dy \, \frac{1}{u^4}\, \Big(1-\Big(\frac{1+v^2-u^2}{2v}\Big)^2\Big)^2  \nonumber\\&\times \Delta^2_{\Phi}(uk)\Delta^2_{\Phi}(vk)\Bigg\{\Theta\left(\frac{u+v}{\sqrt{3}}-1\right)\frac{9\pi^2}{4}n^4
       +\left(\frac{3n^2}{2}\ln{\frac{1+n}{1-n}}-3n\right)^2\Bigg\} \,.
       \end{align}
       For a Dirac delta input scalar spectrum, we have
        \begin{align}\label{monosigw}
       &\langle \Delta^2_{\gamma_1}(k)\rangle= 16 A^2_{\Phi}\,  (k_*/k)^2\, \left[1-\frac{k^2}{4k_*^2}\right]^2\langle I^2\rangle_{u=v=k_*/k}\,\Theta(2k_*-k)\,.
   \end{align}

\section{Explicit formulae for the general kernel\label{app:expressionsgeneralkernel}}
Here we present the explicit formulas for the general kernel in \S~\ref{sec:4}. These are given by:
\begin{align}
G_0[{\rm Si} [x]]\equiv  {\rm Si}\left[(1 + c_s u - v) x\right] + 
 {\rm Si}\left[(1 - c_s u + v) x\right] - {\rm Si}\left[(1 + c_s u + v) x\right]-{\rm Si}\left[(1 - c_s u - v) x\right]\,,
\end{align}
\begin{align}
G_1[\cos x]&\equiv  \frac{\cos\left[(1 + c_s u - v) x\right]}{1+c_su-v}+\frac{\cos\left[(1 - c_s u + v) x\right]}{1-c_su+v}  -\frac{\cos\left[(1 + c_s u + v) x\right]}{1+c_su+v}-\frac{\cos\left[(1 - c_s u - v) x\right]}{1-c_su-v}\,,
\end{align}
\begin{align}
G_2[\sin x]&\equiv\left(1+\frac{(1+c_su+v)^2}{4c_suv}\right)\sin\left[(1 - c_s u - v) x\right] + \left(1-\frac{(1-c_su+v)^2}{4c_suv}\right)\sin\left[(1 + c_s u - v) x\right] \nonumber\\&+\left(1-\frac{(1+c_su-v)^2}{4c_suv}\right) 
 \sin\left[(1 - c_s u + v) x\right] +\left(1+\frac{(1-c_su-v)^2}{4c_suv}\right) \sin\left[(1 + c_s u + v) x\right]\,,
\end{align}
\begin{align}
G_3[\cos x]\equiv & -(1 + 3 c_s u +3 v) \cos[(1 - c_s u - v) x] + (1 - 3 c_s u + 
    3 v) \cos[(1 + c_s u - v) x] \nonumber\\&+ (1 + 3 c_s u - 
    3 v) \cos[(1 - c_s u + v) x] - (1 - 3 c_s u - 
    3 v) \cos[(1 + c_s u + v) x]\,,
\end{align}

\begin{align}
G_4[\sin x]\equiv  \sin\left[(1 + c_s u - v) x\right] + 
 \sin\left[(1 - c_s u + v) x\right] - \sin\left[(1 + c_s u + v) x\right]-\sin\left[(1 - c_s u - v) x\right]\,,
\end{align}

\begin{align}
    F_1[\sin x]&\equiv  \frac{\sin\left[(1 + c_s u - v) x\right]}{1+c_su-v}+\frac{\sin\left[(1 - c_s u + v) x\right]}{1-c_su+v}  -\frac{\sin\left[(1 + c_s u + v) x\right]}{1+c_su+v}-\frac{\sin\left[(1 - c_s u - v) x\right]}{1-c_su-v}\,,
\end{align}

\begin{align}
F_2[\cos x]&\equiv\left(1+\frac{(1+c_su+v)^2}{4c_suv}\right)\cos\left[(1 - c_s u - v) x\right] + \left(1-\frac{(1-c_su+v)^2}{4c_suv}\right)\cos\left[(1 + c_s u - v) x\right] \nonumber\\&+\left(1-\frac{(1+c_su-v)^2}{4c_suv}\right) 
 \cos\left[(1 - c_s u + v) x\right] +\left(1+\frac{(1-c_su-v)^2}{4c_suv}\right) \cos\left[(1 + c_s u + v) x\right]\,,
\end{align}
\begin{align}
F_3[\sin x]\equiv & (1 + 3 c_s u +3 v) \sin[(1 - c_s u - v) x] - (1 - 3 c_s u + 
    3 v) \sin[(1 + c_s u - v) x] \nonumber\\&- (1 + 3 c_s u - 
    3 v) \sin[(1 - c_s u + v) x] + (1 - 3 c_s u - 
    3 v) \sin[(1 + c_s u + v) x]\,,
\end{align}
\begin{align}
F_4[\cos x]\equiv  \cos\left[(1 + c_s u - v) x\right] + 
 \cos\left[(1 - c_s u + v) x\right] - \cos\left[(1 + c_s u + v) x\right]-\cos\left[(1 - c_s u - v) x\right]\,.
\end{align}

\bibliography{refgwscalar.bib} 

\begin{thebibliography}{100}%
\makeatletter
\providecommand \@ifxundefined [1]{%
 \@ifx{#1\undefined}
}%
\providecommand \@ifnum [1]{%
 \ifnum #1\expandafter \@firstoftwo
 \else \expandafter \@secondoftwo
 \fi
}%
\providecommand \@ifx [1]{%
 \ifx #1\expandafter \@firstoftwo
 \else \expandafter \@secondoftwo
 \fi
}%
\providecommand \natexlab [1]{#1}%
\providecommand \enquote  [1]{``#1''}%
\providecommand \bibnamefont  [1]{#1}%
\providecommand \bibfnamefont [1]{#1}%
\providecommand \citenamefont [1]{#1}%
\providecommand \href@noop [0]{\@secondoftwo}%
\providecommand \href [0]{\begingroup \@sanitize@url \@href}%
\providecommand \@href[1]{\@@startlink{#1}\@@href}%
\providecommand \@@href[1]{\endgroup#1\@@endlink}%
\providecommand \@sanitize@url [0]{\catcode `\\12\catcode `\$12\catcode
  `\&12\catcode `\#12\catcode `\^12\catcode `\_12\catcode `\%12\relax}%
\providecommand \@@startlink[1]{}%
\providecommand \@@endlink[0]{}%
\providecommand \url  [0]{\begingroup\@sanitize@url \@url }%
\providecommand \@url [1]{\endgroup\@href {#1}{\urlprefix }}%
\providecommand \urlprefix  [0]{URL }%
\providecommand \Eprint [0]{\href }%
\providecommand \doibase [0]{https://doi.org/}%
\providecommand \selectlanguage [0]{\@gobble}%
\providecommand \bibinfo  [0]{\@secondoftwo}%
\providecommand \bibfield  [0]{\@secondoftwo}%
\providecommand \translation [1]{[#1]}%
\providecommand \BibitemOpen [0]{}%
\providecommand \bibitemStop [0]{}%
\providecommand \bibitemNoStop [0]{.\EOS\space}%
\providecommand \EOS [0]{\spacefactor3000\relax}%
\providecommand \BibitemShut  [1]{\csname bibitem#1\endcsname}%
\let\auto@bib@innerbib\@empty
\bibitem [{\citenamefont {Afzal}\ \emph {et~al.}(2023)\citenamefont {Afzal}
  \emph {et~al.}}]{NANOGrav:2023hvm}%
  \BibitemOpen
  \bibfield  {author} {\bibinfo {author} {\bibfnamefont {A.}~\bibnamefont
  {Afzal}} \emph {et~al.} (\bibinfo {collaboration} {NANOGrav}),\ }\bibfield
  {title} {\bibinfo {title} {{The NANOGrav 15 yr Data Set: Search for Signals
  from New Physics}},\ }\href {https://doi.org/10.3847/2041-8213/acdc91}
  {\bibfield  {journal} {\bibinfo  {journal} {Astrophys. J. Lett.}\ }\textbf
  {\bibinfo {volume} {951}},\ \bibinfo {pages} {L11} (\bibinfo {year}
  {2023})},\ \Eprint {https://arxiv.org/abs/2306.16219} {arXiv:2306.16219
  [astro-ph.HE]} \BibitemShut {NoStop}%
\bibitem [{\citenamefont {{Antoniadis}}\ \emph {et~al.}(2023)\citenamefont
  {{Antoniadis}}, \citenamefont {{Arumugam}}, \citenamefont {{Arumugam}},
  \citenamefont {{Auclair}}, \citenamefont {{Babak}}, \citenamefont {{Bagchi}},
  \citenamefont {{Bak Nielsen}}, \citenamefont {{Barausse}}, \citenamefont
  {{Bassa}}, \citenamefont {{Bathula}}, \citenamefont {{Berthereau}},
  \citenamefont {{Bonetti}}, \citenamefont {{Bortolas}}, \citenamefont
  {{Brook}}, \citenamefont {{Burgay}}, \citenamefont {{Caballero}},
  \citenamefont {{Caprini}}, \citenamefont {{Chalumeau}}, \citenamefont
  {{Champion}}, \citenamefont {{Chanlaridis}}, \citenamefont {{Chen}},
  \citenamefont {{Cognard}}, \citenamefont {{Crisostomi}}, \citenamefont
  {{Dandapat}}, \citenamefont {{Deb}}, \citenamefont {{Desai}}, \citenamefont
  {{Desvignes}}, \citenamefont {{Dhanda-Batra}}, \citenamefont {{Dwivedi}},
  \citenamefont {{Falxa}}, \citenamefont {{Fastidio}}, \citenamefont
  {{Ferdman}}, \citenamefont {{Franchini}}, \citenamefont {{Gair}},
  \citenamefont {{Goncharov}}, \citenamefont {{Gopakumar}}, \citenamefont
  {{Graikou}}, \citenamefont {{Grie{\ss}meier}}, \citenamefont {{Gualandris}},
  \citenamefont {{Guillemot}}, \citenamefont {{Guo}}, \citenamefont {{Gupta}},
  \citenamefont {{Hisano}}, \citenamefont {{Hu}}, \citenamefont {{Iraci}},
  \citenamefont {{Izquierdo-Villalba}}, \citenamefont {{Jang}}, \citenamefont
  {{Jawor}}, \citenamefont {{Janssen}}, \citenamefont {{Jessner}},
  \citenamefont {{Joshi}}, \citenamefont {{Kareem}}, \citenamefont
  {{Karuppusamy}}, \citenamefont {{Keane}}, \citenamefont {{Keith}},
  \citenamefont {{Kharbanda}}, \citenamefont {{Khizriev}}, \citenamefont
  {{Kikunaga}}, \citenamefont {{Kolhe}}, \citenamefont {{Kramer}},
  \citenamefont {{Krishnakumar}}, \citenamefont {{Lackeos}}, \citenamefont
  {{Lee}}, \citenamefont {{Liu}}, \citenamefont {{Liu}}, \citenamefont
  {{Lyne}}, \citenamefont {{McKee}}, \citenamefont {{Maan}}, \citenamefont
  {{Main}}, \citenamefont {{Mickaliger}}, \citenamefont {{Middleton}},
  \citenamefont {{Neronov}}, \citenamefont {{Nitu}}, \citenamefont
  {{Nobleson}}, \citenamefont {{Paladi}}, \citenamefont {{Parthasarathy}},
  \citenamefont {{Perera}}, \citenamefont {{Perrodin}}, \citenamefont
  {{Petiteau}}, \citenamefont {{Porayko}}, \citenamefont {{Possenti}},
  \citenamefont {{Prabu}}, \citenamefont {{Postnov}}, \citenamefont
  {{Quelquejay Leclere}}, \citenamefont {{Rana}}, \citenamefont {{Roper Pol}},
  \citenamefont {{Samajdar}}, \citenamefont {{Sanidas}}, \citenamefont
  {{Semikoz}}, \citenamefont {{Sesana}}, \citenamefont {{Shaifullah}},
  \citenamefont {{Singha}}, \citenamefont {{Smarra}}, \citenamefont {{Speri}},
  \citenamefont {{Spiewak}}, \citenamefont {{Srivastava}}, \citenamefont
  {{Stappers}}, \citenamefont {{Steer}}, \citenamefont {{Surnis}},
  \citenamefont {{Susarla}}, \citenamefont {{Susobhanan}}, \citenamefont
  {{Takahashi}}, \citenamefont {{Tarafdar}}, \citenamefont {{Theureau}},
  \citenamefont {{Tiburzi}}, \citenamefont {{Truant}}, \citenamefont {{van der
  Wateren}}, \citenamefont {{Valtolina}}, \citenamefont {{Vecchio}},
  \citenamefont {{Venkatraman Krishnan}}, \citenamefont {{Verbiest}},
  \citenamefont {{Wang}}, \citenamefont {{Wang}},\ and\ \citenamefont
  {{Wu}}}]{2023arXiv230616227A}%
  \BibitemOpen
  \bibfield  {author} {\bibinfo {author} {\bibfnamefont {J.}~\bibnamefont
  {{Antoniadis}}}, \bibinfo {author} {\bibfnamefont {P.}~\bibnamefont
  {{Arumugam}}}, \bibinfo {author} {\bibfnamefont {S.}~\bibnamefont
  {{Arumugam}}}, \bibinfo {author} {\bibfnamefont {P.}~\bibnamefont
  {{Auclair}}}, \bibinfo {author} {\bibfnamefont {S.}~\bibnamefont {{Babak}}},
  \bibinfo {author} {\bibfnamefont {M.}~\bibnamefont {{Bagchi}}}, \bibinfo
  {author} {\bibfnamefont {A.~S.}\ \bibnamefont {{Bak Nielsen}}}, \bibinfo
  {author} {\bibfnamefont {E.}~\bibnamefont {{Barausse}}}, \bibinfo {author}
  {\bibfnamefont {C.~G.}\ \bibnamefont {{Bassa}}}, \bibinfo {author}
  {\bibfnamefont {A.}~\bibnamefont {{Bathula}}}, \bibinfo {author}
  {\bibfnamefont {A.}~\bibnamefont {{Berthereau}}}, \bibinfo {author}
  {\bibfnamefont {M.}~\bibnamefont {{Bonetti}}}, \bibinfo {author}
  {\bibfnamefont {E.}~\bibnamefont {{Bortolas}}}, \bibinfo {author}
  {\bibfnamefont {P.~R.}\ \bibnamefont {{Brook}}}, \bibinfo {author}
  {\bibfnamefont {M.}~\bibnamefont {{Burgay}}}, \bibinfo {author}
  {\bibfnamefont {R.~N.}\ \bibnamefont {{Caballero}}}, \bibinfo {author}
  {\bibfnamefont {C.}~\bibnamefont {{Caprini}}}, \bibinfo {author}
  {\bibfnamefont {A.}~\bibnamefont {{Chalumeau}}}, \bibinfo {author}
  {\bibfnamefont {D.~J.}\ \bibnamefont {{Champion}}}, \bibinfo {author}
  {\bibfnamefont {S.}~\bibnamefont {{Chanlaridis}}}, \bibinfo {author}
  {\bibfnamefont {S.}~\bibnamefont {{Chen}}}, \bibinfo {author} {\bibfnamefont
  {I.}~\bibnamefont {{Cognard}}}, \bibinfo {author} {\bibfnamefont
  {M.}~\bibnamefont {{Crisostomi}}}, \bibinfo {author} {\bibfnamefont
  {S.}~\bibnamefont {{Dandapat}}}, \bibinfo {author} {\bibfnamefont
  {D.}~\bibnamefont {{Deb}}}, \bibinfo {author} {\bibfnamefont
  {S.}~\bibnamefont {{Desai}}}, \bibinfo {author} {\bibfnamefont
  {G.}~\bibnamefont {{Desvignes}}}, \bibinfo {author} {\bibfnamefont
  {N.}~\bibnamefont {{Dhanda-Batra}}}, \bibinfo {author} {\bibfnamefont
  {C.}~\bibnamefont {{Dwivedi}}}, \bibinfo {author} {\bibfnamefont
  {M.}~\bibnamefont {{Falxa}}}, \bibinfo {author} {\bibfnamefont
  {F.}~\bibnamefont {{Fastidio}}}, \bibinfo {author} {\bibfnamefont {R.~D.}\
  \bibnamefont {{Ferdman}}}, \bibinfo {author} {\bibfnamefont {A.}~\bibnamefont
  {{Franchini}}}, \bibinfo {author} {\bibfnamefont {J.~R.}\ \bibnamefont
  {{Gair}}}, \bibinfo {author} {\bibfnamefont {B.}~\bibnamefont {{Goncharov}}},
  \bibinfo {author} {\bibfnamefont {A.}~\bibnamefont {{Gopakumar}}}, \bibinfo
  {author} {\bibfnamefont {E.}~\bibnamefont {{Graikou}}}, \bibinfo {author}
  {\bibfnamefont {J.~M.}\ \bibnamefont {{Grie{\ss}meier}}}, \bibinfo {author}
  {\bibfnamefont {A.}~\bibnamefont {{Gualandris}}}, \bibinfo {author}
  {\bibfnamefont {L.}~\bibnamefont {{Guillemot}}}, \bibinfo {author}
  {\bibfnamefont {Y.~J.}\ \bibnamefont {{Guo}}}, \bibinfo {author}
  {\bibfnamefont {Y.}~\bibnamefont {{Gupta}}}, \bibinfo {author} {\bibfnamefont
  {S.}~\bibnamefont {{Hisano}}}, \bibinfo {author} {\bibfnamefont
  {H.}~\bibnamefont {{Hu}}}, \bibinfo {author} {\bibfnamefont {F.}~\bibnamefont
  {{Iraci}}}, \bibinfo {author} {\bibfnamefont {D.}~\bibnamefont
  {{Izquierdo-Villalba}}}, \bibinfo {author} {\bibfnamefont {J.}~\bibnamefont
  {{Jang}}}, \bibinfo {author} {\bibfnamefont {J.}~\bibnamefont {{Jawor}}},
  \bibinfo {author} {\bibfnamefont {G.~H.}\ \bibnamefont {{Janssen}}}, \bibinfo
  {author} {\bibfnamefont {A.}~\bibnamefont {{Jessner}}}, \bibinfo {author}
  {\bibfnamefont {B.~C.}\ \bibnamefont {{Joshi}}}, \bibinfo {author}
  {\bibfnamefont {F.}~\bibnamefont {{Kareem}}}, \bibinfo {author}
  {\bibfnamefont {R.}~\bibnamefont {{Karuppusamy}}}, \bibinfo {author}
  {\bibfnamefont {E.~F.}\ \bibnamefont {{Keane}}}, \bibinfo {author}
  {\bibfnamefont {M.~J.}\ \bibnamefont {{Keith}}}, \bibinfo {author}
  {\bibfnamefont {D.}~\bibnamefont {{Kharbanda}}}, \bibinfo {author}
  {\bibfnamefont {T.}~\bibnamefont {{Khizriev}}}, \bibinfo {author}
  {\bibfnamefont {T.}~\bibnamefont {{Kikunaga}}}, \bibinfo {author}
  {\bibfnamefont {N.}~\bibnamefont {{Kolhe}}}, \bibinfo {author} {\bibfnamefont
  {M.}~\bibnamefont {{Kramer}}}, \bibinfo {author} {\bibfnamefont {M.~A.}\
  \bibnamefont {{Krishnakumar}}}, \bibinfo {author} {\bibfnamefont
  {K.}~\bibnamefont {{Lackeos}}}, \bibinfo {author} {\bibfnamefont {K.~J.}\
  \bibnamefont {{Lee}}}, \bibinfo {author} {\bibfnamefont {K.}~\bibnamefont
  {{Liu}}}, \bibinfo {author} {\bibfnamefont {Y.}~\bibnamefont {{Liu}}},
  \bibinfo {author} {\bibfnamefont {A.~G.}\ \bibnamefont {{Lyne}}}, \bibinfo
  {author} {\bibfnamefont {J.~W.}\ \bibnamefont {{McKee}}}, \bibinfo {author}
  {\bibfnamefont {Y.}~\bibnamefont {{Maan}}}, \bibinfo {author} {\bibfnamefont
  {R.~A.}\ \bibnamefont {{Main}}}, \bibinfo {author} {\bibfnamefont {M.~B.}\
  \bibnamefont {{Mickaliger}}}, \bibinfo {author} {\bibfnamefont
  {H.}~\bibnamefont {{Middleton}}}, \bibinfo {author} {\bibfnamefont
  {A.}~\bibnamefont {{Neronov}}}, \bibinfo {author} {\bibfnamefont {I.~C.}\
  \bibnamefont {{Nitu}}}, \bibinfo {author} {\bibfnamefont {K.}~\bibnamefont
  {{Nobleson}}}, \bibinfo {author} {\bibfnamefont {A.~K.}\ \bibnamefont
  {{Paladi}}}, \bibinfo {author} {\bibfnamefont {A.}~\bibnamefont
  {{Parthasarathy}}}, \bibinfo {author} {\bibfnamefont {B.~B.~P.}\ \bibnamefont
  {{Perera}}}, \bibinfo {author} {\bibfnamefont {D.}~\bibnamefont
  {{Perrodin}}}, \bibinfo {author} {\bibfnamefont {A.}~\bibnamefont
  {{Petiteau}}}, \bibinfo {author} {\bibfnamefont {N.~K.}\ \bibnamefont
  {{Porayko}}}, \bibinfo {author} {\bibfnamefont {A.}~\bibnamefont
  {{Possenti}}}, \bibinfo {author} {\bibfnamefont {T.}~\bibnamefont {{Prabu}}},
  \bibinfo {author} {\bibfnamefont {K.}~\bibnamefont {{Postnov}}}, \bibinfo
  {author} {\bibfnamefont {H.}~\bibnamefont {{Quelquejay Leclere}}}, \bibinfo
  {author} {\bibfnamefont {P.}~\bibnamefont {{Rana}}}, \bibinfo {author}
  {\bibfnamefont {A.}~\bibnamefont {{Roper Pol}}}, \bibinfo {author}
  {\bibfnamefont {A.}~\bibnamefont {{Samajdar}}}, \bibinfo {author}
  {\bibfnamefont {S.~A.}\ \bibnamefont {{Sanidas}}}, \bibinfo {author}
  {\bibfnamefont {D.}~\bibnamefont {{Semikoz}}}, \bibinfo {author}
  {\bibfnamefont {A.}~\bibnamefont {{Sesana}}}, \bibinfo {author}
  {\bibfnamefont {G.}~\bibnamefont {{Shaifullah}}}, \bibinfo {author}
  {\bibfnamefont {J.}~\bibnamefont {{Singha}}}, \bibinfo {author}
  {\bibfnamefont {C.}~\bibnamefont {{Smarra}}}, \bibinfo {author}
  {\bibfnamefont {L.}~\bibnamefont {{Speri}}}, \bibinfo {author} {\bibfnamefont
  {R.}~\bibnamefont {{Spiewak}}}, \bibinfo {author} {\bibfnamefont
  {A.}~\bibnamefont {{Srivastava}}}, \bibinfo {author} {\bibfnamefont {B.~W.}\
  \bibnamefont {{Stappers}}}, \bibinfo {author} {\bibfnamefont {D.~A.}\
  \bibnamefont {{Steer}}}, \bibinfo {author} {\bibfnamefont {M.}~\bibnamefont
  {{Surnis}}}, \bibinfo {author} {\bibfnamefont {S.~C.}\ \bibnamefont
  {{Susarla}}}, \bibinfo {author} {\bibfnamefont {A.}~\bibnamefont
  {{Susobhanan}}}, \bibinfo {author} {\bibfnamefont {K.}~\bibnamefont
  {{Takahashi}}}, \bibinfo {author} {\bibfnamefont {P.}~\bibnamefont
  {{Tarafdar}}}, \bibinfo {author} {\bibfnamefont {G.}~\bibnamefont
  {{Theureau}}}, \bibinfo {author} {\bibfnamefont {C.}~\bibnamefont
  {{Tiburzi}}}, \bibinfo {author} {\bibfnamefont {R.~J.}\ \bibnamefont
  {{Truant}}}, \bibinfo {author} {\bibfnamefont {E.}~\bibnamefont {{van der
  Wateren}}}, \bibinfo {author} {\bibfnamefont {S.}~\bibnamefont
  {{Valtolina}}}, \bibinfo {author} {\bibfnamefont {A.}~\bibnamefont
  {{Vecchio}}}, \bibinfo {author} {\bibfnamefont {V.}~\bibnamefont
  {{Venkatraman Krishnan}}}, \bibinfo {author} {\bibfnamefont {J.~P.~W.}\
  \bibnamefont {{Verbiest}}}, \bibinfo {author} {\bibfnamefont
  {J.}~\bibnamefont {{Wang}}}, \bibinfo {author} {\bibfnamefont
  {L.}~\bibnamefont {{Wang}}},\ and\ \bibinfo {author} {\bibfnamefont
  {Z.}~\bibnamefont {{Wu}}},\ }\bibfield  {title} {\bibinfo {title} {{The
  second data release from the European Pulsar Timing Array: V. Implications
  for massive black holes, dark matter and the early Universe}},\ }\href
  {https://doi.org/10.48550/arXiv.2306.16227} {\bibfield  {journal} {\bibinfo
  {journal} {arXiv e-prints}\ ,\ \bibinfo {eid} {arXiv:2306.16227}} (\bibinfo
  {year} {2023})},\ \Eprint {https://arxiv.org/abs/2306.16227}
  {arXiv:2306.16227 [astro-ph.CO]} \BibitemShut {NoStop}%
\bibitem [{\citenamefont {Xu}\ \emph {et~al.}(2023)\citenamefont {Xu} \emph
  {et~al.}}]{Xu:2023wog}%
  \BibitemOpen
  \bibfield  {author} {\bibinfo {author} {\bibfnamefont {H.}~\bibnamefont {Xu}}
  \emph {et~al.},\ }\bibfield  {title} {\bibinfo {title} {{Searching for the
  Nano-Hertz Stochastic Gravitational Wave Background with the Chinese Pulsar
  Timing Array Data Release I}},\ }\href
  {https://doi.org/10.1088/1674-4527/acdfa5} {\bibfield  {journal} {\bibinfo
  {journal} {Res. Astron. Astrophys.}\ }\textbf {\bibinfo {volume} {23}},\
  \bibinfo {pages} {075024} (\bibinfo {year} {2023})},\ \Eprint
  {https://arxiv.org/abs/2306.16216} {arXiv:2306.16216 [astro-ph.HE]}
  \BibitemShut {NoStop}%
\bibitem [{\citenamefont {Reardon}\ \emph {et~al.}(2023)\citenamefont {Reardon}
  \emph {et~al.}}]{Reardon:2023zen}%
  \BibitemOpen
  \bibfield  {author} {\bibinfo {author} {\bibfnamefont {D.~J.}\ \bibnamefont
  {Reardon}} \emph {et~al.},\ }\bibfield  {title} {\bibinfo {title} {{The
  Gravitational-wave Background Null Hypothesis: Characterizing Noise in
  Millisecond Pulsar Arrival Times with the Parkes Pulsar Timing Array}},\
  }\href {https://doi.org/10.3847/2041-8213/acdd03} {\bibfield  {journal}
  {\bibinfo  {journal} {Astrophys. J. Lett.}\ }\textbf {\bibinfo {volume}
  {951}},\ \bibinfo {pages} {L7} (\bibinfo {year} {2023})},\ \Eprint
  {https://arxiv.org/abs/2306.16229} {arXiv:2306.16229 [astro-ph.HE]}
  \BibitemShut {NoStop}%
\bibitem [{\citenamefont {Abbott}\ \emph {et~al.}(2016)\citenamefont {Abbott}
  \emph {et~al.}}]{LIGOScientific:2016aoc}%
  \BibitemOpen
  \bibfield  {author} {\bibinfo {author} {\bibfnamefont {B.~P.}\ \bibnamefont
  {Abbott}} \emph {et~al.} (\bibinfo {collaboration} {LIGO Scientific,
  Virgo}),\ }\bibfield  {title} {\bibinfo {title} {{Observation of
  Gravitational Waves from a Binary Black Hole Merger}},\ }\href
  {https://doi.org/10.1103/PhysRevLett.116.061102} {\bibfield  {journal}
  {\bibinfo  {journal} {Phys. Rev. Lett.}\ }\textbf {\bibinfo {volume} {116}},\
  \bibinfo {pages} {061102} (\bibinfo {year} {2016})},\ \Eprint
  {https://arxiv.org/abs/1602.03837} {arXiv:1602.03837 [gr-qc]} \BibitemShut
  {NoStop}%
\bibitem [{\citenamefont {Buonanno}\ and\ \citenamefont
  {Sathyaprakash}(2014)}]{Buonanno:2014aza}%
  \BibitemOpen
  \bibfield  {author} {\bibinfo {author} {\bibfnamefont {A.}~\bibnamefont
  {Buonanno}}\ and\ \bibinfo {author} {\bibfnamefont {B.~S.}\ \bibnamefont
  {Sathyaprakash}},\ }\bibinfo {title} {{Sources of Gravitational Waves: Theory
  and Observations}}\ (\bibinfo {year} {2014})\ \Eprint
  {https://arxiv.org/abs/1410.7832} {arXiv:1410.7832 [gr-qc]} \BibitemShut
  {NoStop}%
\bibitem [{\citenamefont {Phinney}(2001)}]{phinney2001practical}%
  \BibitemOpen
  \bibfield  {author} {\bibinfo {author} {\bibfnamefont {E.}~\bibnamefont
  {Phinney}},\ }\bibfield  {title} {\bibinfo {title} {A practical theorem on
  gravitational wave backgrounds},\ }\href@noop {} {\bibfield  {journal}
  {\bibinfo  {journal} {arXiv preprint astro-ph/0108028}\ } (\bibinfo {year}
  {2001})}\BibitemShut {NoStop}%
\bibitem [{\citenamefont {Guzzetti}\ \emph {et~al.}(2016)\citenamefont
  {Guzzetti}, \citenamefont {Bartolo}, \citenamefont {Liguori},\ and\
  \citenamefont {Matarrese}}]{Guzzetti:2016mkm}%
  \BibitemOpen
  \bibfield  {author} {\bibinfo {author} {\bibfnamefont {M.~C.}\ \bibnamefont
  {Guzzetti}}, \bibinfo {author} {\bibfnamefont {N.}~\bibnamefont {Bartolo}},
  \bibinfo {author} {\bibfnamefont {M.}~\bibnamefont {Liguori}},\ and\ \bibinfo
  {author} {\bibfnamefont {S.}~\bibnamefont {Matarrese}},\ }\bibfield  {title}
  {\bibinfo {title} {{Gravitational waves from inflation}},\ }\href
  {https://doi.org/10.1393/ncr/i2016-10127-1} {\bibfield  {journal} {\bibinfo
  {journal} {Riv. Nuovo Cim.}\ }\textbf {\bibinfo {volume} {39}},\ \bibinfo
  {pages} {399} (\bibinfo {year} {2016})},\ \Eprint
  {https://arxiv.org/abs/1605.01615} {arXiv:1605.01615 [astro-ph.CO]}
  \BibitemShut {NoStop}%
\bibitem [{\citenamefont {Kamionkowski}\ \emph {et~al.}(1994)\citenamefont
  {Kamionkowski}, \citenamefont {Kosowsky},\ and\ \citenamefont
  {Turner}}]{Kamionkowski:1993fg}%
  \BibitemOpen
  \bibfield  {author} {\bibinfo {author} {\bibfnamefont {M.}~\bibnamefont
  {Kamionkowski}}, \bibinfo {author} {\bibfnamefont {A.}~\bibnamefont
  {Kosowsky}},\ and\ \bibinfo {author} {\bibfnamefont {M.~S.}\ \bibnamefont
  {Turner}},\ }\bibfield  {title} {\bibinfo {title} {{Gravitational radiation
  from first order phase transitions}},\ }\href
  {https://doi.org/10.1103/PhysRevD.49.2837} {\bibfield  {journal} {\bibinfo
  {journal} {Phys. Rev. D}\ }\textbf {\bibinfo {volume} {49}},\ \bibinfo
  {pages} {2837} (\bibinfo {year} {1994})},\ \Eprint
  {https://arxiv.org/abs/astro-ph/9310044} {arXiv:astro-ph/9310044}
  \BibitemShut {NoStop}%
\bibitem [{\citenamefont {Caprini}\ and\ \citenamefont
  {Figueroa}(2018)}]{Caprini:2018mtu}%
  \BibitemOpen
  \bibfield  {author} {\bibinfo {author} {\bibfnamefont {C.}~\bibnamefont
  {Caprini}}\ and\ \bibinfo {author} {\bibfnamefont {D.~G.}\ \bibnamefont
  {Figueroa}},\ }\bibfield  {title} {\bibinfo {title} {{Cosmological
  Backgrounds of Gravitational Waves}},\ }\href
  {https://doi.org/10.1088/1361-6382/aac608} {\bibfield  {journal} {\bibinfo
  {journal} {Class. Quant. Grav.}\ }\textbf {\bibinfo {volume} {35}},\ \bibinfo
  {pages} {163001} (\bibinfo {year} {2018})},\ \Eprint
  {https://arxiv.org/abs/1801.04268} {arXiv:1801.04268 [astro-ph.CO]}
  \BibitemShut {NoStop}%
\bibitem [{\citenamefont {Guth}(1981)}]{Guth:1980zm}%
  \BibitemOpen
  \bibfield  {author} {\bibinfo {author} {\bibfnamefont {A.~H.}\ \bibnamefont
  {Guth}},\ }\bibfield  {title} {\bibinfo {title} {{The Inflationary Universe:
  A Possible Solution to the Horizon and Flatness Problems}},\ }\href
  {https://doi.org/10.1103/PhysRevD.23.347} {\bibfield  {journal} {\bibinfo
  {journal} {Phys. Rev. D}\ }\textbf {\bibinfo {volume} {23}},\ \bibinfo
  {pages} {347} (\bibinfo {year} {1981})}\BibitemShut {NoStop}%
\bibitem [{\citenamefont {Starobinsky}(1980)}]{Starobinsky:1980te}%
  \BibitemOpen
  \bibfield  {author} {\bibinfo {author} {\bibfnamefont {A.~A.}\ \bibnamefont
  {Starobinsky}},\ }\bibfield  {title} {\bibinfo {title} {{A New Type of
  Isotropic Cosmological Models Without Singularity}},\ }\href
  {https://doi.org/10.1016/0370-2693(80)90670-X} {\bibfield  {journal}
  {\bibinfo  {journal} {Phys. Lett. B}\ }\textbf {\bibinfo {volume} {91}},\
  \bibinfo {pages} {99} (\bibinfo {year} {1980})}\BibitemShut {NoStop}%
\bibitem [{\citenamefont {Lyth}\ and\ \citenamefont
  {Riotto}(1999)}]{Lyth:1998xn}%
  \BibitemOpen
  \bibfield  {author} {\bibinfo {author} {\bibfnamefont {D.~H.}\ \bibnamefont
  {Lyth}}\ and\ \bibinfo {author} {\bibfnamefont {A.}~\bibnamefont {Riotto}},\
  }\bibfield  {title} {\bibinfo {title} {{Particle physics models of inflation
  and the cosmological density perturbation}},\ }\href
  {https://doi.org/10.1016/S0370-1573(98)00128-8} {\bibfield  {journal}
  {\bibinfo  {journal} {Phys. Rept.}\ }\textbf {\bibinfo {volume} {314}},\
  \bibinfo {pages} {1} (\bibinfo {year} {1999})},\ \Eprint
  {https://arxiv.org/abs/hep-ph/9807278} {arXiv:hep-ph/9807278} \BibitemShut
  {NoStop}%
\bibitem [{\citenamefont {Boyle}\ and\ \citenamefont
  {Buonanno}(2008)}]{Boyle:2007zx}%
  \BibitemOpen
  \bibfield  {author} {\bibinfo {author} {\bibfnamefont {L.~A.}\ \bibnamefont
  {Boyle}}\ and\ \bibinfo {author} {\bibfnamefont {A.}~\bibnamefont
  {Buonanno}},\ }\bibfield  {title} {\bibinfo {title} {{Relating gravitational
  wave constraints from primordial nucleosynthesis, pulsar timing, laser
  interferometers, and the CMB: Implications for the early Universe}},\ }\href
  {https://doi.org/10.1103/PhysRevD.78.043531} {\bibfield  {journal} {\bibinfo
  {journal} {Phys. Rev. D}\ }\textbf {\bibinfo {volume} {78}},\ \bibinfo
  {pages} {043531} (\bibinfo {year} {2008})},\ \Eprint
  {https://arxiv.org/abs/0708.2279} {arXiv:0708.2279 [astro-ph]} \BibitemShut
  {NoStop}%
\bibitem [{\citenamefont {Inomata}\ \emph {et~al.}(2020)\citenamefont
  {Inomata}, \citenamefont {Kawasaki}, \citenamefont {Mukaida}, \citenamefont
  {Terada},\ and\ \citenamefont {Yanagida}}]{PhysRevD.101.123533}%
  \BibitemOpen
  \bibfield  {author} {\bibinfo {author} {\bibfnamefont {K.}~\bibnamefont
  {Inomata}}, \bibinfo {author} {\bibfnamefont {M.}~\bibnamefont {Kawasaki}},
  \bibinfo {author} {\bibfnamefont {K.}~\bibnamefont {Mukaida}}, \bibinfo
  {author} {\bibfnamefont {T.}~\bibnamefont {Terada}},\ and\ \bibinfo {author}
  {\bibfnamefont {T.~T.}\ \bibnamefont {Yanagida}},\ }\bibfield  {title}
  {\bibinfo {title} {Gravitational wave production right after a primordial
  black hole evaporation},\ }\href
  {https://doi.org/10.1103/PhysRevD.101.123533} {\bibfield  {journal} {\bibinfo
   {journal} {Phys. Rev. D}\ }\textbf {\bibinfo {volume} {101}},\ \bibinfo
  {pages} {123533} (\bibinfo {year} {2020})}\BibitemShut {NoStop}%
\bibitem [{\citenamefont {An}\ \emph {et~al.}(2022)\citenamefont {An},
  \citenamefont {Lyu}, \citenamefont {Wang},\ and\ \citenamefont
  {Zhou}}]{An:2020fff}%
  \BibitemOpen
  \bibfield  {author} {\bibinfo {author} {\bibfnamefont {H.}~\bibnamefont
  {An}}, \bibinfo {author} {\bibfnamefont {K.-F.}\ \bibnamefont {Lyu}},
  \bibinfo {author} {\bibfnamefont {L.-T.}\ \bibnamefont {Wang}},\ and\
  \bibinfo {author} {\bibfnamefont {S.}~\bibnamefont {Zhou}},\ }\bibfield
  {title} {\bibinfo {title} {{A unique gravitational wave signal from phase
  transition during inflation*}},\ }\href
  {https://doi.org/10.1088/1674-1137/ac76a7} {\bibfield  {journal} {\bibinfo
  {journal} {Chin. Phys. C}\ }\textbf {\bibinfo {volume} {46}},\ \bibinfo
  {pages} {101001} (\bibinfo {year} {2022})},\ \Eprint
  {https://arxiv.org/abs/2009.12381} {arXiv:2009.12381 [astro-ph.CO]}
  \BibitemShut {NoStop}%
\bibitem [{\citenamefont {{Berlin}}\ \emph {et~al.}(2022)\citenamefont
  {{Berlin}}, \citenamefont {{Belomestnykh}}, \citenamefont {{Blas}},
  \citenamefont {{Frolov}}, \citenamefont {{Brady}}, \citenamefont {{Braggio}},
  \citenamefont {{Carena}}, \citenamefont {{Cervantes}}, \citenamefont
  {{Checchin}}, \citenamefont {{Contreras-Martinez}}, \citenamefont {{Tito
  D'Agnolo}}, \citenamefont {{Ellis}}, \citenamefont {{Eremeev}}, \citenamefont
  {{Gao}}, \citenamefont {{Giaccone}}, \citenamefont {{Grassellino}},
  \citenamefont {{Harnik}}, \citenamefont {{Hollister}}, \citenamefont
  {{Janish}}, \citenamefont {{Kahn}}, \citenamefont {{Kazakov}}, \citenamefont
  {{Murat Kurkcuoglu}}, \citenamefont {{Liu}}, \citenamefont {{Lunin}},
  \citenamefont {{Netepenko}}, \citenamefont {{Melnychuk}}, \citenamefont
  {{Pilipenko}}, \citenamefont {{Pischalnikov}}, \citenamefont {{Posen}},
  \citenamefont {{Romanenko}}, \citenamefont {{Schutte-Engel}}, \citenamefont
  {{Wang}}, \citenamefont {{Yakovlev}}, \citenamefont {{Zhou}}, \citenamefont
  {{Zorzetti}},\ and\ \citenamefont {{Zhuang}}}]{2022arXiv220312714B}%
  \BibitemOpen
  \bibfield  {author} {\bibinfo {author} {\bibfnamefont {A.}~\bibnamefont
  {{Berlin}}}, \bibinfo {author} {\bibfnamefont {S.}~\bibnamefont
  {{Belomestnykh}}}, \bibinfo {author} {\bibfnamefont {D.}~\bibnamefont
  {{Blas}}}, \bibinfo {author} {\bibfnamefont {D.}~\bibnamefont {{Frolov}}},
  \bibinfo {author} {\bibfnamefont {A.~J.}\ \bibnamefont {{Brady}}}, \bibinfo
  {author} {\bibfnamefont {C.}~\bibnamefont {{Braggio}}}, \bibinfo {author}
  {\bibfnamefont {M.}~\bibnamefont {{Carena}}}, \bibinfo {author}
  {\bibfnamefont {R.}~\bibnamefont {{Cervantes}}}, \bibinfo {author}
  {\bibfnamefont {M.}~\bibnamefont {{Checchin}}}, \bibinfo {author}
  {\bibfnamefont {C.}~\bibnamefont {{Contreras-Martinez}}}, \bibinfo {author}
  {\bibfnamefont {R.}~\bibnamefont {{Tito D'Agnolo}}}, \bibinfo {author}
  {\bibfnamefont {S.~A.~R.}\ \bibnamefont {{Ellis}}}, \bibinfo {author}
  {\bibfnamefont {G.}~\bibnamefont {{Eremeev}}}, \bibinfo {author}
  {\bibfnamefont {C.}~\bibnamefont {{Gao}}}, \bibinfo {author} {\bibfnamefont
  {B.}~\bibnamefont {{Giaccone}}}, \bibinfo {author} {\bibfnamefont
  {A.}~\bibnamefont {{Grassellino}}}, \bibinfo {author} {\bibfnamefont
  {R.}~\bibnamefont {{Harnik}}}, \bibinfo {author} {\bibfnamefont
  {M.}~\bibnamefont {{Hollister}}}, \bibinfo {author} {\bibfnamefont
  {R.}~\bibnamefont {{Janish}}}, \bibinfo {author} {\bibfnamefont
  {Y.}~\bibnamefont {{Kahn}}}, \bibinfo {author} {\bibfnamefont
  {S.}~\bibnamefont {{Kazakov}}}, \bibinfo {author} {\bibfnamefont
  {D.}~\bibnamefont {{Murat Kurkcuoglu}}}, \bibinfo {author} {\bibfnamefont
  {Z.}~\bibnamefont {{Liu}}}, \bibinfo {author} {\bibfnamefont
  {A.}~\bibnamefont {{Lunin}}}, \bibinfo {author} {\bibfnamefont
  {A.}~\bibnamefont {{Netepenko}}}, \bibinfo {author} {\bibfnamefont
  {O.}~\bibnamefont {{Melnychuk}}}, \bibinfo {author} {\bibfnamefont
  {R.}~\bibnamefont {{Pilipenko}}}, \bibinfo {author} {\bibfnamefont
  {Y.}~\bibnamefont {{Pischalnikov}}}, \bibinfo {author} {\bibfnamefont
  {S.}~\bibnamefont {{Posen}}}, \bibinfo {author} {\bibfnamefont
  {A.}~\bibnamefont {{Romanenko}}}, \bibinfo {author} {\bibfnamefont
  {J.}~\bibnamefont {{Schutte-Engel}}}, \bibinfo {author} {\bibfnamefont
  {C.}~\bibnamefont {{Wang}}}, \bibinfo {author} {\bibfnamefont
  {V.}~\bibnamefont {{Yakovlev}}}, \bibinfo {author} {\bibfnamefont
  {K.}~\bibnamefont {{Zhou}}}, \bibinfo {author} {\bibfnamefont
  {S.}~\bibnamefont {{Zorzetti}}},\ and\ \bibinfo {author} {\bibfnamefont
  {Q.}~\bibnamefont {{Zhuang}}},\ }\bibfield  {title} {\bibinfo {title}
  {{Searches for New Particles, Dark Matter, and Gravitational Waves with SRF
  Cavities}},\ }\href {https://doi.org/10.48550/arXiv.2203.12714} {\bibfield
  {journal} {\bibinfo  {journal} {arXiv e-prints}\ ,\ \bibinfo {eid}
  {arXiv:2203.12714}} (\bibinfo {year} {2022})},\ \Eprint
  {https://arxiv.org/abs/2203.12714} {arXiv:2203.12714 [hep-ph]} \BibitemShut
  {NoStop}%
\bibitem [{\citenamefont {Haque}\ \emph {et~al.}(2021)\citenamefont {Haque},
  \citenamefont {Maity}, \citenamefont {Paul},\ and\ \citenamefont
  {Sriramkumar}}]{PhysRevD.104.063513}%
  \BibitemOpen
  \bibfield  {author} {\bibinfo {author} {\bibfnamefont {M.~R.}\ \bibnamefont
  {Haque}}, \bibinfo {author} {\bibfnamefont {D.}~\bibnamefont {Maity}},
  \bibinfo {author} {\bibfnamefont {T.}~\bibnamefont {Paul}},\ and\ \bibinfo
  {author} {\bibfnamefont {L.}~\bibnamefont {Sriramkumar}},\ }\bibfield
  {title} {\bibinfo {title} {Decoding the phases of early and late time
  reheating through imprints on primordial gravitational waves},\ }\href
  {https://doi.org/10.1103/PhysRevD.104.063513} {\bibfield  {journal} {\bibinfo
   {journal} {Phys. Rev. D}\ }\textbf {\bibinfo {volume} {104}},\ \bibinfo
  {pages} {063513} (\bibinfo {year} {2021})}\BibitemShut {NoStop}%
\bibitem [{\citenamefont {Maggiore}(2000)}]{Maggiore:1999vm}%
  \BibitemOpen
  \bibfield  {author} {\bibinfo {author} {\bibfnamefont {M.}~\bibnamefont
  {Maggiore}},\ }\bibfield  {title} {\bibinfo {title} {{Gravitational wave
  experiments and early universe cosmology}},\ }\href
  {https://doi.org/10.1016/S0370-1573(99)00102-7} {\bibfield  {journal}
  {\bibinfo  {journal} {Phys. Rept.}\ }\textbf {\bibinfo {volume} {331}},\
  \bibinfo {pages} {283} (\bibinfo {year} {2000})},\ \Eprint
  {https://arxiv.org/abs/gr-qc/9909001} {arXiv:gr-qc/9909001} \BibitemShut
  {NoStop}%
\bibitem [{\citenamefont {Kamionkowski}\ and\ \citenamefont
  {Kovetz}(2016)}]{Kamionkowski:2015yta}%
  \BibitemOpen
  \bibfield  {author} {\bibinfo {author} {\bibfnamefont {M.}~\bibnamefont
  {Kamionkowski}}\ and\ \bibinfo {author} {\bibfnamefont {E.~D.}\ \bibnamefont
  {Kovetz}},\ }\bibfield  {title} {\bibinfo {title} {{The Quest for B Modes
  from Inflationary Gravitational Waves}},\ }\href
  {https://doi.org/10.1146/annurev-astro-081915-023433} {\bibfield  {journal}
  {\bibinfo  {journal} {Ann. Rev. Astron. Astrophys.}\ }\textbf {\bibinfo
  {volume} {54}},\ \bibinfo {pages} {227} (\bibinfo {year} {2016})},\ \Eprint
  {https://arxiv.org/abs/1510.06042} {arXiv:1510.06042 [astro-ph.CO]}
  \BibitemShut {NoStop}%
\bibitem [{\citenamefont {Watanabe}\ and\ \citenamefont
  {Komatsu}(2006)}]{Watanabe:2006qe}%
  \BibitemOpen
  \bibfield  {author} {\bibinfo {author} {\bibfnamefont {Y.}~\bibnamefont
  {Watanabe}}\ and\ \bibinfo {author} {\bibfnamefont {E.}~\bibnamefont
  {Komatsu}},\ }\bibfield  {title} {\bibinfo {title} {{Improved Calculation of
  the Primordial Gravitational Wave Spectrum in the Standard Model}},\ }\href
  {https://doi.org/10.1103/PhysRevD.73.123515} {\bibfield  {journal} {\bibinfo
  {journal} {Phys. Rev. D}\ }\textbf {\bibinfo {volume} {73}},\ \bibinfo
  {pages} {123515} (\bibinfo {year} {2006})},\ \Eprint
  {https://arxiv.org/abs/astro-ph/0604176} {arXiv:astro-ph/0604176}
  \BibitemShut {NoStop}%
\bibitem [{\citenamefont {Sakamoto}\ \emph {et~al.}(2022)\citenamefont
  {Sakamoto}, \citenamefont {Ahn}, \citenamefont {Ichiki}, \citenamefont
  {Moon},\ and\ \citenamefont {Hasegawa}}]{sakamoto2022probing}%
  \BibitemOpen
  \bibfield  {author} {\bibinfo {author} {\bibfnamefont {H.}~\bibnamefont
  {Sakamoto}}, \bibinfo {author} {\bibfnamefont {K.}~\bibnamefont {Ahn}},
  \bibinfo {author} {\bibfnamefont {K.}~\bibnamefont {Ichiki}}, \bibinfo
  {author} {\bibfnamefont {H.}~\bibnamefont {Moon}},\ and\ \bibinfo {author}
  {\bibfnamefont {K.}~\bibnamefont {Hasegawa}},\ }\bibfield  {title} {\bibinfo
  {title} {Probing the early history of cosmic reionization by future cosmic
  microwave background experiments},\ }\href@noop {} {\bibfield  {journal}
  {\bibinfo  {journal} {The Astrophysical Journal}\ }\textbf {\bibinfo {volume}
  {930}},\ \bibinfo {pages} {140} (\bibinfo {year} {2022})}\BibitemShut
  {NoStop}%
\bibitem [{\citenamefont {Abazajian}\ \emph {et~al.}(2022)\citenamefont
  {Abazajian} \emph {et~al.}}]{CMB-S4:2020lpa}%
  \BibitemOpen
  \bibfield  {author} {\bibinfo {author} {\bibfnamefont {K.}~\bibnamefont
  {Abazajian}} \emph {et~al.} (\bibinfo {collaboration} {CMB-S4}),\ }\bibfield
  {title} {\bibinfo {title} {{CMB-S4: Forecasting Constraints on Primordial
  Gravitational Waves}},\ }\href {https://doi.org/10.3847/1538-4357/ac1596}
  {\bibfield  {journal} {\bibinfo  {journal} {Astrophys. J.}\ }\textbf
  {\bibinfo {volume} {926}},\ \bibinfo {pages} {54} (\bibinfo {year} {2022})},\
  \Eprint {https://arxiv.org/abs/2008.12619} {arXiv:2008.12619 [astro-ph.CO]}
  \BibitemShut {NoStop}%
\bibitem [{\citenamefont {Campeti}\ \emph {et~al.}(2021)\citenamefont
  {Campeti}, \citenamefont {Komatsu}, \citenamefont {Poletti},\ and\
  \citenamefont {Baccigalupi}}]{Campeti:2020xwn}%
  \BibitemOpen
  \bibfield  {author} {\bibinfo {author} {\bibfnamefont {P.}~\bibnamefont
  {Campeti}}, \bibinfo {author} {\bibfnamefont {E.}~\bibnamefont {Komatsu}},
  \bibinfo {author} {\bibfnamefont {D.}~\bibnamefont {Poletti}},\ and\ \bibinfo
  {author} {\bibfnamefont {C.}~\bibnamefont {Baccigalupi}},\ }\bibfield
  {title} {\bibinfo {title} {{Measuring the spectrum of primordial
  gravitational waves with CMB, PTA and Laser Interferometers}},\ }\href
  {https://doi.org/10.1088/1475-7516/2021/01/012} {\bibfield  {journal}
  {\bibinfo  {journal} {JCAP}\ }\textbf {\bibinfo {volume} {01}},\ \bibinfo
  {pages} {012}},\ \Eprint {https://arxiv.org/abs/2007.04241} {arXiv:2007.04241
  [astro-ph.CO]} \BibitemShut {NoStop}%
\bibitem [{\citenamefont {Flauger}\ \emph {et~al.}(2021)\citenamefont
  {Flauger}, \citenamefont {Karnesis}, \citenamefont {Nardini}, \citenamefont
  {Pieroni}, \citenamefont {Ricciardone},\ and\ \citenamefont
  {Torrado}}]{Flauger:2020qyi}%
  \BibitemOpen
  \bibfield  {author} {\bibinfo {author} {\bibfnamefont {R.}~\bibnamefont
  {Flauger}}, \bibinfo {author} {\bibfnamefont {N.}~\bibnamefont {Karnesis}},
  \bibinfo {author} {\bibfnamefont {G.}~\bibnamefont {Nardini}}, \bibinfo
  {author} {\bibfnamefont {M.}~\bibnamefont {Pieroni}}, \bibinfo {author}
  {\bibfnamefont {A.}~\bibnamefont {Ricciardone}},\ and\ \bibinfo {author}
  {\bibfnamefont {J.}~\bibnamefont {Torrado}},\ }\bibfield  {title} {\bibinfo
  {title} {{Improved reconstruction of a stochastic gravitational wave
  background with LISA}},\ }\href
  {https://doi.org/10.1088/1475-7516/2021/01/059} {\bibfield  {journal}
  {\bibinfo  {journal} {JCAP}\ }\textbf {\bibinfo {volume} {01}},\ \bibinfo
  {pages} {059}},\ \Eprint {https://arxiv.org/abs/2009.11845} {arXiv:2009.11845
  [astro-ph.CO]} \BibitemShut {NoStop}%
\bibitem [{\citenamefont {Allys}\ \emph {et~al.}(2023)\citenamefont {Allys}
  \emph {et~al.}}]{LiteBIRD:2022cnt}%
  \BibitemOpen
  \bibfield  {author} {\bibinfo {author} {\bibfnamefont {E.}~\bibnamefont
  {Allys}} \emph {et~al.} (\bibinfo {collaboration} {LiteBIRD}),\ }\bibfield
  {title} {\bibinfo {title} {{Probing Cosmic Inflation with the LiteBIRD Cosmic
  Microwave Background Polarization Survey}},\ }\href
  {https://doi.org/10.1093/ptep/ptac150} {\bibfield  {journal} {\bibinfo
  {journal} {PTEP}\ }\textbf {\bibinfo {volume} {2023}},\ \bibinfo {pages}
  {042F01} (\bibinfo {year} {2023})},\ \Eprint
  {https://arxiv.org/abs/2202.02773} {arXiv:2202.02773 [astro-ph.IM]}
  \BibitemShut {NoStop}%
\bibitem [{\citenamefont {Vilenkin}(1981)}]{Vilenkin:1981bx}%
  \BibitemOpen
  \bibfield  {author} {\bibinfo {author} {\bibfnamefont {A.}~\bibnamefont
  {Vilenkin}},\ }\bibfield  {title} {\bibinfo {title} {{Gravitational radiation
  from cosmic strings}},\ }\href {https://doi.org/10.1016/0370-2693(81)91144-8}
  {\bibfield  {journal} {\bibinfo  {journal} {Phys. Lett. B}\ }\textbf
  {\bibinfo {volume} {107}},\ \bibinfo {pages} {47} (\bibinfo {year}
  {1981})}\BibitemShut {NoStop}%
\bibitem [{\citenamefont {Figueroa}\ \emph {et~al.}(2013)\citenamefont
  {Figueroa}, \citenamefont {Hindmarsh},\ and\ \citenamefont
  {Urrestilla}}]{Figueroa:2012kw}%
  \BibitemOpen
  \bibfield  {author} {\bibinfo {author} {\bibfnamefont {D.~G.}\ \bibnamefont
  {Figueroa}}, \bibinfo {author} {\bibfnamefont {M.}~\bibnamefont
  {Hindmarsh}},\ and\ \bibinfo {author} {\bibfnamefont {J.}~\bibnamefont
  {Urrestilla}},\ }\bibfield  {title} {\bibinfo {title} {{Exact Scale-Invariant
  Background of Gravitational Waves from Cosmic Defects}},\ }\href
  {https://doi.org/10.1103/PhysRevLett.110.101302} {\bibfield  {journal}
  {\bibinfo  {journal} {Phys. Rev. Lett.}\ }\textbf {\bibinfo {volume} {110}},\
  \bibinfo {pages} {101302} (\bibinfo {year} {2013})},\ \Eprint
  {https://arxiv.org/abs/1212.5458} {arXiv:1212.5458 [astro-ph.CO]}
  \BibitemShut {NoStop}%
\bibitem [{\citenamefont {Kosowsky}\ \emph {et~al.}(1992)\citenamefont
  {Kosowsky}, \citenamefont {Turner},\ and\ \citenamefont
  {Watkins}}]{Kosowsky:1991ua}%
  \BibitemOpen
  \bibfield  {author} {\bibinfo {author} {\bibfnamefont {A.}~\bibnamefont
  {Kosowsky}}, \bibinfo {author} {\bibfnamefont {M.~S.}\ \bibnamefont
  {Turner}},\ and\ \bibinfo {author} {\bibfnamefont {R.}~\bibnamefont
  {Watkins}},\ }\bibfield  {title} {\bibinfo {title} {{Gravitational radiation
  from colliding vacuum bubbles}},\ }\href
  {https://doi.org/10.1103/PhysRevD.45.4514} {\bibfield  {journal} {\bibinfo
  {journal} {Phys. Rev. D}\ }\textbf {\bibinfo {volume} {45}},\ \bibinfo
  {pages} {4514} (\bibinfo {year} {1992})}\BibitemShut {NoStop}%
\bibitem [{\citenamefont {{Tomita}}(1967)}]{1967PThPh..37..831T}%
  \BibitemOpen
  \bibfield  {author} {\bibinfo {author} {\bibfnamefont {K.}~\bibnamefont
  {{Tomita}}},\ }\bibfield  {title} {\bibinfo {title} {{Non-Linear Theory of
  Gravitational Instability in the Expanding Universe}},\ }\href
  {https://doi.org/10.1143/PTP.37.831} {\bibfield  {journal} {\bibinfo
  {journal} {Progress of Theoretical Physics}\ }\textbf {\bibinfo {volume}
  {37}},\ \bibinfo {pages} {831} (\bibinfo {year} {1967})}\BibitemShut
  {NoStop}%
\bibitem [{\citenamefont {Matarrese}\ \emph {et~al.}(1994)\citenamefont
  {Matarrese}, \citenamefont {Pantano},\ and\ \citenamefont
  {Saez}}]{Matarrese:1993zf}%
  \BibitemOpen
  \bibfield  {author} {\bibinfo {author} {\bibfnamefont {S.}~\bibnamefont
  {Matarrese}}, \bibinfo {author} {\bibfnamefont {O.}~\bibnamefont {Pantano}},\
  and\ \bibinfo {author} {\bibfnamefont {D.}~\bibnamefont {Saez}},\ }\bibfield
  {title} {\bibinfo {title} {{General relativistic dynamics of irrotational
  dust: Cosmological implications}},\ }\href
  {https://doi.org/10.1103/PhysRevLett.72.320} {\bibfield  {journal} {\bibinfo
  {journal} {Phys. Rev. Lett.}\ }\textbf {\bibinfo {volume} {72}},\ \bibinfo
  {pages} {320} (\bibinfo {year} {1994})},\ \Eprint
  {https://arxiv.org/abs/astro-ph/9310036} {arXiv:astro-ph/9310036}
  \BibitemShut {NoStop}%
\bibitem [{\citenamefont {Matarrese}\ \emph {et~al.}(1998)\citenamefont
  {Matarrese}, \citenamefont {Mollerach},\ and\ \citenamefont
  {Bruni}}]{Matarrese_1998}%
  \BibitemOpen
  \bibfield  {author} {\bibinfo {author} {\bibfnamefont {S.}~\bibnamefont
  {Matarrese}}, \bibinfo {author} {\bibfnamefont {S.}~\bibnamefont
  {Mollerach}},\ and\ \bibinfo {author} {\bibfnamefont {M.}~\bibnamefont
  {Bruni}},\ }\bibfield  {title} {\bibinfo {title} {Relativistic second-order
  perturbations of the einstein–de sitter universe},\ }\bibfield  {journal}
  {\bibinfo  {journal} {Physical Review D}\ }\textbf {\bibinfo {volume} {58}},\
  \href {https://doi.org/10.1103/physrevd.58.043504}
  {10.1103/physrevd.58.043504} (\bibinfo {year} {1998})\BibitemShut {NoStop}%
\bibitem [{\citenamefont {Dom\`enech}\ \emph {et~al.}(2020)\citenamefont
  {Dom\`enech}, \citenamefont {Pi},\ and\ \citenamefont
  {Sasaki}}]{Domenech:2020kqm}%
  \BibitemOpen
  \bibfield  {author} {\bibinfo {author} {\bibfnamefont {G.}~\bibnamefont
  {Dom\`enech}}, \bibinfo {author} {\bibfnamefont {S.}~\bibnamefont {Pi}},\
  and\ \bibinfo {author} {\bibfnamefont {M.}~\bibnamefont {Sasaki}},\
  }\bibfield  {title} {\bibinfo {title} {{Induced gravitational waves as a
  probe of thermal history of the universe}},\ }\href
  {https://doi.org/10.1088/1475-7516/2020/08/017} {\bibfield  {journal}
  {\bibinfo  {journal} {JCAP}\ }\textbf {\bibinfo {volume} {08}},\ \bibinfo
  {pages} {017}},\ \Eprint {https://arxiv.org/abs/2005.12314} {arXiv:2005.12314
  [gr-qc]} \BibitemShut {NoStop}%
\bibitem [{\citenamefont {Espinosa}\ \emph {et~al.}(2018)\citenamefont
  {Espinosa}, \citenamefont {Racco},\ and\ \citenamefont
  {Riotto}}]{Espinosa_2018}%
  \BibitemOpen
  \bibfield  {author} {\bibinfo {author} {\bibfnamefont {J.}~\bibnamefont
  {Espinosa}}, \bibinfo {author} {\bibfnamefont {D.}~\bibnamefont {Racco}},\
  and\ \bibinfo {author} {\bibfnamefont {A.}~\bibnamefont {Riotto}},\
  }\bibfield  {title} {\bibinfo {title} {A cosmological signature of the {SM}
  higgs instability: gravitational waves},\ }\href
  {https://doi.org/10.1088/1475-7516/2018/09/012} {\bibfield  {journal}
  {\bibinfo  {journal} {Journal of Cosmology and Astroparticle Physics}\
  }\textbf {\bibinfo {volume} {2018}}\bibinfo  {number} { (09)},\ \bibinfo
  {pages} {012}}\BibitemShut {NoStop}%
\bibitem [{\citenamefont {Matarrese}\ and\ \citenamefont
  {Mollerach}(1996)}]{Matarrese:1996pp}%
  \BibitemOpen
\bibfield  {number} {  }\bibfield  {author} {\bibinfo {author} {\bibfnamefont
  {S.}~\bibnamefont {Matarrese}}\ and\ \bibinfo {author} {\bibfnamefont
  {S.}~\bibnamefont {Mollerach}},\ }\bibfield  {title} {\bibinfo {title} {{The
  Stochastic gravitational wave background produced by nonlinear cosmological
  perturbations}},\ }in\ \href@noop {} {\emph {\bibinfo {booktitle} {{ERE -
  Spanish Relativity Conference}}}}\ (\bibinfo {year} {1996})\ \Eprint
  {https://arxiv.org/abs/astro-ph/9705168} {arXiv:astro-ph/9705168}
  \BibitemShut {NoStop}%
\bibitem [{\citenamefont {Bartolo}\ \emph {et~al.}(2007)\citenamefont
  {Bartolo}, \citenamefont {Matarrese}, \citenamefont {Riotto},\ and\
  \citenamefont {Vaihkonen}}]{Bartolo:2007vp}%
  \BibitemOpen
  \bibfield  {author} {\bibinfo {author} {\bibfnamefont {N.}~\bibnamefont
  {Bartolo}}, \bibinfo {author} {\bibfnamefont {S.}~\bibnamefont {Matarrese}},
  \bibinfo {author} {\bibfnamefont {A.}~\bibnamefont {Riotto}},\ and\ \bibinfo
  {author} {\bibfnamefont {A.}~\bibnamefont {Vaihkonen}},\ }\bibfield  {title}
  {\bibinfo {title} {{The Maximal Amount of Gravitational Waves in the Curvaton
  Scenario}},\ }\href {https://doi.org/10.1103/PhysRevD.76.061302} {\bibfield
  {journal} {\bibinfo  {journal} {Phys. Rev. D}\ }\textbf {\bibinfo {volume}
  {76}},\ \bibinfo {pages} {061302} (\bibinfo {year} {2007})},\ \Eprint
  {https://arxiv.org/abs/0705.4240} {arXiv:0705.4240 [astro-ph]} \BibitemShut
  {NoStop}%
\bibitem [{\citenamefont {Baumann}\ \emph {et~al.}(2007)\citenamefont
  {Baumann}, \citenamefont {Steinhardt}, \citenamefont {Takahashi},\ and\
  \citenamefont {Ichiki}}]{Baumann_2007}%
  \BibitemOpen
  \bibfield  {author} {\bibinfo {author} {\bibfnamefont {D.}~\bibnamefont
  {Baumann}}, \bibinfo {author} {\bibfnamefont {P.}~\bibnamefont {Steinhardt}},
  \bibinfo {author} {\bibfnamefont {K.}~\bibnamefont {Takahashi}},\ and\
  \bibinfo {author} {\bibfnamefont {K.}~\bibnamefont {Ichiki}},\ }\bibfield
  {title} {\bibinfo {title} {Gravitational wave spectrum induced by primordial
  scalar perturbations},\ }\bibfield  {journal} {\bibinfo  {journal} {Physical
  Review D}\ }\textbf {\bibinfo {volume} {76}},\ \href
  {https://doi.org/10.1103/physrevd.76.084019} {10.1103/physrevd.76.084019}
  (\bibinfo {year} {2007})\BibitemShut {NoStop}%
\bibitem [{\citenamefont {Inomata}\ and\ \citenamefont
  {Terada}(2020)}]{Inomata:2019yww}%
  \BibitemOpen
  \bibfield  {author} {\bibinfo {author} {\bibfnamefont {K.}~\bibnamefont
  {Inomata}}\ and\ \bibinfo {author} {\bibfnamefont {T.}~\bibnamefont
  {Terada}},\ }\bibfield  {title} {\bibinfo {title} {{Gauge Independence of
  Induced Gravitational Waves}},\ }\href
  {https://doi.org/10.1103/PhysRevD.101.023523} {\bibfield  {journal} {\bibinfo
   {journal} {Phys. Rev. D}\ }\textbf {\bibinfo {volume} {101}},\ \bibinfo
  {pages} {023523} (\bibinfo {year} {2020})},\ \Eprint
  {https://arxiv.org/abs/1912.00785} {arXiv:1912.00785 [gr-qc]} \BibitemShut
  {NoStop}%
\bibitem [{\citenamefont {Yuan}\ \emph {et~al.}(2020)\citenamefont {Yuan},
  \citenamefont {Chen},\ and\ \citenamefont {Huang}}]{Yuan_2020}%
  \BibitemOpen
  \bibfield  {author} {\bibinfo {author} {\bibfnamefont {C.}~\bibnamefont
  {Yuan}}, \bibinfo {author} {\bibfnamefont {Z.-C.}\ \bibnamefont {Chen}},\
  and\ \bibinfo {author} {\bibfnamefont {Q.-G.}\ \bibnamefont {Huang}},\
  }\bibfield  {title} {\bibinfo {title} {Scalar induced gravitational waves in
  different gauges},\ }\bibfield  {journal} {\bibinfo  {journal} {Physical
  Review D}\ }\textbf {\bibinfo {volume} {101}},\ \href
  {https://doi.org/10.1103/physrevd.101.063018} {10.1103/physrevd.101.063018}
  (\bibinfo {year} {2020})\BibitemShut {NoStop}%
\bibitem [{\citenamefont {Ananda}\ \emph {et~al.}(2007)\citenamefont {Ananda},
  \citenamefont {Clarkson},\ and\ \citenamefont {Wands}}]{Ananda_2007}%
  \BibitemOpen
  \bibfield  {author} {\bibinfo {author} {\bibfnamefont {K.~N.}\ \bibnamefont
  {Ananda}}, \bibinfo {author} {\bibfnamefont {C.}~\bibnamefont {Clarkson}},\
  and\ \bibinfo {author} {\bibfnamefont {D.}~\bibnamefont {Wands}},\ }\bibfield
   {title} {\bibinfo {title} {Cosmological gravitational wave background from
  primordial density perturbations},\ }\bibfield  {journal} {\bibinfo
  {journal} {Physical Review D}\ }\textbf {\bibinfo {volume} {75}},\ \href
  {https://doi.org/10.1103/physrevd.75.123518} {10.1103/physrevd.75.123518}
  (\bibinfo {year} {2007})\BibitemShut {NoStop}%
\bibitem [{\citenamefont {Kohri}\ and\ \citenamefont
  {Terada}(2018)}]{Kohri:2018awv}%
  \BibitemOpen
  \bibfield  {author} {\bibinfo {author} {\bibfnamefont {K.}~\bibnamefont
  {Kohri}}\ and\ \bibinfo {author} {\bibfnamefont {T.}~\bibnamefont {Terada}},\
  }\bibfield  {title} {\bibinfo {title} {{Semianalytic calculation of
  gravitational wave spectrum nonlinearly induced from primordial curvature
  perturbations}},\ }\href {https://doi.org/10.1103/PhysRevD.97.123532}
  {\bibfield  {journal} {\bibinfo  {journal} {Phys. Rev. D}\ }\textbf {\bibinfo
  {volume} {97}},\ \bibinfo {pages} {123532} (\bibinfo {year} {2018})},\
  \Eprint {https://arxiv.org/abs/1804.08577} {arXiv:1804.08577 [gr-qc]}
  \BibitemShut {NoStop}%
\bibitem [{\citenamefont {Dom\`enech}(2020)}]{Domenech:2019quo}%
  \BibitemOpen
  \bibfield  {author} {\bibinfo {author} {\bibfnamefont {G.}~\bibnamefont
  {Dom\`enech}},\ }\bibfield  {title} {\bibinfo {title} {{Induced gravitational
  waves in a general cosmological background}},\ }\href
  {https://doi.org/10.1142/S0218271820500285} {\bibfield  {journal} {\bibinfo
  {journal} {Int. J. Mod. Phys. D}\ }\textbf {\bibinfo {volume} {29}},\
  \bibinfo {pages} {2050028} (\bibinfo {year} {2020})},\ \Eprint
  {https://arxiv.org/abs/1912.05583} {arXiv:1912.05583 [gr-qc]} \BibitemShut
  {NoStop}%
\bibitem [{\citenamefont {{Dom{\`e}nech}}(2021)}]{2021Univ....7..398D}%
  \BibitemOpen
  \bibfield  {author} {\bibinfo {author} {\bibfnamefont {G.}~\bibnamefont
  {{Dom{\`e}nech}}},\ }\bibfield  {title} {\bibinfo {title} {{Scalar Induced
  Gravitational Waves Review}},\ }\href
  {https://doi.org/10.3390/universe7110398} {\bibfield  {journal} {\bibinfo
  {journal} {Universe}\ }\textbf {\bibinfo {volume} {7}},\ \bibinfo {pages}
  {398} (\bibinfo {year} {2021})},\ \Eprint {https://arxiv.org/abs/2109.01398}
  {arXiv:2109.01398 [gr-qc]} \BibitemShut {NoStop}%
\bibitem [{\citenamefont {{Agazie}}\ \emph {et~al.}(2023)\citenamefont
  {{Agazie}}, \citenamefont {{Anumarlapudi}}, \citenamefont {{Archibald}},
  \citenamefont {{Baker}}, \citenamefont {{B{\'e}csy}}, \citenamefont
  {{Blecha}}, \citenamefont {{Bonilla}}, \citenamefont {{Brazier}},
  \citenamefont {{Brook}}, \citenamefont {{Burke-Spolaor}}, \citenamefont
  {{Burnette}}, \citenamefont {{Case}}, \citenamefont {{Casey-Clyde}},
  \citenamefont {{Charisi}}, \citenamefont {{Chatterjee}}, \citenamefont
  {{Chatziioannou}}, \citenamefont {{Cheeseboro}}, \citenamefont {{Chen}},
  \citenamefont {{Cohen}}, \citenamefont {{Cordes}}, \citenamefont {{Cornish}},
  \citenamefont {{Crawford}}, \citenamefont {{Cromartie}}, \citenamefont
  {{Crowter}}, \citenamefont {{Cutler}}, \citenamefont {{D'Orazio}},
  \citenamefont {{DeCesar}}, \citenamefont {{DeGan}}, \citenamefont
  {{Demorest}}, \citenamefont {{Deng}}, \citenamefont {{Dolch}}, \citenamefont
  {{Drachler}}, \citenamefont {{Ferrara}}, \citenamefont {{Fiore}},
  \citenamefont {{Fonseca}}, \citenamefont {{Freedman}}, \citenamefont
  {{Gardiner}}, \citenamefont {{Garver-Daniels}}, \citenamefont {{Gentile}},
  \citenamefont {{Gersbach}}, \citenamefont {{Glaser}}, \citenamefont {{Good}},
  \citenamefont {{G{\"u}ltekin}}, \citenamefont {{Hazboun}}, \citenamefont
  {{Hourihane}}, \citenamefont {{Islo}}, \citenamefont {{Jennings}},
  \citenamefont {{Johnson}}, \citenamefont {{Jones}}, \citenamefont {{Kaiser}},
  \citenamefont {{Kaplan}}, \citenamefont {{Kelley}}, \citenamefont {{Kerr}},
  \citenamefont {{Key}}, \citenamefont {{Laal}}, \citenamefont {{Lam}},
  \citenamefont {{Lamb}}, \citenamefont {{Lazio}}, \citenamefont
  {{Lewandowska}}, \citenamefont {{Littenberg}}, \citenamefont {{Liu}},
  \citenamefont {{Luo}}, \citenamefont {{Lynch}}, \citenamefont {{Ma}},
  \citenamefont {{Madison}}, \citenamefont {{McEwen}}, \citenamefont {{McKee}},
  \citenamefont {{McLaughlin}}, \citenamefont {{McMann}}, \citenamefont
  {{Meyers}}, \citenamefont {{Meyers}}, \citenamefont {{Mingarelli}},
  \citenamefont {{Mitridate}}, \citenamefont {{Natarajan}}, \citenamefont
  {{Ng}}, \citenamefont {{Nice}}, \citenamefont {{Ocker}}, \citenamefont
  {{Olum}}, \citenamefont {{Pennucci}}, \citenamefont {{Perera}}, \citenamefont
  {{Petrov}}, \citenamefont {{Pol}}, \citenamefont {{Radovan}}, \citenamefont
  {{Ransom}}, \citenamefont {{Ray}}, \citenamefont {{Romano}}, \citenamefont
  {{Runnoe}}, \citenamefont {{Sardesai}}, \citenamefont {{Schmiedekamp}},
  \citenamefont {{Schmiedekamp}}, \citenamefont {{Schmitz}}, \citenamefont
  {{Schult}}, \citenamefont {{Shapiro-Albert}}, \citenamefont {{Siemens}},
  \citenamefont {{Simon}}, \citenamefont {{Siwek}}, \citenamefont {{Stairs}},
  \citenamefont {{Stinebring}}, \citenamefont {{Stovall}}, \citenamefont
  {{Sun}}, \citenamefont {{Susobhanan}}, \citenamefont {{Swiggum}},
  \citenamefont {{Taylor}}, \citenamefont {{Taylor}}, \citenamefont {{Turner}},
  \citenamefont {{Unal}}, \citenamefont {{Vallisneri}}, \citenamefont
  {{Vigeland}}, \citenamefont {{Wachter}}, \citenamefont {{Wahl}},
  \citenamefont {{Wang}}, \citenamefont {{Witt}}, \citenamefont {{Wright}},\
  and\ \citenamefont {{Young}}}]{2023arXiv230616220A}%
  \BibitemOpen
  \bibfield  {author} {\bibinfo {author} {\bibfnamefont {G.}~\bibnamefont
  {{Agazie}}}, \bibinfo {author} {\bibfnamefont {A.}~\bibnamefont
  {{Anumarlapudi}}}, \bibinfo {author} {\bibfnamefont {A.~M.}\ \bibnamefont
  {{Archibald}}}, \bibinfo {author} {\bibfnamefont {P.~T.}\ \bibnamefont
  {{Baker}}}, \bibinfo {author} {\bibfnamefont {B.}~\bibnamefont
  {{B{\'e}csy}}}, \bibinfo {author} {\bibfnamefont {L.}~\bibnamefont
  {{Blecha}}}, \bibinfo {author} {\bibfnamefont {A.}~\bibnamefont {{Bonilla}}},
  \bibinfo {author} {\bibfnamefont {A.}~\bibnamefont {{Brazier}}}, \bibinfo
  {author} {\bibfnamefont {P.~R.}\ \bibnamefont {{Brook}}}, \bibinfo {author}
  {\bibfnamefont {S.}~\bibnamefont {{Burke-Spolaor}}}, \bibinfo {author}
  {\bibfnamefont {R.}~\bibnamefont {{Burnette}}}, \bibinfo {author}
  {\bibfnamefont {R.}~\bibnamefont {{Case}}}, \bibinfo {author} {\bibfnamefont
  {J.~A.}\ \bibnamefont {{Casey-Clyde}}}, \bibinfo {author} {\bibfnamefont
  {M.}~\bibnamefont {{Charisi}}}, \bibinfo {author} {\bibfnamefont
  {S.}~\bibnamefont {{Chatterjee}}}, \bibinfo {author} {\bibfnamefont
  {K.}~\bibnamefont {{Chatziioannou}}}, \bibinfo {author} {\bibfnamefont
  {B.~D.}\ \bibnamefont {{Cheeseboro}}}, \bibinfo {author} {\bibfnamefont
  {S.}~\bibnamefont {{Chen}}}, \bibinfo {author} {\bibfnamefont
  {T.}~\bibnamefont {{Cohen}}}, \bibinfo {author} {\bibfnamefont {J.~M.}\
  \bibnamefont {{Cordes}}}, \bibinfo {author} {\bibfnamefont {N.~J.}\
  \bibnamefont {{Cornish}}}, \bibinfo {author} {\bibfnamefont {F.}~\bibnamefont
  {{Crawford}}}, \bibinfo {author} {\bibfnamefont {H.~T.}\ \bibnamefont
  {{Cromartie}}}, \bibinfo {author} {\bibfnamefont {K.}~\bibnamefont
  {{Crowter}}}, \bibinfo {author} {\bibfnamefont {C.~J.}\ \bibnamefont
  {{Cutler}}}, \bibinfo {author} {\bibfnamefont {D.~J.}\ \bibnamefont
  {{D'Orazio}}}, \bibinfo {author} {\bibfnamefont {M.~E.}\ \bibnamefont
  {{DeCesar}}}, \bibinfo {author} {\bibfnamefont {D.}~\bibnamefont {{DeGan}}},
  \bibinfo {author} {\bibfnamefont {P.~B.}\ \bibnamefont {{Demorest}}},
  \bibinfo {author} {\bibfnamefont {H.}~\bibnamefont {{Deng}}}, \bibinfo
  {author} {\bibfnamefont {T.}~\bibnamefont {{Dolch}}}, \bibinfo {author}
  {\bibfnamefont {B.}~\bibnamefont {{Drachler}}}, \bibinfo {author}
  {\bibfnamefont {E.~C.}\ \bibnamefont {{Ferrara}}}, \bibinfo {author}
  {\bibfnamefont {W.}~\bibnamefont {{Fiore}}}, \bibinfo {author} {\bibfnamefont
  {E.}~\bibnamefont {{Fonseca}}}, \bibinfo {author} {\bibfnamefont {G.~E.}\
  \bibnamefont {{Freedman}}}, \bibinfo {author} {\bibfnamefont
  {E.}~\bibnamefont {{Gardiner}}}, \bibinfo {author} {\bibfnamefont
  {N.}~\bibnamefont {{Garver-Daniels}}}, \bibinfo {author} {\bibfnamefont
  {P.~A.}\ \bibnamefont {{Gentile}}}, \bibinfo {author} {\bibfnamefont {K.~A.}\
  \bibnamefont {{Gersbach}}}, \bibinfo {author} {\bibfnamefont
  {J.}~\bibnamefont {{Glaser}}}, \bibinfo {author} {\bibfnamefont {D.~C.}\
  \bibnamefont {{Good}}}, \bibinfo {author} {\bibfnamefont {K.}~\bibnamefont
  {{G{\"u}ltekin}}}, \bibinfo {author} {\bibfnamefont {J.~S.}\ \bibnamefont
  {{Hazboun}}}, \bibinfo {author} {\bibfnamefont {S.}~\bibnamefont
  {{Hourihane}}}, \bibinfo {author} {\bibfnamefont {K.}~\bibnamefont {{Islo}}},
  \bibinfo {author} {\bibfnamefont {R.~J.}\ \bibnamefont {{Jennings}}},
  \bibinfo {author} {\bibfnamefont {A.}~\bibnamefont {{Johnson}}}, \bibinfo
  {author} {\bibfnamefont {M.~L.}\ \bibnamefont {{Jones}}}, \bibinfo {author}
  {\bibfnamefont {A.~R.}\ \bibnamefont {{Kaiser}}}, \bibinfo {author}
  {\bibfnamefont {D.~L.}\ \bibnamefont {{Kaplan}}}, \bibinfo {author}
  {\bibfnamefont {L.~Z.}\ \bibnamefont {{Kelley}}}, \bibinfo {author}
  {\bibfnamefont {M.}~\bibnamefont {{Kerr}}}, \bibinfo {author} {\bibfnamefont
  {J.~S.}\ \bibnamefont {{Key}}}, \bibinfo {author} {\bibfnamefont
  {N.}~\bibnamefont {{Laal}}}, \bibinfo {author} {\bibfnamefont {M.~T.}\
  \bibnamefont {{Lam}}}, \bibinfo {author} {\bibfnamefont {W.~G.}\ \bibnamefont
  {{Lamb}}}, \bibinfo {author} {\bibfnamefont {T.~J.~W.}\ \bibnamefont
  {{Lazio}}}, \bibinfo {author} {\bibfnamefont {N.}~\bibnamefont
  {{Lewandowska}}}, \bibinfo {author} {\bibfnamefont {T.~B.}\ \bibnamefont
  {{Littenberg}}}, \bibinfo {author} {\bibfnamefont {T.}~\bibnamefont {{Liu}}},
  \bibinfo {author} {\bibfnamefont {J.}~\bibnamefont {{Luo}}}, \bibinfo
  {author} {\bibfnamefont {R.~S.}\ \bibnamefont {{Lynch}}}, \bibinfo {author}
  {\bibfnamefont {C.-P.}\ \bibnamefont {{Ma}}}, \bibinfo {author}
  {\bibfnamefont {D.~R.}\ \bibnamefont {{Madison}}}, \bibinfo {author}
  {\bibfnamefont {A.}~\bibnamefont {{McEwen}}}, \bibinfo {author}
  {\bibfnamefont {J.~W.}\ \bibnamefont {{McKee}}}, \bibinfo {author}
  {\bibfnamefont {M.~A.}\ \bibnamefont {{McLaughlin}}}, \bibinfo {author}
  {\bibfnamefont {N.}~\bibnamefont {{McMann}}}, \bibinfo {author}
  {\bibfnamefont {B.~W.}\ \bibnamefont {{Meyers}}}, \bibinfo {author}
  {\bibfnamefont {P.~M.}\ \bibnamefont {{Meyers}}}, \bibinfo {author}
  {\bibfnamefont {C.~M.~F.}\ \bibnamefont {{Mingarelli}}}, \bibinfo {author}
  {\bibfnamefont {A.}~\bibnamefont {{Mitridate}}}, \bibinfo {author}
  {\bibfnamefont {P.}~\bibnamefont {{Natarajan}}}, \bibinfo {author}
  {\bibfnamefont {C.}~\bibnamefont {{Ng}}}, \bibinfo {author} {\bibfnamefont
  {D.~J.}\ \bibnamefont {{Nice}}}, \bibinfo {author} {\bibfnamefont {S.~K.}\
  \bibnamefont {{Ocker}}}, \bibinfo {author} {\bibfnamefont {K.~D.}\
  \bibnamefont {{Olum}}}, \bibinfo {author} {\bibfnamefont {T.~T.}\
  \bibnamefont {{Pennucci}}}, \bibinfo {author} {\bibfnamefont {B.~B.~P.}\
  \bibnamefont {{Perera}}}, \bibinfo {author} {\bibfnamefont {P.}~\bibnamefont
  {{Petrov}}}, \bibinfo {author} {\bibfnamefont {N.~S.}\ \bibnamefont {{Pol}}},
  \bibinfo {author} {\bibfnamefont {H.~A.}\ \bibnamefont {{Radovan}}}, \bibinfo
  {author} {\bibfnamefont {S.~M.}\ \bibnamefont {{Ransom}}}, \bibinfo {author}
  {\bibfnamefont {P.~S.}\ \bibnamefont {{Ray}}}, \bibinfo {author}
  {\bibfnamefont {J.~D.}\ \bibnamefont {{Romano}}}, \bibinfo {author}
  {\bibfnamefont {J.~C.}\ \bibnamefont {{Runnoe}}}, \bibinfo {author}
  {\bibfnamefont {S.~C.}\ \bibnamefont {{Sardesai}}}, \bibinfo {author}
  {\bibfnamefont {A.}~\bibnamefont {{Schmiedekamp}}}, \bibinfo {author}
  {\bibfnamefont {C.}~\bibnamefont {{Schmiedekamp}}}, \bibinfo {author}
  {\bibfnamefont {K.}~\bibnamefont {{Schmitz}}}, \bibinfo {author}
  {\bibfnamefont {L.}~\bibnamefont {{Schult}}}, \bibinfo {author}
  {\bibfnamefont {B.~J.}\ \bibnamefont {{Shapiro-Albert}}}, \bibinfo {author}
  {\bibfnamefont {X.}~\bibnamefont {{Siemens}}}, \bibinfo {author}
  {\bibfnamefont {J.}~\bibnamefont {{Simon}}}, \bibinfo {author} {\bibfnamefont
  {M.~S.}\ \bibnamefont {{Siwek}}}, \bibinfo {author} {\bibfnamefont {I.~H.}\
  \bibnamefont {{Stairs}}}, \bibinfo {author} {\bibfnamefont {D.~R.}\
  \bibnamefont {{Stinebring}}}, \bibinfo {author} {\bibfnamefont
  {K.}~\bibnamefont {{Stovall}}}, \bibinfo {author} {\bibfnamefont {J.~P.}\
  \bibnamefont {{Sun}}}, \bibinfo {author} {\bibfnamefont {A.}~\bibnamefont
  {{Susobhanan}}}, \bibinfo {author} {\bibfnamefont {J.~K.}\ \bibnamefont
  {{Swiggum}}}, \bibinfo {author} {\bibfnamefont {J.}~\bibnamefont {{Taylor}}},
  \bibinfo {author} {\bibfnamefont {S.~R.}\ \bibnamefont {{Taylor}}}, \bibinfo
  {author} {\bibfnamefont {J.~E.}\ \bibnamefont {{Turner}}}, \bibinfo {author}
  {\bibfnamefont {C.}~\bibnamefont {{Unal}}}, \bibinfo {author} {\bibfnamefont
  {M.}~\bibnamefont {{Vallisneri}}}, \bibinfo {author} {\bibfnamefont {S.~J.}\
  \bibnamefont {{Vigeland}}}, \bibinfo {author} {\bibfnamefont {J.~M.}\
  \bibnamefont {{Wachter}}}, \bibinfo {author} {\bibfnamefont {H.~M.}\
  \bibnamefont {{Wahl}}}, \bibinfo {author} {\bibfnamefont {Q.}~\bibnamefont
  {{Wang}}}, \bibinfo {author} {\bibfnamefont {C.~A.}\ \bibnamefont {{Witt}}},
  \bibinfo {author} {\bibfnamefont {D.}~\bibnamefont {{Wright}}},\ and\
  \bibinfo {author} {\bibfnamefont {O.}~\bibnamefont {{Young}}},\ }\bibfield
  {title} {\bibinfo {title} {{The NANOGrav 15-year Data Set: Constraints on
  Supermassive Black Hole Binaries from the Gravitational Wave Background}},\
  }\href {https://doi.org/10.48550/arXiv.2306.16220} {\bibfield  {journal}
  {\bibinfo  {journal} {arXiv e-prints}\ ,\ \bibinfo {eid} {arXiv:2306.16220}}
  (\bibinfo {year} {2023})},\ \Eprint {https://arxiv.org/abs/2306.16220}
  {arXiv:2306.16220 [astro-ph.HE]} \BibitemShut {NoStop}%
\bibitem [{\citenamefont {{Addazi}}\ \emph {et~al.}(2023)\citenamefont
  {{Addazi}}, \citenamefont {{Cai}}, \citenamefont {{Marciano}},\ and\
  \citenamefont {{Visinelli}}}]{2023arXiv230617205A}%
  \BibitemOpen
  \bibfield  {author} {\bibinfo {author} {\bibfnamefont {A.}~\bibnamefont
  {{Addazi}}}, \bibinfo {author} {\bibfnamefont {Y.-F.}\ \bibnamefont {{Cai}}},
  \bibinfo {author} {\bibfnamefont {A.}~\bibnamefont {{Marciano}}},\ and\
  \bibinfo {author} {\bibfnamefont {L.}~\bibnamefont {{Visinelli}}},\
  }\bibfield  {title} {\bibinfo {title} {{Have pulsar timing array methods
  detected a cosmological phase transition?}},\ }\href
  {https://doi.org/10.48550/arXiv.2306.17205} {\bibfield  {journal} {\bibinfo
  {journal} {arXiv e-prints}\ ,\ \bibinfo {eid} {arXiv:2306.17205}} (\bibinfo
  {year} {2023})},\ \Eprint {https://arxiv.org/abs/2306.17205}
  {arXiv:2306.17205 [astro-ph.CO]} \BibitemShut {NoStop}%
\bibitem [{\citenamefont {{Kitajima}}\ and\ \citenamefont
  {{Nakayama}}(2023)}]{2023arXiv230617390K}%
  \BibitemOpen
  \bibfield  {author} {\bibinfo {author} {\bibfnamefont {N.}~\bibnamefont
  {{Kitajima}}}\ and\ \bibinfo {author} {\bibfnamefont {K.}~\bibnamefont
  {{Nakayama}}},\ }\bibfield  {title} {\bibinfo {title} {{Nanohertz
  gravitational waves from cosmic strings and dark photon dark matter}},\
  }\href {https://doi.org/10.48550/arXiv.2306.17390} {\bibfield  {journal}
  {\bibinfo  {journal} {arXiv e-prints}\ ,\ \bibinfo {eid} {arXiv:2306.17390}}
  (\bibinfo {year} {2023})},\ \Eprint {https://arxiv.org/abs/2306.17390}
  {arXiv:2306.17390 [hep-ph]} \BibitemShut {NoStop}%
\bibitem [{\citenamefont {{Figueroa}}\ \emph {et~al.}(2023)\citenamefont
  {{Figueroa}}, \citenamefont {{Pieroni}}, \citenamefont {{Ricciardone}},\ and\
  \citenamefont {{Simakachorn}}}]{2023arXiv230702399F}%
  \BibitemOpen
  \bibfield  {author} {\bibinfo {author} {\bibfnamefont {D.~G.}\ \bibnamefont
  {{Figueroa}}}, \bibinfo {author} {\bibfnamefont {M.}~\bibnamefont
  {{Pieroni}}}, \bibinfo {author} {\bibfnamefont {A.}~\bibnamefont
  {{Ricciardone}}},\ and\ \bibinfo {author} {\bibfnamefont {P.}~\bibnamefont
  {{Simakachorn}}},\ }\bibfield  {title} {\bibinfo {title} {{Cosmological
  Background Interpretation of Pulsar Timing Array Data}},\ }\href
  {https://doi.org/10.48550/arXiv.2307.02399} {\bibfield  {journal} {\bibinfo
  {journal} {arXiv e-prints}\ ,\ \bibinfo {eid} {arXiv:2307.02399}} (\bibinfo
  {year} {2023})},\ \Eprint {https://arxiv.org/abs/2307.02399}
  {arXiv:2307.02399 [astro-ph.CO]} \BibitemShut {NoStop}%
\bibitem [{\citenamefont {{Inomata}}\ \emph {et~al.}(2023)\citenamefont
  {{Inomata}}, \citenamefont {{Kohri}},\ and\ \citenamefont
  {{Terada}}}]{2023arXiv230617834I}%
  \BibitemOpen
  \bibfield  {author} {\bibinfo {author} {\bibfnamefont {K.}~\bibnamefont
  {{Inomata}}}, \bibinfo {author} {\bibfnamefont {K.}~\bibnamefont {{Kohri}}},\
  and\ \bibinfo {author} {\bibfnamefont {T.}~\bibnamefont {{Terada}}},\
  }\bibfield  {title} {\bibinfo {title} {{The Detected Stochastic Gravitational
  Waves and Sub-Solar Primordial Black Holes}},\ }\href
  {https://doi.org/10.48550/arXiv.2306.17834} {\bibfield  {journal} {\bibinfo
  {journal} {arXiv e-prints}\ ,\ \bibinfo {eid} {arXiv:2306.17834}} (\bibinfo
  {year} {2023})},\ \Eprint {https://arxiv.org/abs/2306.17834}
  {arXiv:2306.17834 [astro-ph.CO]} \BibitemShut {NoStop}%
\bibitem [{\citenamefont {{Franciolini}}\ \emph {et~al.}(2023)\citenamefont
  {{Franciolini}}, \citenamefont {{Junior Iovino}}, \citenamefont
  {{Vaskonen}},\ and\ \citenamefont {{Veermae}}}]{2023arXiv230617149F}%
  \BibitemOpen
  \bibfield  {author} {\bibinfo {author} {\bibfnamefont {G.}~\bibnamefont
  {{Franciolini}}}, \bibinfo {author} {\bibfnamefont {A.}~\bibnamefont {{Junior
  Iovino}}}, \bibinfo {author} {\bibfnamefont {V.}~\bibnamefont {{Vaskonen}}},\
  and\ \bibinfo {author} {\bibfnamefont {H.}~\bibnamefont {{Veermae}}},\
  }\bibfield  {title} {\bibinfo {title} {{The recent gravitational wave
  observation by pulsar timing arrays and primordial black holes: the
  importance of non-gaussianities}},\ }\href
  {https://doi.org/10.48550/arXiv.2306.17149} {\bibfield  {journal} {\bibinfo
  {journal} {arXiv e-prints}\ ,\ \bibinfo {eid} {arXiv:2306.17149}} (\bibinfo
  {year} {2023})},\ \Eprint {https://arxiv.org/abs/2306.17149}
  {arXiv:2306.17149 [astro-ph.CO]} \BibitemShut {NoStop}%
\bibitem [{\citenamefont {Saito}\ and\ \citenamefont
  {Yokoyama}(2009)}]{Saito_2009}%
  \BibitemOpen
  \bibfield  {author} {\bibinfo {author} {\bibfnamefont {R.}~\bibnamefont
  {Saito}}\ and\ \bibinfo {author} {\bibfnamefont {J.}~\bibnamefont
  {Yokoyama}},\ }\bibfield  {title} {\bibinfo {title} {Gravitational-wave
  background as a probe of the primordial black-hole abundance},\ }\bibfield
  {journal} {\bibinfo  {journal} {Physical Review Letters}\ }\textbf {\bibinfo
  {volume} {102}},\ \href {https://doi.org/10.1103/physrevlett.102.161101}
  {10.1103/physrevlett.102.161101} (\bibinfo {year} {2009})\BibitemShut
  {NoStop}%
\bibitem [{\citenamefont {Bartolo}\ \emph
  {et~al.}(2019{\natexlab{a}})\citenamefont {Bartolo}, \citenamefont {De~Luca},
  \citenamefont {Franciolini}, \citenamefont {Lewis}, \citenamefont {Peloso},\
  and\ \citenamefont {Riotto}}]{Bartolo:2018evs}%
  \BibitemOpen
  \bibfield  {author} {\bibinfo {author} {\bibfnamefont {N.}~\bibnamefont
  {Bartolo}}, \bibinfo {author} {\bibfnamefont {V.}~\bibnamefont {De~Luca}},
  \bibinfo {author} {\bibfnamefont {G.}~\bibnamefont {Franciolini}}, \bibinfo
  {author} {\bibfnamefont {A.}~\bibnamefont {Lewis}}, \bibinfo {author}
  {\bibfnamefont {M.}~\bibnamefont {Peloso}},\ and\ \bibinfo {author}
  {\bibfnamefont {A.}~\bibnamefont {Riotto}},\ }\bibfield  {title} {\bibinfo
  {title} {{Primordial Black Hole Dark Matter: LISA Serendipity}},\ }\href
  {https://doi.org/10.1103/PhysRevLett.122.211301} {\bibfield  {journal}
  {\bibinfo  {journal} {Phys. Rev. Lett.}\ }\textbf {\bibinfo {volume} {122}},\
  \bibinfo {pages} {211301} (\bibinfo {year} {2019}{\natexlab{a}})},\ \Eprint
  {https://arxiv.org/abs/1810.12218} {arXiv:1810.12218 [astro-ph.CO]}
  \BibitemShut {NoStop}%
\bibitem [{\citenamefont {Bartolo}\ \emph {et~al.}(2020)\citenamefont
  {Bartolo}, \citenamefont {Bertacca}, \citenamefont {De~Luca}, \citenamefont
  {Franciolini}, \citenamefont {Matarrese}, \citenamefont {Peloso},
  \citenamefont {Ricciardone}, \citenamefont {Riotto},\ and\ \citenamefont
  {Tasinato}}]{Bartolo:2019zvb}%
  \BibitemOpen
  \bibfield  {author} {\bibinfo {author} {\bibfnamefont {N.}~\bibnamefont
  {Bartolo}}, \bibinfo {author} {\bibfnamefont {D.}~\bibnamefont {Bertacca}},
  \bibinfo {author} {\bibfnamefont {V.}~\bibnamefont {De~Luca}}, \bibinfo
  {author} {\bibfnamefont {G.}~\bibnamefont {Franciolini}}, \bibinfo {author}
  {\bibfnamefont {S.}~\bibnamefont {Matarrese}}, \bibinfo {author}
  {\bibfnamefont {M.}~\bibnamefont {Peloso}}, \bibinfo {author} {\bibfnamefont
  {A.}~\bibnamefont {Ricciardone}}, \bibinfo {author} {\bibfnamefont
  {A.}~\bibnamefont {Riotto}},\ and\ \bibinfo {author} {\bibfnamefont
  {G.}~\bibnamefont {Tasinato}},\ }\bibfield  {title} {\bibinfo {title}
  {{Gravitational wave anisotropies from primordial black holes}},\ }\href
  {https://doi.org/10.1088/1475-7516/2020/02/028} {\bibfield  {journal}
  {\bibinfo  {journal} {JCAP}\ }\textbf {\bibinfo {volume} {02}},\ \bibinfo
  {pages} {028}},\ \Eprint {https://arxiv.org/abs/1909.12619} {arXiv:1909.12619
  [astro-ph.CO]} \BibitemShut {NoStop}%
\bibitem [{\citenamefont {Garcia-Bellido}\ \emph {et~al.}(2017)\citenamefont
  {Garcia-Bellido}, \citenamefont {Peloso},\ and\ \citenamefont
  {Unal}}]{Garcia-Bellido:2017aan}%
  \BibitemOpen
  \bibfield  {author} {\bibinfo {author} {\bibfnamefont {J.}~\bibnamefont
  {Garcia-Bellido}}, \bibinfo {author} {\bibfnamefont {M.}~\bibnamefont
  {Peloso}},\ and\ \bibinfo {author} {\bibfnamefont {C.}~\bibnamefont {Unal}},\
  }\bibfield  {title} {\bibinfo {title} {{Gravitational Wave signatures of
  inflationary models from Primordial Black Hole Dark Matter}},\ }\href
  {https://doi.org/10.1088/1475-7516/2017/09/013} {\bibfield  {journal}
  {\bibinfo  {journal} {JCAP}\ }\textbf {\bibinfo {volume} {09}},\ \bibinfo
  {pages} {013}},\ \Eprint {https://arxiv.org/abs/1707.02441} {arXiv:1707.02441
  [astro-ph.CO]} \BibitemShut {NoStop}%
\bibitem [{\citenamefont {Gong}(2022)}]{Gong:2019mui}%
  \BibitemOpen
  \bibfield  {author} {\bibinfo {author} {\bibfnamefont {J.-O.}\ \bibnamefont
  {Gong}},\ }\bibfield  {title} {\bibinfo {title} {{Analytic Integral Solutions
  for Induced Gravitational Waves}},\ }\href
  {https://doi.org/10.3847/1538-4357/ac3a6c} {\bibfield  {journal} {\bibinfo
  {journal} {Astrophys. J.}\ }\textbf {\bibinfo {volume} {925}},\ \bibinfo
  {pages} {102} (\bibinfo {year} {2022})},\ \Eprint
  {https://arxiv.org/abs/1909.12708} {arXiv:1909.12708 [gr-qc]} \BibitemShut
  {NoStop}%
\bibitem [{\citenamefont {Aghanim}\ \emph {et~al.}(2020)\citenamefont {Aghanim}
  \emph {et~al.}}]{Planck:2018vyg}%
  \BibitemOpen
  \bibfield  {author} {\bibinfo {author} {\bibfnamefont {N.}~\bibnamefont
  {Aghanim}} \emph {et~al.} (\bibinfo {collaboration} {Planck}),\ }\bibfield
  {title} {\bibinfo {title} {{Planck 2018 results. VI. Cosmological
  parameters}},\ }\href {https://doi.org/10.1051/0004-6361/201833910}
  {\bibfield  {journal} {\bibinfo  {journal} {Astron. Astrophys.}\ }\textbf
  {\bibinfo {volume} {641}},\ \bibinfo {pages} {A6} (\bibinfo {year} {2020})},\
  \bibinfo {note} {[Erratum: Astron.Astrophys. 652, C4 (2021)]},\ \Eprint
  {https://arxiv.org/abs/1807.06209} {arXiv:1807.06209 [astro-ph.CO]}
  \BibitemShut {NoStop}%
\bibitem [{\citenamefont {{Galloni}}\ \emph {et~al.}(2022)\citenamefont
  {{Galloni}}, \citenamefont {{Bartolo}}, \citenamefont {{Matarrese}},
  \citenamefont {{Migliaccio}}, \citenamefont {{Ricciardone}},\ and\
  \citenamefont {{Vittorio}}}]{2022arXiv220800188G}%
  \BibitemOpen
  \bibfield  {author} {\bibinfo {author} {\bibfnamefont {G.}~\bibnamefont
  {{Galloni}}}, \bibinfo {author} {\bibfnamefont {N.}~\bibnamefont
  {{Bartolo}}}, \bibinfo {author} {\bibfnamefont {S.}~\bibnamefont
  {{Matarrese}}}, \bibinfo {author} {\bibfnamefont {M.}~\bibnamefont
  {{Migliaccio}}}, \bibinfo {author} {\bibfnamefont {A.}~\bibnamefont
  {{Ricciardone}}},\ and\ \bibinfo {author} {\bibfnamefont {N.}~\bibnamefont
  {{Vittorio}}},\ }\bibfield  {title} {\bibinfo {title} {{Updated constraints
  on amplitude and tilt of the tensor primordial spectrum}},\ }\href@noop {}
  {\bibfield  {journal} {\bibinfo  {journal} {arXiv e-prints}\ ,\ \bibinfo
  {eid} {arXiv:2208.00188}} (\bibinfo {year} {2022})},\ \Eprint
  {https://arxiv.org/abs/2208.00188} {arXiv:2208.00188 [astro-ph.CO]}
  \BibitemShut {NoStop}%
\bibitem [{\citenamefont {Bari}\ \emph {et~al.}(2022)\citenamefont {Bari},
  \citenamefont {Ricciardone}, \citenamefont {Bartolo}, \citenamefont
  {Bertacca},\ and\ \citenamefont {Matarrese}}]{PhysRevLett.129.091301}%
  \BibitemOpen
  \bibfield  {author} {\bibinfo {author} {\bibfnamefont {P.}~\bibnamefont
  {Bari}}, \bibinfo {author} {\bibfnamefont {A.}~\bibnamefont {Ricciardone}},
  \bibinfo {author} {\bibfnamefont {N.}~\bibnamefont {Bartolo}}, \bibinfo
  {author} {\bibfnamefont {D.}~\bibnamefont {Bertacca}},\ and\ \bibinfo
  {author} {\bibfnamefont {S.}~\bibnamefont {Matarrese}},\ }\bibfield  {title}
  {\bibinfo {title} {Signatures of primordial gravitational waves on the
  large-scale structure of the universe},\ }\href
  {https://doi.org/10.1103/PhysRevLett.129.091301} {\bibfield  {journal}
  {\bibinfo  {journal} {Phys. Rev. Lett.}\ }\textbf {\bibinfo {volume} {129}},\
  \bibinfo {pages} {091301} (\bibinfo {year} {2022})}\BibitemShut {NoStop}%
\bibitem [{\citenamefont {{Garoffolo}}(2022)}]{2022arXiv221005718G}%
  \BibitemOpen
  \bibfield  {author} {\bibinfo {author} {\bibfnamefont {A.}~\bibnamefont
  {{Garoffolo}}},\ }\bibfield  {title} {\bibinfo {title} {{Wave-optics limit of
  the stochastic gravitational wave background}},\ }\href
  {https://doi.org/10.48550/arXiv.2210.05718} {\bibfield  {journal} {\bibinfo
  {journal} {arXiv e-prints}\ ,\ \bibinfo {eid} {arXiv:2210.05718}} (\bibinfo
  {year} {2022})},\ \Eprint {https://arxiv.org/abs/2210.05718}
  {arXiv:2210.05718 [astro-ph.CO]} \BibitemShut {NoStop}%
\bibitem [{\citenamefont {Nakamura}\ and\ \citenamefont
  {Deguchi}(1999)}]{Nakamura:1999uwi}%
  \BibitemOpen
  \bibfield  {author} {\bibinfo {author} {\bibfnamefont {T.~T.}\ \bibnamefont
  {Nakamura}}\ and\ \bibinfo {author} {\bibfnamefont {S.}~\bibnamefont
  {Deguchi}},\ }\bibfield  {title} {\bibinfo {title} {{Wave Optics in
  Gravitational Lensing}},\ }\href {https://doi.org/10.1143/ptps.133.137}
  {\bibfield  {journal} {\bibinfo  {journal} {Prog. Theor. Phys. Suppl.}\
  }\textbf {\bibinfo {volume} {133}},\ \bibinfo {pages} {137} (\bibinfo {year}
  {1999})}\BibitemShut {NoStop}%
\bibitem [{\citenamefont {Takahashi}\ and\ \citenamefont
  {Nakamura}(2003)}]{Takahashi:2003ix}%
  \BibitemOpen
  \bibfield  {author} {\bibinfo {author} {\bibfnamefont {R.}~\bibnamefont
  {Takahashi}}\ and\ \bibinfo {author} {\bibfnamefont {T.}~\bibnamefont
  {Nakamura}},\ }\bibfield  {title} {\bibinfo {title} {{Wave effects in
  gravitational lensing of gravitational waves from chirping binaries}},\
  }\href {https://doi.org/10.1086/377430} {\bibfield  {journal} {\bibinfo
  {journal} {Astrophys. J.}\ }\textbf {\bibinfo {volume} {595}},\ \bibinfo
  {pages} {1039} (\bibinfo {year} {2003})},\ \Eprint
  {https://arxiv.org/abs/astro-ph/0305055} {arXiv:astro-ph/0305055}
  \BibitemShut {NoStop}%
\bibitem [{\citenamefont {Cusin}\ and\ \citenamefont
  {Lagos}(2020)}]{Cusin:2019rmt}%
  \BibitemOpen
  \bibfield  {author} {\bibinfo {author} {\bibfnamefont {G.}~\bibnamefont
  {Cusin}}\ and\ \bibinfo {author} {\bibfnamefont {M.}~\bibnamefont {Lagos}},\
  }\bibfield  {title} {\bibinfo {title} {{Gravitational wave propagation beyond
  geometric optics}},\ }\href {https://doi.org/10.1103/PhysRevD.101.044041}
  {\bibfield  {journal} {\bibinfo  {journal} {Phys. Rev. D}\ }\textbf {\bibinfo
  {volume} {101}},\ \bibinfo {pages} {044041} (\bibinfo {year} {2020})},\
  \Eprint {https://arxiv.org/abs/1910.13326} {arXiv:1910.13326 [gr-qc]}
  \BibitemShut {NoStop}%
\bibitem [{\citenamefont {Chang}\ \emph {et~al.}(2023)\citenamefont {Chang},
  \citenamefont {Zhang},\ and\ \citenamefont {Zhou}}]{Chang:2022vlv}%
  \BibitemOpen
  \bibfield  {author} {\bibinfo {author} {\bibfnamefont {Z.}~\bibnamefont
  {Chang}}, \bibinfo {author} {\bibfnamefont {X.}~\bibnamefont {Zhang}},\ and\
  \bibinfo {author} {\bibfnamefont {J.-Z.}\ \bibnamefont {Zhou}},\ }\bibfield
  {title} {\bibinfo {title} {{Gravitational waves from primordial scalar and
  tensor perturbations}},\ }\href {https://doi.org/10.1103/PhysRevD.107.063510}
  {\bibfield  {journal} {\bibinfo  {journal} {Phys. Rev. D}\ }\textbf {\bibinfo
  {volume} {107}},\ \bibinfo {pages} {063510} (\bibinfo {year} {2023})},\
  \Eprint {https://arxiv.org/abs/2209.07693} {arXiv:2209.07693 [astro-ph.CO]}
  \BibitemShut {NoStop}%
\bibitem [{\citenamefont {Obata}\ and\ \citenamefont
  {Soda}(2016)}]{Obata:2016tmo}%
  \BibitemOpen
  \bibfield  {author} {\bibinfo {author} {\bibfnamefont {I.}~\bibnamefont
  {Obata}}\ and\ \bibinfo {author} {\bibfnamefont {J.}~\bibnamefont {Soda}},\
  }\bibfield  {title} {\bibinfo {title} {{Chiral primordial Chiral primordial
  gravitational waves from dilaton induced delayed chromonatural inflation}},\
  }\href {https://doi.org/10.1103/PhysRevD.93.123502} {\bibfield  {journal}
  {\bibinfo  {journal} {Phys. Rev. D}\ }\textbf {\bibinfo {volume} {93}},\
  \bibinfo {pages} {123502} (\bibinfo {year} {2016})},\ \bibinfo {note}
  {[Addendum: Phys.Rev.D 95, 109903 (2017)]},\ \Eprint
  {https://arxiv.org/abs/1602.06024} {arXiv:1602.06024 [hep-th]} \BibitemShut
  {NoStop}%
\bibitem [{\citenamefont {Bartolo}\ and\ \citenamefont
  {Orlando}(2017)}]{Bartolo:2017szm}%
  \BibitemOpen
  \bibfield  {author} {\bibinfo {author} {\bibfnamefont {N.}~\bibnamefont
  {Bartolo}}\ and\ \bibinfo {author} {\bibfnamefont {G.}~\bibnamefont
  {Orlando}},\ }\bibfield  {title} {\bibinfo {title} {{Parity breaking
  signatures from a Chern-Simons coupling during inflation: the case of
  non-Gaussian gravitational waves}},\ }\href
  {https://doi.org/10.1088/1475-7516/2017/07/034} {\bibfield  {journal}
  {\bibinfo  {journal} {JCAP}\ }\textbf {\bibinfo {volume} {07}},\ \bibinfo
  {pages} {034}},\ \Eprint {https://arxiv.org/abs/1706.04627} {arXiv:1706.04627
  [astro-ph.CO]} \BibitemShut {NoStop}%
\bibitem [{\citenamefont {Bartolo}\ \emph
  {et~al.}(2019{\natexlab{b}})\citenamefont {Bartolo}, \citenamefont
  {Orlando},\ and\ \citenamefont {Shiraishi}}]{Bartolo:2018elp}%
  \BibitemOpen
  \bibfield  {author} {\bibinfo {author} {\bibfnamefont {N.}~\bibnamefont
  {Bartolo}}, \bibinfo {author} {\bibfnamefont {G.}~\bibnamefont {Orlando}},\
  and\ \bibinfo {author} {\bibfnamefont {M.}~\bibnamefont {Shiraishi}},\
  }\bibfield  {title} {\bibinfo {title} {{Measuring chiral gravitational waves
  in Chern-Simons gravity with CMB bispectra}},\ }\href
  {https://doi.org/10.1088/1475-7516/2019/01/050} {\bibfield  {journal}
  {\bibinfo  {journal} {JCAP}\ }\textbf {\bibinfo {volume} {01}},\ \bibinfo
  {pages} {050}},\ \Eprint {https://arxiv.org/abs/1809.11170} {arXiv:1809.11170
  [astro-ph.CO]} \BibitemShut {NoStop}%
\bibitem [{\citenamefont {Komatsu}(2022)}]{Komatsu:2022nvu}%
  \BibitemOpen
  \bibfield  {author} {\bibinfo {author} {\bibfnamefont {E.}~\bibnamefont
  {Komatsu}},\ }\bibfield  {title} {\bibinfo {title} {{New physics from the
  polarized light of the cosmic microwave background}},\ }\href
  {https://doi.org/10.1038/s42254-022-00452-4} {\bibfield  {journal} {\bibinfo
  {journal} {Nature Rev. Phys.}\ }\textbf {\bibinfo {volume} {4}},\ \bibinfo
  {pages} {452} (\bibinfo {year} {2022})},\ \Eprint
  {https://arxiv.org/abs/2202.13919} {arXiv:2202.13919 [astro-ph.CO]}
  \BibitemShut {NoStop}%
\bibitem [{\citenamefont {Dom\`enech}(2021)}]{Domenech:2021ztg}%
  \BibitemOpen
  \bibfield  {author} {\bibinfo {author} {\bibfnamefont {G.}~\bibnamefont
  {Dom\`enech}},\ }\bibfield  {title} {\bibinfo {title} {{Scalar Induced
  Gravitational Waves Review}},\ }\href
  {https://doi.org/10.3390/universe7110398} {\bibfield  {journal} {\bibinfo
  {journal} {Universe}\ }\textbf {\bibinfo {volume} {7}},\ \bibinfo {pages}
  {398} (\bibinfo {year} {2021})},\ \Eprint {https://arxiv.org/abs/2109.01398}
  {arXiv:2109.01398 [gr-qc]} \BibitemShut {NoStop}%
\bibitem [{\citenamefont {Chen}\ \emph {et~al.}(2023)\citenamefont {Chen},
  \citenamefont {Ota}, \citenamefont {Zhu},\ and\ \citenamefont
  {Zhu}}]{Chen:2022dah}%
  \BibitemOpen
  \bibfield  {author} {\bibinfo {author} {\bibfnamefont {C.}~\bibnamefont
  {Chen}}, \bibinfo {author} {\bibfnamefont {A.}~\bibnamefont {Ota}}, \bibinfo
  {author} {\bibfnamefont {H.-Y.}\ \bibnamefont {Zhu}},\ and\ \bibinfo {author}
  {\bibfnamefont {Y.}~\bibnamefont {Zhu}},\ }\bibfield  {title} {\bibinfo
  {title} {{Missing one-loop contributions in secondary gravitational waves}},\
  }\href {https://doi.org/10.1103/PhysRevD.107.083518} {\bibfield  {journal}
  {\bibinfo  {journal} {Phys. Rev. D}\ }\textbf {\bibinfo {volume} {107}},\
  \bibinfo {pages} {083518} (\bibinfo {year} {2023})},\ \Eprint
  {https://arxiv.org/abs/2210.17176} {arXiv:2210.17176 [astro-ph.CO]}
  \BibitemShut {NoStop}%
\bibitem [{\citenamefont {Garcia-Bellido}\ \emph {et~al.}(2016)\citenamefont
  {Garcia-Bellido}, \citenamefont {Peloso},\ and\ \citenamefont
  {Unal}}]{Garcia-Bellido:2016dkw}%
  \BibitemOpen
  \bibfield  {author} {\bibinfo {author} {\bibfnamefont {J.}~\bibnamefont
  {Garcia-Bellido}}, \bibinfo {author} {\bibfnamefont {M.}~\bibnamefont
  {Peloso}},\ and\ \bibinfo {author} {\bibfnamefont {C.}~\bibnamefont {Unal}},\
  }\bibfield  {title} {\bibinfo {title} {{Gravitational waves at interferometer
  scales and primordial black holes in axion inflation}},\ }\href
  {https://doi.org/10.1088/1475-7516/2016/12/031} {\bibfield  {journal}
  {\bibinfo  {journal} {JCAP}\ }\textbf {\bibinfo {volume} {12}},\ \bibinfo
  {pages} {031}},\ \Eprint {https://arxiv.org/abs/1610.03763} {arXiv:1610.03763
  [astro-ph.CO]} \BibitemShut {NoStop}%
\bibitem [{\citenamefont {Namba}\ \emph {et~al.}(2016)\citenamefont {Namba},
  \citenamefont {Peloso}, \citenamefont {Shiraishi}, \citenamefont {Sorbo},\
  and\ \citenamefont {Unal}}]{Namba:2015gja}%
  \BibitemOpen
  \bibfield  {author} {\bibinfo {author} {\bibfnamefont {R.}~\bibnamefont
  {Namba}}, \bibinfo {author} {\bibfnamefont {M.}~\bibnamefont {Peloso}},
  \bibinfo {author} {\bibfnamefont {M.}~\bibnamefont {Shiraishi}}, \bibinfo
  {author} {\bibfnamefont {L.}~\bibnamefont {Sorbo}},\ and\ \bibinfo {author}
  {\bibfnamefont {C.}~\bibnamefont {Unal}},\ }\bibfield  {title} {\bibinfo
  {title} {{Scale-dependent gravitational waves from a rolling axion}},\ }\href
  {https://doi.org/10.1088/1475-7516/2016/01/041} {\bibfield  {journal}
  {\bibinfo  {journal} {JCAP}\ }\textbf {\bibinfo {volume} {01}},\ \bibinfo
  {pages} {041}},\ \Eprint {https://arxiv.org/abs/1509.07521} {arXiv:1509.07521
  [astro-ph.CO]} \BibitemShut {NoStop}%
\bibitem [{\citenamefont {Thorne}\ \emph {et~al.}(2018)\citenamefont {Thorne},
  \citenamefont {Fujita}, \citenamefont {Hazumi}, \citenamefont {Katayama},
  \citenamefont {Komatsu},\ and\ \citenamefont {Shiraishi}}]{Thorne_2018}%
  \BibitemOpen
  \bibfield  {author} {\bibinfo {author} {\bibfnamefont {B.}~\bibnamefont
  {Thorne}}, \bibinfo {author} {\bibfnamefont {T.}~\bibnamefont {Fujita}},
  \bibinfo {author} {\bibfnamefont {M.}~\bibnamefont {Hazumi}}, \bibinfo
  {author} {\bibfnamefont {N.}~\bibnamefont {Katayama}}, \bibinfo {author}
  {\bibfnamefont {E.}~\bibnamefont {Komatsu}},\ and\ \bibinfo {author}
  {\bibfnamefont {M.}~\bibnamefont {Shiraishi}},\ }\bibfield  {title} {\bibinfo
  {title} {Finding the chiral gravitational wave background of an axion- su(2)
  inflationary model using cmb observations and laser interferometers},\
  }\bibfield  {journal} {\bibinfo  {journal} {Physical Review D}\ }\textbf
  {\bibinfo {volume} {97}},\ \href {https://doi.org/10.1103/physrevd.97.043506}
  {10.1103/physrevd.97.043506} (\bibinfo {year} {2018})\BibitemShut {NoStop}%
\bibitem [{\citenamefont {Shiraishi}\ \emph {et~al.}(2016)\citenamefont
  {Shiraishi}, \citenamefont {Hikage}, \citenamefont {Namba}, \citenamefont
  {Namikawa},\ and\ \citenamefont {Hazumi}}]{Shiraishi:2016yun}%
  \BibitemOpen
  \bibfield  {author} {\bibinfo {author} {\bibfnamefont {M.}~\bibnamefont
  {Shiraishi}}, \bibinfo {author} {\bibfnamefont {C.}~\bibnamefont {Hikage}},
  \bibinfo {author} {\bibfnamefont {R.}~\bibnamefont {Namba}}, \bibinfo
  {author} {\bibfnamefont {T.}~\bibnamefont {Namikawa}},\ and\ \bibinfo
  {author} {\bibfnamefont {M.}~\bibnamefont {Hazumi}},\ }\bibfield  {title}
  {\bibinfo {title} {{Testing statistics of the CMB B -mode polarization toward
  unambiguously establishing quantum fluctuation of the vacuum}},\ }\href
  {https://doi.org/10.1103/PhysRevD.94.043506} {\bibfield  {journal} {\bibinfo
  {journal} {Phys. Rev. D}\ }\textbf {\bibinfo {volume} {94}},\ \bibinfo
  {pages} {043506} (\bibinfo {year} {2016})},\ \Eprint
  {https://arxiv.org/abs/1606.06082} {arXiv:1606.06082 [astro-ph.CO]}
  \BibitemShut {NoStop}%
\bibitem [{\citenamefont {Wang}\ \emph {et~al.}(2019)\citenamefont {Wang},
  \citenamefont {Terada},\ and\ \citenamefont {Kohri}}]{Wang:2019kaf}%
  \BibitemOpen
  \bibfield  {author} {\bibinfo {author} {\bibfnamefont {S.}~\bibnamefont
  {Wang}}, \bibinfo {author} {\bibfnamefont {T.}~\bibnamefont {Terada}},\ and\
  \bibinfo {author} {\bibfnamefont {K.}~\bibnamefont {Kohri}},\ }\bibfield
  {title} {\bibinfo {title} {{Prospective constraints on the primordial black
  hole abundance from the stochastic gravitational-wave backgrounds produced by
  coalescing events and curvature perturbations}},\ }\href
  {https://doi.org/10.1103/PhysRevD.99.103531} {\bibfield  {journal} {\bibinfo
  {journal} {Phys. Rev. D}\ }\textbf {\bibinfo {volume} {99}},\ \bibinfo
  {pages} {103531} (\bibinfo {year} {2019})},\ \bibinfo {note} {[Erratum:
  Phys.Rev.D 101, 069901 (2020)]},\ \Eprint {https://arxiv.org/abs/1903.05924}
  {arXiv:1903.05924 [astro-ph.CO]} \BibitemShut {NoStop}%
\bibitem [{\citenamefont {Byrnes}\ \emph {et~al.}(2019)\citenamefont {Byrnes},
  \citenamefont {Cole},\ and\ \citenamefont {Patil}}]{Byrnes:2018txb}%
  \BibitemOpen
  \bibfield  {author} {\bibinfo {author} {\bibfnamefont {C.~T.}\ \bibnamefont
  {Byrnes}}, \bibinfo {author} {\bibfnamefont {P.~S.}\ \bibnamefont {Cole}},\
  and\ \bibinfo {author} {\bibfnamefont {S.~P.}\ \bibnamefont {Patil}},\
  }\bibfield  {title} {\bibinfo {title} {{Steepest growth of the power spectrum
  and primordial black holes}},\ }\href
  {https://doi.org/10.1088/1475-7516/2019/06/028} {\bibfield  {journal}
  {\bibinfo  {journal} {JCAP}\ }\textbf {\bibinfo {volume} {06}},\ \bibinfo
  {pages} {028}},\ \Eprint {https://arxiv.org/abs/1811.11158} {arXiv:1811.11158
  [astro-ph.CO]} \BibitemShut {NoStop}%
\bibitem [{\citenamefont {Thrane}\ and\ \citenamefont
  {Romano}(2013)}]{Thrane:2013oya}%
  \BibitemOpen
  \bibfield  {author} {\bibinfo {author} {\bibfnamefont {E.}~\bibnamefont
  {Thrane}}\ and\ \bibinfo {author} {\bibfnamefont {J.~D.}\ \bibnamefont
  {Romano}},\ }\bibfield  {title} {\bibinfo {title} {{Sensitivity curves for
  searches for gravitational-wave backgrounds}},\ }\href
  {https://doi.org/10.1103/PhysRevD.88.124032} {\bibfield  {journal} {\bibinfo
  {journal} {Phys. Rev.}\ }\textbf {\bibinfo {volume} {D88}},\ \bibinfo {pages}
  {124032} (\bibinfo {year} {2013})},\ \Eprint
  {https://arxiv.org/abs/1310.5300} {arXiv:1310.5300 [astro-ph.IM]}
  \BibitemShut {NoStop}%
\bibitem [{\citenamefont {Barke}\ \emph {et~al.}(2015)\citenamefont {Barke},
  \citenamefont {Wang}, \citenamefont {Esteban~Delgado}, \citenamefont
  {Tr\"obs}, \citenamefont {Heinzel},\ and\ \citenamefont
  {Danzmann}}]{Barke:2014lsa}%
  \BibitemOpen
  \bibfield  {author} {\bibinfo {author} {\bibfnamefont {S.}~\bibnamefont
  {Barke}}, \bibinfo {author} {\bibfnamefont {Y.}~\bibnamefont {Wang}},
  \bibinfo {author} {\bibfnamefont {J.~J.}\ \bibnamefont {Esteban~Delgado}},
  \bibinfo {author} {\bibfnamefont {M.}~\bibnamefont {Tr\"obs}}, \bibinfo
  {author} {\bibfnamefont {G.}~\bibnamefont {Heinzel}},\ and\ \bibinfo {author}
  {\bibfnamefont {K.}~\bibnamefont {Danzmann}},\ }\bibfield  {title} {\bibinfo
  {title} {{Towards a gravitational wave observatory designer: sensitivity
  limits of spaceborne detectors}},\ }\href
  {https://doi.org/10.1088/0264-9381/32/9/095004} {\bibfield  {journal}
  {\bibinfo  {journal} {Class. Quant. Grav.}\ }\textbf {\bibinfo {volume}
  {32}},\ \bibinfo {pages} {095004} (\bibinfo {year} {2015})},\ \Eprint
  {https://arxiv.org/abs/1411.1260} {arXiv:1411.1260 [physics.ins-det]}
  \BibitemShut {NoStop}%
\bibitem [{\citenamefont {Ruan}\ \emph {et~al.}(2020)\citenamefont {Ruan},
  \citenamefont {Guo}, \citenamefont {Cai},\ and\ \citenamefont
  {Zhang}}]{Ruan:2018tsw}%
  \BibitemOpen
  \bibfield  {author} {\bibinfo {author} {\bibfnamefont {W.-H.}\ \bibnamefont
  {Ruan}}, \bibinfo {author} {\bibfnamefont {Z.-K.}\ \bibnamefont {Guo}},
  \bibinfo {author} {\bibfnamefont {R.-G.}\ \bibnamefont {Cai}},\ and\ \bibinfo
  {author} {\bibfnamefont {Y.-Z.}\ \bibnamefont {Zhang}},\ }\bibfield  {title}
  {\bibinfo {title} {{Taiji program: Gravitational-wave sources}},\ }\href
  {https://doi.org/10.1142/S0217751X2050075X} {\bibfield  {journal} {\bibinfo
  {journal} {Int. J. Mod. Phys. A}\ }\textbf {\bibinfo {volume} {35}},\
  \bibinfo {pages} {2050075} (\bibinfo {year} {2020})},\ \Eprint
  {https://arxiv.org/abs/1807.09495} {arXiv:1807.09495 [gr-qc]} \BibitemShut
  {NoStop}%
\bibitem [{ce()}]{ce}%
  \BibitemOpen
  \href@noop {} {\bibinfo {title} {Cosmic explorer sensitivity curve}},\
  \bibinfo {howpublished} {\url{https://cosmicexplorer.org/sensitivity.html}},\
  \bibinfo {note} {[Online; accessed 05-May-2023]}\BibitemShut {NoStop}%
\bibitem [{A+()}]{A+}%
  \BibitemOpen
  \href@noop {} {\bibinfo {title} {The {A+} design curve}},\ \bibinfo
  {howpublished} {\url{https://dcc.ligo.org/LIGO-T1800042/public}},\ \bibinfo
  {note} {[Online; accessed 05-May-2023]}\BibitemShut {NoStop}%
\bibitem [{voy()}]{voyager}%
  \BibitemOpen
  \href@noop {} {\bibinfo {title} {Ligo unofficial sensitivity curves}},\
  \bibinfo {howpublished} {\url{https://dcc.ligo.org/LIGO-T1500293/public}},\
  \bibinfo {note} {[Online; accessed 05-May-2023]}\BibitemShut {NoStop}%
\bibitem [{\citenamefont {Schmitz}(2021)}]{Schmitz:2020syl}%
  \BibitemOpen
  \bibfield  {author} {\bibinfo {author} {\bibfnamefont {K.}~\bibnamefont
  {Schmitz}},\ }\bibfield  {title} {\bibinfo {title} {{New Sensitivity Curves
  for Gravitational-Wave Signals from Cosmological Phase Transitions}},\ }\href
  {https://doi.org/10.1007/JHEP01(2021)097} {\bibfield  {journal} {\bibinfo
  {journal} {JHEP}\ }\textbf {\bibinfo {volume} {01}},\ \bibinfo {pages}
  {097}},\ \Eprint {https://arxiv.org/abs/2002.04615} {arXiv:2002.04615
  [hep-ph]} \BibitemShut {NoStop}%
\bibitem [{\citenamefont {Abbott}\ \emph {et~al.}(2021)\citenamefont {Abbott}
  \emph {et~al.}}]{KAGRA:2021kbb}%
  \BibitemOpen
  \bibfield  {author} {\bibinfo {author} {\bibfnamefont {R.}~\bibnamefont
  {Abbott}} \emph {et~al.} (\bibinfo {collaboration} {KAGRA, Virgo, LIGO
  Scientific}),\ }\bibfield  {title} {\bibinfo {title} {{Upper limits on the
  isotropic gravitational-wave background from Advanced LIGO and Advanced
  Virgo\textquoteright{}s third observing run}},\ }\href
  {https://doi.org/10.1103/PhysRevD.104.022004} {\bibfield  {journal} {\bibinfo
   {journal} {Phys. Rev. D}\ }\textbf {\bibinfo {volume} {104}},\ \bibinfo
  {pages} {022004} (\bibinfo {year} {2021})},\ \Eprint
  {https://arxiv.org/abs/2101.12130} {arXiv:2101.12130 [gr-qc]} \BibitemShut
  {NoStop}%
\bibitem [{\citenamefont {Cyburt}\ \emph {et~al.}(2005)\citenamefont {Cyburt},
  \citenamefont {Fields}, \citenamefont {Olive},\ and\ \citenamefont
  {Skillman}}]{Cyburt:2004yc}%
  \BibitemOpen
  \bibfield  {author} {\bibinfo {author} {\bibfnamefont {R.~H.}\ \bibnamefont
  {Cyburt}}, \bibinfo {author} {\bibfnamefont {B.~D.}\ \bibnamefont {Fields}},
  \bibinfo {author} {\bibfnamefont {K.~A.}\ \bibnamefont {Olive}},\ and\
  \bibinfo {author} {\bibfnamefont {E.}~\bibnamefont {Skillman}},\ }\bibfield
  {title} {\bibinfo {title} {{New BBN limits on physics beyond the standard
  model from $^4He$}},\ }\href
  {https://doi.org/10.1016/j.astropartphys.2005.01.005} {\bibfield  {journal}
  {\bibinfo  {journal} {Astropart. Phys.}\ }\textbf {\bibinfo {volume} {23}},\
  \bibinfo {pages} {313} (\bibinfo {year} {2005})},\ \Eprint
  {https://arxiv.org/abs/astro-ph/0408033} {arXiv:astro-ph/0408033}
  \BibitemShut {NoStop}%
\bibitem [{\citenamefont {Arbey}\ \emph {et~al.}(2021)\citenamefont {Arbey},
  \citenamefont {Auffinger}, \citenamefont {Sandick}, \citenamefont {Shams
  Es~Haghi},\ and\ \citenamefont {Sinha}}]{Arbey:2021ysg}%
  \BibitemOpen
  \bibfield  {author} {\bibinfo {author} {\bibfnamefont {A.}~\bibnamefont
  {Arbey}}, \bibinfo {author} {\bibfnamefont {J.}~\bibnamefont {Auffinger}},
  \bibinfo {author} {\bibfnamefont {P.}~\bibnamefont {Sandick}}, \bibinfo
  {author} {\bibfnamefont {B.}~\bibnamefont {Shams Es~Haghi}},\ and\ \bibinfo
  {author} {\bibfnamefont {K.}~\bibnamefont {Sinha}},\ }\bibfield  {title}
  {\bibinfo {title} {{Precision calculation of dark radiation from spinning
  primordial black holes and early matter-dominated eras}},\ }\href
  {https://doi.org/10.1103/PhysRevD.103.123549} {\bibfield  {journal} {\bibinfo
   {journal} {Phys. Rev. D}\ }\textbf {\bibinfo {volume} {103}},\ \bibinfo
  {pages} {123549} (\bibinfo {year} {2021})},\ \Eprint
  {https://arxiv.org/abs/2104.04051} {arXiv:2104.04051 [astro-ph.CO]}
  \BibitemShut {NoStop}%
\bibitem [{\citenamefont {{Grohs}}\ and\ \citenamefont
  {{Fuller}}(2023)}]{2023arXiv230112299G}%
  \BibitemOpen
  \bibfield  {author} {\bibinfo {author} {\bibfnamefont {E.}~\bibnamefont
  {{Grohs}}}\ and\ \bibinfo {author} {\bibfnamefont {G.~M.}\ \bibnamefont
  {{Fuller}}},\ }\bibfield  {title} {\bibinfo {title} {{Big Bang
  Nucleosynthesis}},\ }\href {https://doi.org/10.48550/arXiv.2301.12299}
  {\bibfield  {journal} {\bibinfo  {journal} {arXiv e-prints}\ ,\ \bibinfo
  {eid} {arXiv:2301.12299}} (\bibinfo {year} {2023})},\ \Eprint
  {https://arxiv.org/abs/2301.12299} {arXiv:2301.12299 [astro-ph.CO]}
  \BibitemShut {NoStop}%
\bibitem [{\citenamefont {{Abazajian}}\ \emph {et~al.}(2016)\citenamefont
  {{Abazajian}}, \citenamefont {{Adshead}}, \citenamefont {{Ahmed}},
  \citenamefont {{Allen}}, \citenamefont {{Alonso}}, \citenamefont {{Arnold}},
  \citenamefont {{Baccigalupi}}, \citenamefont {{Bartlett}}, \citenamefont
  {{Battaglia}}, \citenamefont {{Benson}}, \citenamefont {{Bischoff}},
  \citenamefont {{Borrill}}, \citenamefont {{Buza}}, \citenamefont
  {{Calabrese}}, \citenamefont {{Caldwell}}, \citenamefont {{Carlstrom}},
  \citenamefont {{Chang}}, \citenamefont {{Crawford}}, \citenamefont
  {{Cyr-Racine}}, \citenamefont {{De Bernardis}}, \citenamefont {{de Haan}},
  \citenamefont {{di Serego Alighieri}}, \citenamefont {{Dunkley}},
  \citenamefont {{Dvorkin}}, \citenamefont {{Errard}}, \citenamefont
  {{Fabbian}}, \citenamefont {{Feeney}}, \citenamefont {{Ferraro}},
  \citenamefont {{Filippini}}, \citenamefont {{Flauger}}, \citenamefont
  {{Fuller}}, \citenamefont {{Gluscevic}}, \citenamefont {{Green}},
  \citenamefont {{Grin}}, \citenamefont {{Grohs}}, \citenamefont {{Henning}},
  \citenamefont {{Hill}}, \citenamefont {{Hlozek}}, \citenamefont {{Holder}},
  \citenamefont {{Holzapfel}}, \citenamefont {{Hu}}, \citenamefont
  {{Huffenberger}}, \citenamefont {{Keskitalo}}, \citenamefont {{Knox}},
  \citenamefont {{Kosowsky}}, \citenamefont {{Kovac}}, \citenamefont
  {{Kovetz}}, \citenamefont {{Kuo}}, \citenamefont {{Kusaka}}, \citenamefont
  {{Le Jeune}}, \citenamefont {{Lee}}, \citenamefont {{Lilley}}, \citenamefont
  {{Loverde}}, \citenamefont {{Madhavacheril}}, \citenamefont {{Mantz}},
  \citenamefont {{Marsh}}, \citenamefont {{McMahon}}, \citenamefont
  {{Meerburg}}, \citenamefont {{Meyers}}, \citenamefont {{Miller}},
  \citenamefont {{Munoz}}, \citenamefont {{Nguyen}}, \citenamefont {{Niemack}},
  \citenamefont {{Peloso}}, \citenamefont {{Peloton}}, \citenamefont
  {{Pogosian}}, \citenamefont {{Pryke}}, \citenamefont {{Raveri}},
  \citenamefont {{Reichardt}}, \citenamefont {{Rocha}}, \citenamefont
  {{Rotti}}, \citenamefont {{Schaan}}, \citenamefont {{Schmittfull}},
  \citenamefont {{Scott}}, \citenamefont {{Sehgal}}, \citenamefont
  {{Shandera}}, \citenamefont {{Sherwin}}, \citenamefont {{Smith}},
  \citenamefont {{Sorbo}}, \citenamefont {{Starkman}}, \citenamefont {{Story}},
  \citenamefont {{van Engelen}}, \citenamefont {{Vieira}}, \citenamefont
  {{Watson}}, \citenamefont {{Whitehorn}},\ and\ \citenamefont {{Kimmy
  Wu}}}]{2016arXiv161002743A}%
  \BibitemOpen
  \bibfield  {author} {\bibinfo {author} {\bibfnamefont {K.~N.}\ \bibnamefont
  {{Abazajian}}}, \bibinfo {author} {\bibfnamefont {P.}~\bibnamefont
  {{Adshead}}}, \bibinfo {author} {\bibfnamefont {Z.}~\bibnamefont {{Ahmed}}},
  \bibinfo {author} {\bibfnamefont {S.~W.}\ \bibnamefont {{Allen}}}, \bibinfo
  {author} {\bibfnamefont {D.}~\bibnamefont {{Alonso}}}, \bibinfo {author}
  {\bibfnamefont {K.~S.}\ \bibnamefont {{Arnold}}}, \bibinfo {author}
  {\bibfnamefont {C.}~\bibnamefont {{Baccigalupi}}}, \bibinfo {author}
  {\bibfnamefont {J.~G.}\ \bibnamefont {{Bartlett}}}, \bibinfo {author}
  {\bibfnamefont {N.}~\bibnamefont {{Battaglia}}}, \bibinfo {author}
  {\bibfnamefont {B.~A.}\ \bibnamefont {{Benson}}}, \bibinfo {author}
  {\bibfnamefont {C.~A.}\ \bibnamefont {{Bischoff}}}, \bibinfo {author}
  {\bibfnamefont {J.}~\bibnamefont {{Borrill}}}, \bibinfo {author}
  {\bibfnamefont {V.}~\bibnamefont {{Buza}}}, \bibinfo {author} {\bibfnamefont
  {E.}~\bibnamefont {{Calabrese}}}, \bibinfo {author} {\bibfnamefont
  {R.}~\bibnamefont {{Caldwell}}}, \bibinfo {author} {\bibfnamefont {J.~E.}\
  \bibnamefont {{Carlstrom}}}, \bibinfo {author} {\bibfnamefont {C.~L.}\
  \bibnamefont {{Chang}}}, \bibinfo {author} {\bibfnamefont {T.~M.}\
  \bibnamefont {{Crawford}}}, \bibinfo {author} {\bibfnamefont {F.-Y.}\
  \bibnamefont {{Cyr-Racine}}}, \bibinfo {author} {\bibfnamefont
  {F.}~\bibnamefont {{De Bernardis}}}, \bibinfo {author} {\bibfnamefont
  {T.}~\bibnamefont {{de Haan}}}, \bibinfo {author} {\bibfnamefont
  {S.}~\bibnamefont {{di Serego Alighieri}}}, \bibinfo {author} {\bibfnamefont
  {J.}~\bibnamefont {{Dunkley}}}, \bibinfo {author} {\bibfnamefont
  {C.}~\bibnamefont {{Dvorkin}}}, \bibinfo {author} {\bibfnamefont
  {J.}~\bibnamefont {{Errard}}}, \bibinfo {author} {\bibfnamefont
  {G.}~\bibnamefont {{Fabbian}}}, \bibinfo {author} {\bibfnamefont
  {S.}~\bibnamefont {{Feeney}}}, \bibinfo {author} {\bibfnamefont
  {S.}~\bibnamefont {{Ferraro}}}, \bibinfo {author} {\bibfnamefont {J.~P.}\
  \bibnamefont {{Filippini}}}, \bibinfo {author} {\bibfnamefont
  {R.}~\bibnamefont {{Flauger}}}, \bibinfo {author} {\bibfnamefont {G.~M.}\
  \bibnamefont {{Fuller}}}, \bibinfo {author} {\bibfnamefont {V.}~\bibnamefont
  {{Gluscevic}}}, \bibinfo {author} {\bibfnamefont {D.}~\bibnamefont
  {{Green}}}, \bibinfo {author} {\bibfnamefont {D.}~\bibnamefont {{Grin}}},
  \bibinfo {author} {\bibfnamefont {E.}~\bibnamefont {{Grohs}}}, \bibinfo
  {author} {\bibfnamefont {J.~W.}\ \bibnamefont {{Henning}}}, \bibinfo {author}
  {\bibfnamefont {J.~C.}\ \bibnamefont {{Hill}}}, \bibinfo {author}
  {\bibfnamefont {R.}~\bibnamefont {{Hlozek}}}, \bibinfo {author}
  {\bibfnamefont {G.}~\bibnamefont {{Holder}}}, \bibinfo {author}
  {\bibfnamefont {W.}~\bibnamefont {{Holzapfel}}}, \bibinfo {author}
  {\bibfnamefont {W.}~\bibnamefont {{Hu}}}, \bibinfo {author} {\bibfnamefont
  {K.~M.}\ \bibnamefont {{Huffenberger}}}, \bibinfo {author} {\bibfnamefont
  {R.}~\bibnamefont {{Keskitalo}}}, \bibinfo {author} {\bibfnamefont
  {L.}~\bibnamefont {{Knox}}}, \bibinfo {author} {\bibfnamefont
  {A.}~\bibnamefont {{Kosowsky}}}, \bibinfo {author} {\bibfnamefont
  {J.}~\bibnamefont {{Kovac}}}, \bibinfo {author} {\bibfnamefont {E.~D.}\
  \bibnamefont {{Kovetz}}}, \bibinfo {author} {\bibfnamefont {C.-L.}\
  \bibnamefont {{Kuo}}}, \bibinfo {author} {\bibfnamefont {A.}~\bibnamefont
  {{Kusaka}}}, \bibinfo {author} {\bibfnamefont {M.}~\bibnamefont {{Le
  Jeune}}}, \bibinfo {author} {\bibfnamefont {A.~T.}\ \bibnamefont {{Lee}}},
  \bibinfo {author} {\bibfnamefont {M.}~\bibnamefont {{Lilley}}}, \bibinfo
  {author} {\bibfnamefont {M.}~\bibnamefont {{Loverde}}}, \bibinfo {author}
  {\bibfnamefont {M.~S.}\ \bibnamefont {{Madhavacheril}}}, \bibinfo {author}
  {\bibfnamefont {A.}~\bibnamefont {{Mantz}}}, \bibinfo {author} {\bibfnamefont
  {D.~J.~E.}\ \bibnamefont {{Marsh}}}, \bibinfo {author} {\bibfnamefont
  {J.}~\bibnamefont {{McMahon}}}, \bibinfo {author} {\bibfnamefont {P.~D.}\
  \bibnamefont {{Meerburg}}}, \bibinfo {author} {\bibfnamefont
  {J.}~\bibnamefont {{Meyers}}}, \bibinfo {author} {\bibfnamefont {A.~D.}\
  \bibnamefont {{Miller}}}, \bibinfo {author} {\bibfnamefont {J.~B.}\
  \bibnamefont {{Munoz}}}, \bibinfo {author} {\bibfnamefont {H.~N.}\
  \bibnamefont {{Nguyen}}}, \bibinfo {author} {\bibfnamefont {M.~D.}\
  \bibnamefont {{Niemack}}}, \bibinfo {author} {\bibfnamefont {M.}~\bibnamefont
  {{Peloso}}}, \bibinfo {author} {\bibfnamefont {J.}~\bibnamefont {{Peloton}}},
  \bibinfo {author} {\bibfnamefont {L.}~\bibnamefont {{Pogosian}}}, \bibinfo
  {author} {\bibfnamefont {C.}~\bibnamefont {{Pryke}}}, \bibinfo {author}
  {\bibfnamefont {M.}~\bibnamefont {{Raveri}}}, \bibinfo {author}
  {\bibfnamefont {C.~L.}\ \bibnamefont {{Reichardt}}}, \bibinfo {author}
  {\bibfnamefont {G.}~\bibnamefont {{Rocha}}}, \bibinfo {author} {\bibfnamefont
  {A.}~\bibnamefont {{Rotti}}}, \bibinfo {author} {\bibfnamefont
  {E.}~\bibnamefont {{Schaan}}}, \bibinfo {author} {\bibfnamefont {M.~M.}\
  \bibnamefont {{Schmittfull}}}, \bibinfo {author} {\bibfnamefont
  {D.}~\bibnamefont {{Scott}}}, \bibinfo {author} {\bibfnamefont
  {N.}~\bibnamefont {{Sehgal}}}, \bibinfo {author} {\bibfnamefont
  {S.}~\bibnamefont {{Shandera}}}, \bibinfo {author} {\bibfnamefont {B.~D.}\
  \bibnamefont {{Sherwin}}}, \bibinfo {author} {\bibfnamefont {T.~L.}\
  \bibnamefont {{Smith}}}, \bibinfo {author} {\bibfnamefont {L.}~\bibnamefont
  {{Sorbo}}}, \bibinfo {author} {\bibfnamefont {G.~D.}\ \bibnamefont
  {{Starkman}}}, \bibinfo {author} {\bibfnamefont {K.~T.}\ \bibnamefont
  {{Story}}}, \bibinfo {author} {\bibfnamefont {A.}~\bibnamefont {{van
  Engelen}}}, \bibinfo {author} {\bibfnamefont {J.~D.}\ \bibnamefont
  {{Vieira}}}, \bibinfo {author} {\bibfnamefont {S.}~\bibnamefont {{Watson}}},
  \bibinfo {author} {\bibfnamefont {N.}~\bibnamefont {{Whitehorn}}},\ and\
  \bibinfo {author} {\bibfnamefont {W.~L.}\ \bibnamefont {{Kimmy Wu}}},\
  }\bibfield  {title} {\bibinfo {title} {{CMB-S4 Science Book, First
  Edition}},\ }\href {https://doi.org/10.48550/arXiv.1610.02743} {\bibfield
  {journal} {\bibinfo  {journal} {arXiv e-prints}\ ,\ \bibinfo {eid}
  {arXiv:1610.02743}} (\bibinfo {year} {2016})},\ \Eprint
  {https://arxiv.org/abs/1610.02743} {arXiv:1610.02743 [astro-ph.CO]}
  \BibitemShut {NoStop}%
\bibitem [{\citenamefont {Domcke}\ \emph {et~al.}(2020)\citenamefont {Domcke},
  \citenamefont {Garcia-Bellido}, \citenamefont {Peloso}, \citenamefont
  {Pieroni}, \citenamefont {Ricciardone}, \citenamefont {Sorbo},\ and\
  \citenamefont {Tasinato}}]{Domcke:2019zls}%
  \BibitemOpen
  \bibfield  {author} {\bibinfo {author} {\bibfnamefont {V.}~\bibnamefont
  {Domcke}}, \bibinfo {author} {\bibfnamefont {J.}~\bibnamefont
  {Garcia-Bellido}}, \bibinfo {author} {\bibfnamefont {M.}~\bibnamefont
  {Peloso}}, \bibinfo {author} {\bibfnamefont {M.}~\bibnamefont {Pieroni}},
  \bibinfo {author} {\bibfnamefont {A.}~\bibnamefont {Ricciardone}}, \bibinfo
  {author} {\bibfnamefont {L.}~\bibnamefont {Sorbo}},\ and\ \bibinfo {author}
  {\bibfnamefont {G.}~\bibnamefont {Tasinato}},\ }\bibfield  {title} {\bibinfo
  {title} {{Measuring the net circular polarization of the stochastic
  gravitational wave background with interferometers}},\ }\href
  {https://doi.org/10.1088/1475-7516/2020/05/028} {\bibfield  {journal}
  {\bibinfo  {journal} {JCAP}\ }\textbf {\bibinfo {volume} {05}},\ \bibinfo
  {pages} {028}},\ \Eprint {https://arxiv.org/abs/1910.08052} {arXiv:1910.08052
  [astro-ph.CO]} \BibitemShut {NoStop}%
\bibitem [{\citenamefont {Orlando}\ \emph {et~al.}(2021)\citenamefont
  {Orlando}, \citenamefont {Pieroni},\ and\ \citenamefont
  {Ricciardone}}]{Orlando:2020oko}%
  \BibitemOpen
  \bibfield  {author} {\bibinfo {author} {\bibfnamefont {G.}~\bibnamefont
  {Orlando}}, \bibinfo {author} {\bibfnamefont {M.}~\bibnamefont {Pieroni}},\
  and\ \bibinfo {author} {\bibfnamefont {A.}~\bibnamefont {Ricciardone}},\
  }\bibfield  {title} {\bibinfo {title} {{Measuring Parity Violation in the
  Stochastic Gravitational Wave Background with the LISA-Taiji network}},\
  }\href {https://doi.org/10.1088/1475-7516/2021/03/069} {\bibfield  {journal}
  {\bibinfo  {journal} {JCAP}\ }\textbf {\bibinfo {volume} {03}},\ \bibinfo
  {pages} {069}},\ \Eprint {https://arxiv.org/abs/2011.07059} {arXiv:2011.07059
  [astro-ph.CO]} \BibitemShut {NoStop}%
\bibitem [{\citenamefont {Seto}(2020)}]{Seto:2020zxw}%
  \BibitemOpen
  \bibfield  {author} {\bibinfo {author} {\bibfnamefont {N.}~\bibnamefont
  {Seto}},\ }\bibfield  {title} {\bibinfo {title} {{Measuring Parity Asymmetry
  of Gravitational Wave Backgrounds with a Heliocentric Detector Network in the
  mHz Band}},\ }\href {https://doi.org/10.1103/PhysRevLett.125.251101}
  {\bibfield  {journal} {\bibinfo  {journal} {Phys. Rev. Lett.}\ }\textbf
  {\bibinfo {volume} {125}},\ \bibinfo {pages} {251101} (\bibinfo {year}
  {2020})},\ \Eprint {https://arxiv.org/abs/2009.02928} {arXiv:2009.02928
  [gr-qc]} \BibitemShut {NoStop}%
\bibitem [{\citenamefont {Sesana}\ \emph {et~al.}(2021)\citenamefont {Sesana}
  \emph {et~al.}}]{Sesana:2019vho}%
  \BibitemOpen
  \bibfield  {author} {\bibinfo {author} {\bibfnamefont {A.}~\bibnamefont
  {Sesana}} \emph {et~al.},\ }\bibfield  {title} {\bibinfo {title} {{Unveiling
  the gravitational universe at $\mu$-Hz frequencies}},\ }\href
  {https://doi.org/10.1007/s10686-021-09709-9} {\bibfield  {journal} {\bibinfo
  {journal} {Exper. Astron.}\ }\textbf {\bibinfo {volume} {51}},\ \bibinfo
  {pages} {1333} (\bibinfo {year} {2021})},\ \Eprint
  {https://arxiv.org/abs/1908.11391} {arXiv:1908.11391 [astro-ph.IM]}
  \BibitemShut {NoStop}%
\bibitem [{\citenamefont {{Blas}}\ and\ \citenamefont
  {{Jenkins}}(2022)}]{2022PhRvL.128j1103B}%
  \BibitemOpen
  \bibfield  {author} {\bibinfo {author} {\bibfnamefont {D.}~\bibnamefont
  {{Blas}}}\ and\ \bibinfo {author} {\bibfnamefont {A.~C.}\ \bibnamefont
  {{Jenkins}}},\ }\bibfield  {title} {\bibinfo {title} {{Bridging the
  {\ensuremath{\mu}} Hz Gap in the Gravitational-Wave Landscape with Binary
  Resonances}},\ }\href {https://doi.org/10.1103/PhysRevLett.128.101103}
  {\bibfield  {journal} {\bibinfo  {journal} {\prl}\ }\textbf {\bibinfo
  {volume} {128}},\ \bibinfo {eid} {101103} (\bibinfo {year} {2022})},\ \Eprint
  {https://arxiv.org/abs/2107.04601} {arXiv:2107.04601 [astro-ph.CO]}
  \BibitemShut {NoStop}%
\bibitem [{\citenamefont {Fedderke}\ \emph {et~al.}(2022)\citenamefont
  {Fedderke}, \citenamefont {Graham},\ and\ \citenamefont
  {Rajendran}}]{Fedderke:2021kuy}%
  \BibitemOpen
  \bibfield  {author} {\bibinfo {author} {\bibfnamefont {M.~A.}\ \bibnamefont
  {Fedderke}}, \bibinfo {author} {\bibfnamefont {P.~W.}\ \bibnamefont
  {Graham}},\ and\ \bibinfo {author} {\bibfnamefont {S.}~\bibnamefont
  {Rajendran}},\ }\bibfield  {title} {\bibinfo {title} {{Asteroids for
  \ensuremath{\mu}Hz gravitational-wave detection}},\ }\href
  {https://doi.org/10.1103/PhysRevD.105.103018} {\bibfield  {journal} {\bibinfo
   {journal} {Phys. Rev. D}\ }\textbf {\bibinfo {volume} {105}},\ \bibinfo
  {pages} {103018} (\bibinfo {year} {2022})},\ \Eprint
  {https://arxiv.org/abs/2112.11431} {arXiv:2112.11431 [gr-qc]} \BibitemShut
  {NoStop}%
\bibitem [{\citenamefont {Salehian}\ \emph {et~al.}(2021)\citenamefont
  {Salehian}, \citenamefont {Gorji}, \citenamefont {Mukohyama},\ and\
  \citenamefont {Firouzjahi}}]{Salehian:2020dsf}%
  \BibitemOpen
  \bibfield  {author} {\bibinfo {author} {\bibfnamefont {B.}~\bibnamefont
  {Salehian}}, \bibinfo {author} {\bibfnamefont {M.~A.}\ \bibnamefont {Gorji}},
  \bibinfo {author} {\bibfnamefont {S.}~\bibnamefont {Mukohyama}},\ and\
  \bibinfo {author} {\bibfnamefont {H.}~\bibnamefont {Firouzjahi}},\ }\bibfield
   {title} {\bibinfo {title} {{Analytic study of dark photon and gravitational
  wave production from axion}},\ }\href
  {https://doi.org/10.1007/JHEP05(2021)043} {\bibfield  {journal} {\bibinfo
  {journal} {JHEP}\ }\textbf {\bibinfo {volume} {05}},\ \bibinfo {pages}
  {043}},\ \Eprint {https://arxiv.org/abs/2007.08148} {arXiv:2007.08148
  [hep-ph]} \BibitemShut {NoStop}%
\bibitem [{\citenamefont {Machado}\ \emph {et~al.}(2019)\citenamefont
  {Machado}, \citenamefont {Ratzinger}, \citenamefont {Schwaller},\ and\
  \citenamefont {Stefanek}}]{Machado:2018nqk}%
  \BibitemOpen
  \bibfield  {author} {\bibinfo {author} {\bibfnamefont {C.~S.}\ \bibnamefont
  {Machado}}, \bibinfo {author} {\bibfnamefont {W.}~\bibnamefont {Ratzinger}},
  \bibinfo {author} {\bibfnamefont {P.}~\bibnamefont {Schwaller}},\ and\
  \bibinfo {author} {\bibfnamefont {B.~A.}\ \bibnamefont {Stefanek}},\
  }\bibfield  {title} {\bibinfo {title} {{Audible Axions}},\ }\href
  {https://doi.org/10.1007/JHEP01(2019)053} {\bibfield  {journal} {\bibinfo
  {journal} {JHEP}\ }\textbf {\bibinfo {volume} {01}},\ \bibinfo {pages}
  {053}},\ \Eprint {https://arxiv.org/abs/1811.01950} {arXiv:1811.01950
  [hep-ph]} \BibitemShut {NoStop}%
\bibitem [{\citenamefont {Bartolo}\ \emph
  {et~al.}(2019{\natexlab{c}})\citenamefont {Bartolo}, \citenamefont {De~Luca},
  \citenamefont {Franciolini}, \citenamefont {Peloso}, \citenamefont {Racco},\
  and\ \citenamefont {Riotto}}]{Bartolo:2018rku}%
  \BibitemOpen
  \bibfield  {author} {\bibinfo {author} {\bibfnamefont {N.}~\bibnamefont
  {Bartolo}}, \bibinfo {author} {\bibfnamefont {V.}~\bibnamefont {De~Luca}},
  \bibinfo {author} {\bibfnamefont {G.}~\bibnamefont {Franciolini}}, \bibinfo
  {author} {\bibfnamefont {M.}~\bibnamefont {Peloso}}, \bibinfo {author}
  {\bibfnamefont {D.}~\bibnamefont {Racco}},\ and\ \bibinfo {author}
  {\bibfnamefont {A.}~\bibnamefont {Riotto}},\ }\bibfield  {title} {\bibinfo
  {title} {{Testing primordial black holes as dark matter with LISA}},\ }\href
  {https://doi.org/10.1103/PhysRevD.99.103521} {\bibfield  {journal} {\bibinfo
  {journal} {Phys. Rev. D}\ }\textbf {\bibinfo {volume} {99}},\ \bibinfo
  {pages} {103521} (\bibinfo {year} {2019}{\natexlab{c}})},\ \Eprint
  {https://arxiv.org/abs/1810.12224} {arXiv:1810.12224 [astro-ph.CO]}
  \BibitemShut {NoStop}%
\bibitem [{\citenamefont {Dai}\ \emph {et~al.}(2015)\citenamefont {Dai},
  \citenamefont {Pajer},\ and\ \citenamefont {Schmidt}}]{Dai:2015rda}%
  \BibitemOpen
  \bibfield  {author} {\bibinfo {author} {\bibfnamefont {L.}~\bibnamefont
  {Dai}}, \bibinfo {author} {\bibfnamefont {E.}~\bibnamefont {Pajer}},\ and\
  \bibinfo {author} {\bibfnamefont {F.}~\bibnamefont {Schmidt}},\ }\bibfield
  {title} {\bibinfo {title} {{Conformal Fermi Coordinates}},\ }\href
  {https://doi.org/10.1088/1475-7516/2015/11/043} {\bibfield  {journal}
  {\bibinfo  {journal} {JCAP}\ }\textbf {\bibinfo {volume} {11}},\ \bibinfo
  {pages} {043}},\ \Eprint {https://arxiv.org/abs/1502.02011} {arXiv:1502.02011
  [gr-qc]} \BibitemShut {NoStop}%
\bibitem [{\citenamefont {Balaji}\ \emph {et~al.}(2022)\citenamefont {Balaji},
  \citenamefont {Domenech},\ and\ \citenamefont {Silk}}]{Balaji:2022dbi}%
  \BibitemOpen
  \bibfield  {author} {\bibinfo {author} {\bibfnamefont {S.}~\bibnamefont
  {Balaji}}, \bibinfo {author} {\bibfnamefont {G.}~\bibnamefont {Domenech}},\
  and\ \bibinfo {author} {\bibfnamefont {J.}~\bibnamefont {Silk}},\ }\bibfield
  {title} {\bibinfo {title} {{Induced gravitational waves from slow-roll
  inflation after an enhancing phase}},\ }\href
  {https://doi.org/10.1088/1475-7516/2022/09/016} {\bibfield  {journal}
  {\bibinfo  {journal} {JCAP}\ }\textbf {\bibinfo {volume} {09}},\ \bibinfo
  {pages} {016}},\ \Eprint {https://arxiv.org/abs/2205.01696} {arXiv:2205.01696
  [astro-ph.CO]} \BibitemShut {NoStop}%
\bibitem [{\citenamefont {{Witkowski}}(2022)}]{2022arXiv220905296W}%
  \BibitemOpen
  \bibfield  {author} {\bibinfo {author} {\bibfnamefont {L.~T.}\ \bibnamefont
  {{Witkowski}}},\ }\bibfield  {title} {\bibinfo {title} {{SIGWfast: a python
  package for the computation of scalar-induced gravitational wave spectra}},\
  }\href {https://doi.org/10.48550/arXiv.2209.05296} {\bibfield  {journal}
  {\bibinfo  {journal} {arXiv e-prints}\ ,\ \bibinfo {eid} {arXiv:2209.05296}}
  (\bibinfo {year} {2022})},\ \Eprint {https://arxiv.org/abs/2209.05296}
  {arXiv:2209.05296 [astro-ph.CO]} \BibitemShut {NoStop}%
\bibitem [{\citenamefont {Salopek}\ and\ \citenamefont
  {Bond}(1990)}]{Salopek:1990jq}%
  \BibitemOpen
  \bibfield  {author} {\bibinfo {author} {\bibfnamefont {D.}~\bibnamefont
  {Salopek}}\ and\ \bibinfo {author} {\bibfnamefont {J.}~\bibnamefont {Bond}},\
  }\bibfield  {title} {\bibinfo {title} {{Nonlinear evolution of long
  wavelength metric fluctuations in inflationary models}},\ }\href
  {https://doi.org/10.1103/PhysRevD.42.3936} {\bibfield  {journal} {\bibinfo
  {journal} {Phys. Rev. D}\ }\textbf {\bibinfo {volume} {42}},\ \bibinfo
  {pages} {3936} (\bibinfo {year} {1990})}\BibitemShut {NoStop}%
\bibitem [{\citenamefont {{Gervois}}\ and\ \citenamefont
  {{Navelet}}(1985)}]{1985JMP....26..633G}%
  \BibitemOpen
  \bibfield  {author} {\bibinfo {author} {\bibfnamefont {A.}~\bibnamefont
  {{Gervois}}}\ and\ \bibinfo {author} {\bibfnamefont {H.}~\bibnamefont
  {{Navelet}}},\ }\bibfield  {title} {\bibinfo {title} {{Integrals of three
  Bessel functions and Legendre functions. I}},\ }\href
  {https://doi.org/10.1063/1.526600} {\bibfield  {journal} {\bibinfo  {journal}
  {Journal of Mathematical Physics}\ }\textbf {\bibinfo {volume} {26}},\
  \bibinfo {pages} {633} (\bibinfo {year} {1985})}\BibitemShut {NoStop}%
\end{thebibliography}%
\end{document}